\documentclass[useAMS,usenatbib, usedcolumn]{mn2e}
\usepackage{amsmath}
\usepackage{wasysym}
\usepackage{graphicx}
\usepackage{aas_macros}
\usepackage{dcolumn}
\usepackage{bm}

\begin{document}
\bibliographystyle{mn2e}
\title[GAMA: The photometric pipeline]
{Galaxy and Mass Assembly (GAMA): FUV, NUV, $ugrizYJHK$ Petrosian, Kron and S\'{e}rsic photometry}

\author[Hill et al.]{David T.~Hill$^{1}$\thanks{E-mail:dth4@st-andrews.ac.uk (DTH)}, Lee S.~Kelvin$^{1}$, Simon P.~Driver$^{1}$, Aaron S.G.~Robotham$^{1}$, \newauthor Ewan~Cameron$^{1,2}$, Nicholas~Cross$^{3}$, Ellen~Andrae$^{4}$, Ivan K.~Baldry$^5$, Steven P.~Bamford$^6$, \newauthor Joss~Bland-Hawthorn$^7$, Sarah~Brough$^8$, Christopher J.~Conselice$^6$, Simon ~Dye$^9$,\newauthor Andrew M.~Hopkins$^8$, Jochen~Liske$^{10}$, Jon~Loveday$^{11}$, Peder~Norberg$^3$, John A.~Peacock$^3$,\newauthor Scott M.~Croom$^7$, Carlos S.~Frenk$^{12}$, Alister W.~Graham$^{13}$, D. Heath~Jones$^8$, \newauthor Konrad~Kuijken$^{14}$, Barry F.~Madore$^{15}$, Robert C.~Nichol$^{16}$, Hannah R.~Parkinson$^3$, \newauthor Steven~Phillipps$^{17}$, Kevin A. Pimbblet$^{18}$, Cristina C.~Popescu$^{19}$, Matthew~Prescott$^5$,\newauthor Mark~Seibert$^{15}$, Rob G.~Sharp$^8$, Will J.~Sutherland$^{20}$, Daniel~Thomas$^{16}$, \newauthor Richard J.~Tuffs$^{4}$, and Elco~van~Kampen$^{10}$.\\
$^1$School of Physics \& Astronomy, University of St Andrews, North Haugh, St Andrews, KY16 9SS, UK; SUPA\\
$^2$Department of Physics, Swiss Federal Institute of Technology (ETH-Z{\" u}rich), 8093 Z{\" u}rich, Switzerland\\
$^3$Institute for Astronomy, University of Edinburgh, Royal Observatory, Blackford Hill, Edinburgh EH9 3HJ, UK; SUPA\\
$^4$Max Planck Institute for Nuclear Physics (MPIK), Saupfercheckweg,1, 69117 Heidelberg, Germany\\
$^5$Astrophysics Research Institute, Liverpool John Moores University, Twelve Quays House, Egerton Wharf, Birkenhead, CH41 1LD, UK\\
$^6$Centre for Astronomy and Particle Theory, University of Nottingham, University Park, Nottingham NG7 2RD, UK\\
$^7$Sydney Institute for Astronomy, School of Physics, University of Sydney, NSW 2006, Australia\\
$^8$Anglo Australian Observatory, PO Box 296, Epping, NSW 1710, Australia\\
$^9$School of Physics and Astronomy, Cardiff University, The Parade, Cardiff CF24 3AA, UK\\ 
$^{10}$European Southern Observatory, Karl-Schwarzschild-Str.~2, 85748, Garching, Germany\\
$^{11}$Astronomy Centre, University of Sussex, Falmer, Brighton BN1 9QH, UK\\
$^{12}$Institute for Computational Cosmology, Department of Physics, Durham University, South Road, Durham DH1 3LE, UK\\
$^{13}$Centre for Astrophysics and Supercomputing, Swinburne University of Technology, Hawthorn, Victoria 3122, Australia\\
$^{14}$Leiden University, P.O.~Box 9500, 2300 RA Leiden, The Netherlands\\
$^{15}$Observatories of the Carnegie Institution of Washington, 813 Santa Barbara Street, Pasadena, CA91101, USA\\
$^{16}$Institute of Cosmology and Gravitation (ICG), University of Portsmouth, Dennis Sciama Building, Burnaby Road, Portsmouth PO1 3FX, UK\\
$^{17}$Astrophysics Group, H.H. Wills Physics Laboratory, University of Bristol, Tyndall Avenue, Bristol, BS8 1TL, UK\\
$^{18}$School of Physics, Monash University, Clayton, VIC 3800, Australia\\
$^{19}$Jeremiah Horrocks Institute, University of Central Lancashire, Preston PR1 2HE, UK\\
$^{20}$Astronomy Unit, Queen Mary University London, Mile End Rd, London E1 4NS, UK
}

\date{Received XXXX; Accepted XXXX}
\pubyear{2010} \volume{000}
\pagerange{\pageref{firstpage}--\pageref{lastpage}}
\maketitle
\label{firstpage}
\begin{abstract}
In order to generate credible $0.1$-$2\mu$m SEDs, the GAMA project requires many Gigabytes of imaging data from a number of instruments to be re-processed into a standard format. In this paper we discuss the software infrastructure we use, and create self-consistent $ugrizYJHK$ photometry for all sources within the GAMA sample. Using UKIDSS and SDSS archive data, we outline the pre-processing necessary to standardise all images to a common zeropoint, the steps taken to correct for seeing bias across the dataset, and the creation of Gigapixel-scale mosaics of the three 4x12 deg GAMA regions in each filter. From these mosaics, we extract source catalogues for the GAMA regions using elliptical Kron and Petrosian matched apertures. We also calculate S\'{e}rsic magnitudes for all galaxies within the GAMA sample using \texttt{SIGMA}, a galaxy component modelling wrapper for \textsc{GALFIT 3}. We compare the resultant photometry directly, and also calculate the $r$ band galaxy LF for all photometric datasets to highlight the uncertainty introduced by the photometric method. We find that (1) Changing the object detection threshold has a minor effect on the best-fitting Schechter parameters of the overall population ($M^{*} \pm 0.055$\,mag, $\alpha \pm 0.014$, $\phi^{*} \pm 0.0005$\,$\rm h^3 \text{} Mpc^{-3}$). (2) An offset between datasets that use Kron or Petrosian photometry regardless of the filter. (3) The decision to use circular or elliptical apertures causes an offset in $M^{*}$ of $0.20$\,mag. (4) The best-fitting Schechter parameters from total-magnitude photometric systems (such as SDSS \textsc{modelmag} or S\'{e}rsic magnitudes) have a steeper faint-end slope than photometry dependent on Kron or Petrosian magnitudes. (5) Our Universe's total luminosity density, when calculated using Kron or Petrosian $r$-band photometry, is underestimated by at least 15\%.
\end{abstract}

\begin{keywords}
galaxies: fundamental parameters --- 
surveys ---
techniques: photometric ---
methods: observational ---
methods: data analysis ---
techniques: image processing
\end{keywords}

\section{Introduction}
When calculating any statistic it is essential that the sample used to generate it is both numerous and without systematic bias. For a number of fundamental parameters in cosmology, for example the galaxy stellar mass function or the total luminosity density, the dataset used will be made up of a large sample of galaxies, and contain a measure of the flux from each galaxy (e.g., \citealt{tex:hill}). Unfortunately, our ability to accurately calculate the flux of any galaxy is imprecise; at some distance from its centre the luminosity of the galaxy will drop into the background noise and the quantification of the missing light beyond that point is problematic with no obviously correct procedure. Even using deep photometry ($\mu_{B}>29$\,mag\,arcsec$^2$), \citet{tex:caon1990} did not reveal the presence of an elliptical galaxy light profile truncation. \\ 
A number of methods to calculate the flux from a galaxy have been proposed. They tend to be either simple and impractical, such as setting the aperture to be a fixed constant size, or limit it using a detection threshold (ignoring the missing light issue completely), or complex and subject to bias, such as using the light distribution of the easily detected part of the object to calculate the size the aperture should be set to \citep{tex:petrosian,tex:kron}, which will return a different fraction of the total light emission depending on whether the galaxy follows an exponential \citep{tex:patterson,tex:exponentialprof} or \citet{tex:devauprof} light profile. \citet{tex:nickbbd} discuss the use of different missing-light estimators and their inherent selection effects. A third option is to attempt to fit a light profile, such as the aforementioned deVaucouleur and exponential profiles (i.e. SDSS model magnitudes, \citealt{tex:sdssedr}), or the more general S\'{e}rsic profile (\citealt{tex:Sersicprofile}, \citealt{tex:grahamdriver}), to the available data, and integrate that profile to infinity to calculate a total-magnitude for the galaxy. \citet{tex:grahampetrosian} investigate the discrepancy between the S\'{e}rsic and SDSS Petrosian magnitudes for different light profiles, providing a simple correction for SDSS data.\\
Unfortunately, no standard, efficacious photometric formula is used in all surveys. If one looks at 3 of the largest photometric surveys, 2MASS, SDSS and UKIDSS, one finds a variety of methods. The 2MASS survey dataset contains Isophotal and Kron circularised, elliptical aperture magnitudes (elliptical apertures with a fixed minimum semi-minor axis), and an elliptical S\'{e}rsic total magnitude. SDSS use two methods for their extended source photometry: \textsc{petromag}, which fits a circular Petrosian aperture to an object, and \textsc{modelmag}, which chooses whether an exponential or deVaucouleur profile is the more accurate fit and returns a magnitude determined by integrating the chosen profile to a specified number of effective radii (profiles are smoothly truncated between 7 and 8 $R_e$ for a deVaucouleur profile, between 3 to 4 $R_e$ for an exponential profile). The \textsc{modelmags} used within this paper specify the profile type in the $r$ band, and use that profile in each passband. UKIDSS catalogues were designed to have multiple methods: again a circular Petrosian magnitude and a 2D S\'{e}rsic magnitude, calculated by fitting the best S\'{e}rsic profile to the source. The 2D S\'{e}rsic magnitude has not yet been implemented. As these surveys then form the basis for photometric calibration of other studies it is important to understand any biases that may be introduced by the photometric method.\\ 
The GAMA survey \citep{tex:gama} is a multi-wavelength ($151.6$\,$\rm{nm} - \sim 6$\,$\rm{m}$), spectroscopic survey of galaxies within three 4 x 12\,deg regions of equatorial sky centred around 9h, 12h and 14.5h (with aspirations to establish further blocks in the SGP). Amongst other legacy goals, the survey team will create a complete, magnitude limited sample of galaxies with redshift and colour information from the FUV to Radio passbands, in order to accurately model the AGN, stellar, dust and gas contents of each individual galaxy. This requires the combination of observations from many surveys, each with different instrument resolutions, observational conditions and detection technologies. As the luminous output of different stellar populations peaks in different parts of the EM spectrum, it is not always a simple task to match an extended source across surveys. SDSS, which covers only a relatively modest wavelength range ($300-900$\,nm), detects objects using a combination of all filters, defines apertures using the $r$ band and then applies them to $ugiz$ observations to negate this problem. This ensures a consistent deblending outcome and accurate colours. The UKIDSS extraction pipeline generates independent detection lists separately in each frame (i.e., for every filter) and merges these lists together for frames that cover the same region of sky (a frame set). Sources are then defined as detections within a certain tolerance. This process is detailed in \citet{tex:wfcam}. Unfortunately, it is susceptible to differing deblending outcomes that may produce less reliable colours. As a key focus of GAMA is the production of optimal SEDs, it is necessary for us to internally standardise the photometry so that is immune to aperture bias from $u$-$K$. The pipeline outlined in this paper is the result of these efforts.\\ Imaging data is taken from UKIDSS DR4/SDSS DR6 observations. In section \ref{sec:surveys} we briefly outline the surveys that acquired the data we use in this work. In section \ref{sec:constr} we describe how we standardised our data and formed image mosaics for each filter/region combination. Section \ref{sec:photomethods} discusses the photometric methods we use, and in sections \ref{sec:sourcecomp} and \ref{sec:catprops} we discuss the source catalogues produced following source extraction on these mosaics. We define the photometry we are using as the GAMA standard in section \ref{sec:gamaphotometry}. Finally, in order to quantify the systematic bias introduced by the choice of photometric method, we present $r$ band luminosity functions, calculated using a series of different photometric methods, in Section \ref{sec:photomimpact}. Throughout we adopt an $h=1, \Omega_{M} = 0.3, \Omega_{\Lambda} = 0.7$ cosmological model. All magnitudes are quoted in the AB system unless otherwise stated. Execution speeds provided are from a run of the pipeline on a 16 processor server built in 2009. As other processes were running simultaneously, processing speed will vary and these parameters should only be taken as approximate timescales.

\section{Survey Data} \label{sec:surveys}
\subsection{GAMA} \label{sec:gamasurvey}
\begin{figure*}
\includegraphics[width=440pt]{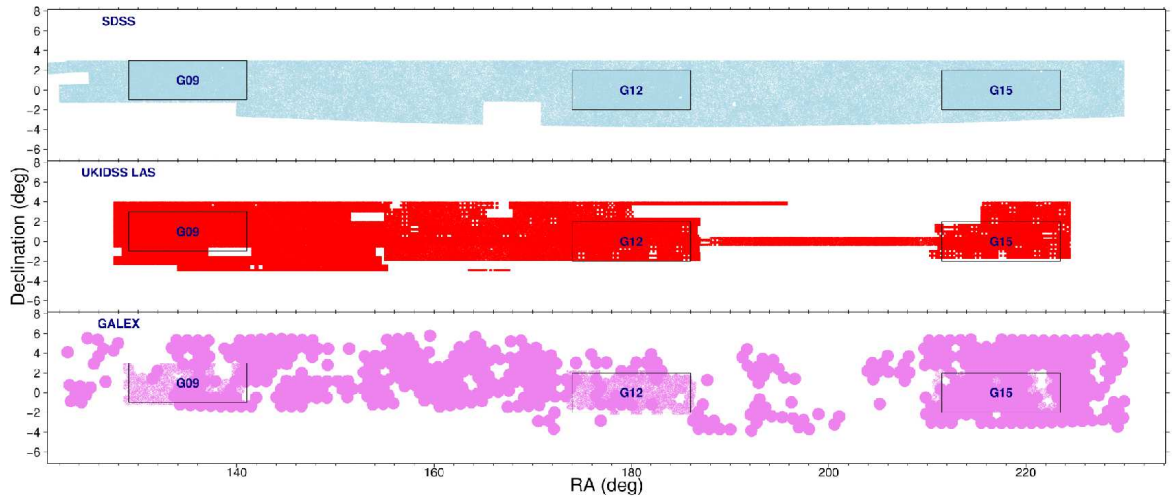}
\caption{Coverage of the equatorial region of sky that contains the GAMA regions, by SDSS DR6 (blue), UKIDSS LAS (red) and GALEX (violet) imaging. The three rectangular boxes contain the regions of sky surveyed by GAMA.}
\label{fig:coverage}
\end{figure*}

The GAMA project \citep{tex:gama} aims to study galaxy formation and evolution using a range of cutting-edge instruments (AAT, VST, VISTA, ASKAP, HERSCHEL WISE, GALEX and GMRT), creating a database of $\sim$350 thousand galaxies observed from UV to radio wavelengths. The first stage of the GAMA project, GAMA I, covers 144\,deg$^{\rm 2}$ of equatorial sky, in three separate $4$\,deg $\times 12$\,deg regions centred at 9h +1d (GAMA9), 12h +0d (GAMA12) and 14h30m +0d (GAMA15). These areas have complete SDSS coverage, and we are in the process of obtaining complete UKIDSS-LAS coverage (Figure \ref{fig:coverage}). Between 2008 and 2010, the GAMA project was allocated 66 nights on the AAT to use the AAOmega spectrograph in order to carry out the GAMA I spectroscopic campaign.\\
A complete description of the input catalogue for the spectroscopic campaign can be found in \citet{tex:gamaic}. To summarise: the aim is to provide spectroscopy of all galaxies in the GAMA I regions brighter than $r_{\rm{petro,SDSS}}=19.4$\,mag, $z_{\rm{model,SDSS}}=18.2$\,mag and $K_{\rm{kron, AB}}=17.5$\,mag, with the sample extended to $r_{\rm{petro,SDSS}}<19.8$\,mag in the GAMA12 region.  Where a galaxy would not be selected by its $r$ magnitude, but would be selected using the $K$ or $z$ magnitude cut, the galaxy must also have $r_{\rm{petro,SDSS}}<20.5$\,mag. This ensures that the galaxy is credible and the likelihood of obtaining a redshift is reasonable. In order to guarantee a complete sample of galaxies, including compact objects, the GAMA input catalogue utilises a star-galaxy selection algorithm that includes optical ($r_{psf}$-$r_{model}$, $g$-$i$) and infrared colour selections ($J$-$K$). The latter uses colours taken from sources extracted using this pipeline.\\ 
The 2008 and 2009 observations make up a sample of 100051 reliable redshifts, of which 82696 come from the AAOmega spectrograph. The tiling strategy used to allocate objects to AAOmega fibres is detailed in \citet{tex:gamatiling}. A breakdown of redshift completeness by luminosity and colour selection of the year 2 observations is shown in Table 5 of \citeauthor{tex:gamaic}, and in Table 3 of the same paper there is a list of spectra used from external surveys.
\subsection{SDSS}
The Sloan Digital Sky Survey (SDSS, \citealt{tex:sdssyork}) is the largest combined photometric and spectroscopic survey ever undertaken, and contains spectra of 930\,000 galaxies spread over $8423$\,square degrees of sky, with imaging in five filters with effective wavelengths between $300$ and $1000$\,nm ($ugriz$). SDSS data has been publicly released in a series of 7 data releases. \citet{tex:sdssdr2} state that SDSS imaging is 95\% complete to $u=22.0$\,mag, $g=22.2$\,mag, $r=22.2$\,mag, $i=21.3$\,mag, and $z=20.5$\,mag (all depths measured for point sources in typical seeing using the SDSS photometric system, which is comparable in all bands to the AB system $\pm 0.05$\,mag).\\ 
Images are taken using an imaging array of 30 $2048$x$2048$ Tektronix CCDs with a pixel size of $0.396$\,arcsec, but only on nights where the seeing is $<1.5$\,arcsec and there is less than $1\%$ uncertainty in the zeropoint. When such conditions are not reached, spectroscopy is attempted instead. SDSS catalogues can be accessed through the SDSS Catalog Archive Server (\textit{CAS}), and imaging data through the Data Archive Server (\textit{DAS}). \\
Astrometry for SDSS-DR6 \citep{tex:sdssastrom} is undertaken by comparing $r$ band observations to the USNO CCD Astrograph Catalog (UCAC, \citealt{tex:ucac}), where it had coverage at time of release, or Tycho-2 \citep{tex:tycho}, when UCAC did not have coverage. For sources brighter than $r=20$\,mag, the astrometric accuracy when comparing to UCAC is $45$\,mas, and when comparing to Tycho-2 is $75$\,mas. In both cases, there is a further $30$\,mas systematic error, and a relative error between filters (i.e., in $ugiz$) of $25$-$35$\,mas.\\
The GAMA input catalogue is defined using data from the sixth data release catalogue (SDSS DR6, \citealt{tex:sdssdr6}). The GAMA regions fall within the SDSS DR6 area of coverage, in SDSS stripes 9 to 12. 

\subsection{UKIDSS}
UKIDSS \citep{tex:ukirt} is a seven year near-infrared (NIR) survey programme that will cover several thousand degrees of sky. The programme utilises the Wide Field Camera (WFCAM) on the 3.8\,m United Kingdom Infra-Red Telescope (UKIRT). The UKIDSS program consists of five separate surveys, each probing to a different depth and for a different scientific purpose. One of these surveys, the UKIDSS Large Area Survey (LAS) will cover 4000\,deg$^{\rm 2}$ and will overlap with the SDSS stripes 9 to 16, 26 to 33 and part of stripe 82. As the GAMA survey regions are within SDSS stripes 9 to 12, the LAS survey will provide high density near-IR photometric coverage over the entire GAMA area. The UKIDSS-LAS survey observes to a far greater depth ($K_{LAS}=18.2$\,mag using the Vega magnitude system) than the previous Two-Micron All-Sky Survey (2MASS, $K_{2MASS} = 15.50$\,mag using the Vega magnitude system).\\
When complete, the LAS has target depths of $K=18.2$\,mag, $H=18.6$\,mag, $J=19.9$\,mag (after two passes; this paper uses only the first $J$ pass which is complete to $19.5$\,mag) and $Y=20.3$\,mag (all depths measured use the Vega system for a point source with 5$\sigma$ detection within a 2\,arcsec aperture). Currently, observations have been conducted in the equatorial regions, and will soon cover large swathes of the Northern Sky. It is designed to have a seeing FWHM of $<1.2$\,arcsec, photometric rms uncertainty of $<\pm0.02$\,mag and astrometric rms of $<\pm0.1$\,arcsec. Each position on the sky will be viewed for 40s per pass. All survey data for this paper is taken from the fourth data release (DR4). \\
The WFCAM Science Archive (WSA, \citealt{tex:wfcam}) is the storage facility for post-pipeline, calibrated UKIDSS data. It provides users with access to fits images and CASU-generated object catalogues for all five UKIDSS surveys. We do not use the CASU generated catalogues for a few reasons. Firstly, the CASU catalogues for early UKIDSS data releases suffer from a fault where deblended objects are significantly brighter than their parent object, in some cases by several magnitudes \citep{tex:ajsmith,tex:hill}. Secondly, the CASU catalogues only contain circular aperture fluxes. Thirdly, CASU decisions (e.g., deblending and aperture sizes) are not consistent between filters. For instance, the aperture radius and centre used to calculate \textsc{kpetromag} of a source is not necessarily the same as the aperture radius and centre used to calculate \textsc{ypetromag} or \textsc{hpetromag}. We require accurate extended-aperture colours; the CASU catalogues do not provide this.

\section{Construction of the mosaics from SDSS and UKIDSS imaging}  \label{sec:constr}
\begin{figure}
\includegraphics[width=220pt]{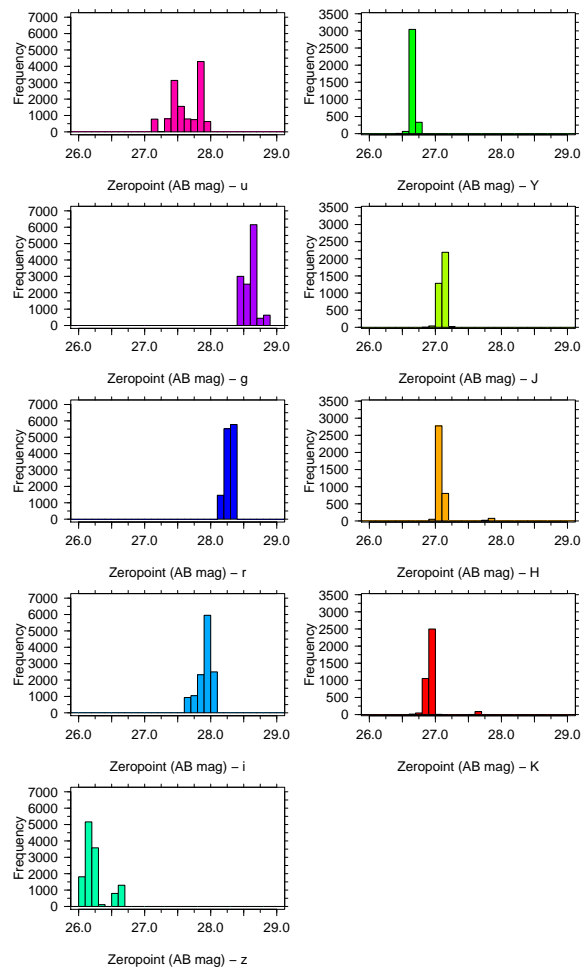}
\caption{A histogram of the calculated total zeropoints for the fields used to create our master region mosaics.}
\label{fig:zp}
\end{figure}

One of GAMA's priorities is the accurate measurement of SEDs from as broad a wavelength range as possible. This is non-trivial when combining data from multiple surveys. While each survey may be internally consistent with data collected contemporaneously, conditions between surveys can vary. In particular, seeing and zeropoint parameters may greatly differ between frames. When matching between surveys one may find an object in the centre of the frame in one survey is split across two frames in another survey. There may also be variation in the angular scale of a pixel between different instruments, and even when two instruments have the same pixel size, a shift of half of a pixel between two frames can cause significant difficulties in calculating colours for small, low surface brightness objects. Furthermore, in order to use \texttt{SExtractor} in dual frame mode, the source-detection and the observation frame pixels must be calibrated to the same world coordinate system. In the GAMA survey, we have attempted to circumvent these difficulties by creating Gigapixel scale mosaics with a common zeropoint and consistent WCS calibration. The construction process is outlined within this section.\\ 
To generate our image mosaics, we use the Terapix \texttt{SWARP} \citep{tex:swarp} utility. This is a mosaic generation tool, and how we utilise it is described in subsection \ref{sec:mosaic}. Before we can use this software, however, it is necessary for us to normalise the contributing SDSS and UKIDSS data to take into account differences in sky conditions and exposure times between observations. For every file we must identify its current zeropoint (see the distribution in Figure \ref{fig:zp}), and transform it to a defined standard. This process is described in subsections \ref{sec:ukren} and \ref{sec:sdren}.

\subsection{UKIDSS: Acquisition of data and renormalisation to a common zeropoint}\label{sec:ukren}
UKIDSS imaging is stored within the WFCAM Science Archive (WSA). We downloaded 862 $Y$, 883 $J$, 931$H$ and 928 $K$ band compressed UKIDSS-LAS fits files that contained images of sky from the GAMA regions. These files were decompressed using the \texttt{imcopy} utility. The files for each band are stored and treated separately.\\ 
A specially designed pipeline accesses each file, reads the \textit{MAGZPT} ($ZP_{\rm mag}$), \textit{EXP\_TIME} ($t$), airmass ($0.5*(AMSTART+AMEND)=\rm{sec} \chi_{\rm mean}$) and \textit{EXTINCT} (Ext) keywords from the fits header and creates a total AB magnitude zeropoint for the file using Equation \ref{eqn:zpukidss}.\\
\begin{equation}\label{eqn:zpukidss}
ZP_{\rm total} = ZP_{\rm mag} -2.5 \rm{log}(\frac{1}{t}) - \rm{Ext} \times(\rm{sec} \chi_{\rm{mean}}-1) + ABV_{X}
\end{equation}
where $ABV_{X}$ is the AB magnitude of Vega in the $X$ band (Table \ref{tab:ABoffsets}).\\ 
To correct each frame to a standard zeropoint ($30$), the value of each pixel is multiplied by a factor, calculated using Equation \ref{eqn:pixmod}. Whilst we show the distribution of frame zeropoints in Figure \ref{fig:zp} in bins of 0.1\,mag, we use the actual zeropoint of each frame to calculate the total AB magnitude zeropoint. This has a far smaller variation (e.g., $0.02$\,mag in photometric conditions in the $JHK$ filters; \citealt{tex:ukidssdr1}).\\ 
\begin{equation}\label{eqn:pixmod}
\rm{pixelmodifier}=10^{-0.4(ZP_{\rm total}-30)}
\end{equation} 
A new file is created to store the corrected pixel table, and the \textit{MAGZPT} fits header parameter is updated. The \textit{SKYLEVEL} and \textit{SKYNOISE} parameters are then scaled using the same multiplying factor. This process takes 3 seconds per file.

\begin{table}
\begin{center}

\begin{tabular}{cc} \hline \hline
Band& AB offset (mag)\\ \hline
u&-0.04\\
g&0\\
r&0\\
i&0\\
z&+0.02\\
Y&0.634\\
J&0.938\\
H&1.379\\
K&1.900\\
\hline
\end{tabular}
\end{center}
\caption{Conversion to AB magnitudes. The SDSS photometric system is roughly equivalent to the AB magnitude system, with only small offsets in the $u$ and $z$ passbands. UKIDSS photometry is calculated on the Vega magnitude system, and our conversions are from \citet{tex:hewettpassband}. Whilst we convert UKIDSS data using a high precision parameter, it should be noted that the conversion uncertainty is only known to $\sim \pm 0.02$\,mag \citep{tex:cohen}. }
\label{tab:ABoffsets}
\end{table}

\subsection{SDSS: Acquisition of data and renormalisation to a common zeropoint} \label{sec:sdren}
The tsField and fpC files for the 12757 SDSS fields that cover the GAMA regions were downloaded from the SDSS data archive server (\textit{das.sdss.org}) for all five passbands. Again, the files for each passband are stored and treated separately.\\

We use a specially designed pipeline that brings in the \textit{aa} (zeropoint), \textit{kk} (extinction coefficient) and \textit{airmass} keywords from a field's tsField file, and the \textit{EXPTIME} ($t$) keyword from the same field's fpC file. Combining these using Equation \ref{eqn:zpsdss} we calculate the current total AB magnitude zeropoint of the field ($ZP_{\rm total}$).
\begin{equation}\label{eqn:zpsdss}
ZP_{\rm total} = -aa - 2.5 \rm{log}(1/t) - kk \times \rm{airmass} + sAo
\end{equation} 
where $sAo$ is the offset between the SDSS magnitude system and the actual AB magnitude system ($-0.04$\,mag for $u$, $0.02$\,mag for $z$, otherwise 0). The SDSS photometric zeropoint uncertainty is estimated to be no larger than $0.03$\,mag in any band \citep{tex:sdssphotoquality}. We calculate the multiplier required to transform every pixel in the field (again using Equation \ref{eqn:pixmod}) to a standard zeropoint ($30$). As every pixel must be modified by the same factor, we utilise the \texttt{fcarith} program (part of the \texttt{Ftools} package), to multiply the entire image by pixelmodifier. \texttt{fcarith} can normalise an SDSS image every 1.25 seconds. 

\begin{figure}
\includegraphics[width=220pt]{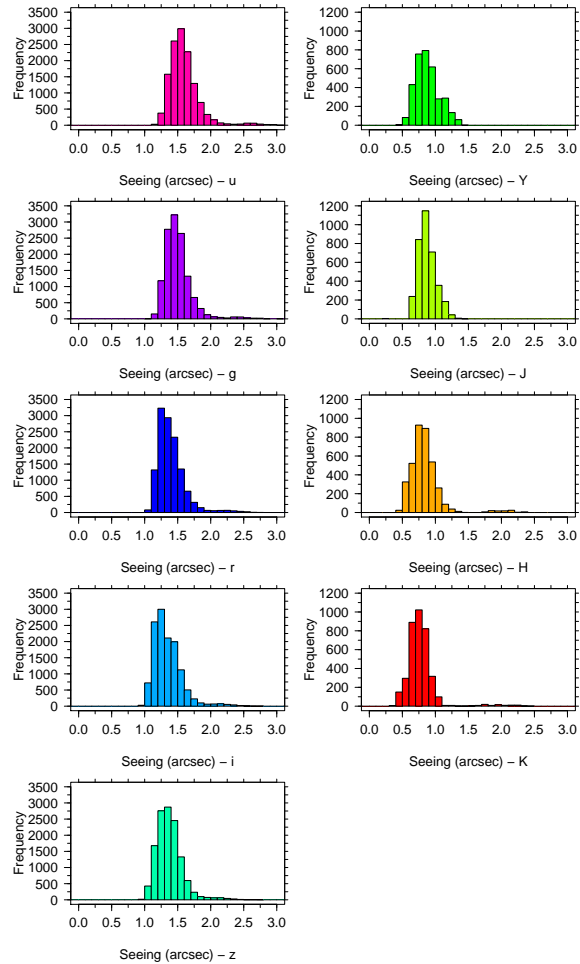}
\caption{A histogram of the calculated seeing of the fields used to create our master region mosaics.}
\label{fig:seeing}
\end{figure}

\subsection{Correction of seeing bias}\label{sec:seeingconv}
As observations were taken in different conditions there is an intrinsic seeing bias between different input images, and between different filters (Figure \ref{fig:seeing}). This could cause inaccuracies in photometric colour measurements that use apertures defined in one filter to derive magnitudes in a second filter. To rectify this problem, it is necessary for us to degrade the better quality images to a lower seeing. However, if we degrade all images to our lowest quality seeing ($3.12$\,arcsec), we should lose the ability to resolve the smallest galaxies in our sample. Therefore, we elect to degrade our normalised images to $2$\,arcsec seeing. The fraction of images with seeing worse than $2$\,arcsec is $4.4$\%, $2.7$\%, $2.5$\%, $2.1$\%, $1.7$\%, $0$\%, $0$\%, $1.3$\% and $0.9$\% in $u$, $g$, $r$, $i$, $z$, $Y$, $J$, $H$ and $K$, respectively. Images with worse seeing than $2$\,arcsec are included in our degraded seeing mosaics. We do not attempt to modify their seeing. Although each survey uses a different method of calculating the seeing within their data (SDSS use a double gaussian to model their PSF, UKIDSS use the average FWHM of the stellar sources within the image), we assume that the seeing provided for every frame is correct.\\
To achieve a final PSF FWHM of $2$\,arcsec ($\sigma_{final}$) we assume that the seeing within an image follows a perfect Gaussian distribution, $\sigma_{initial}$. Theoretically, a Gaussian distribution can be generated from the convolution of two Gaussian distributions. The \texttt{fgauss} utility (also part of the \texttt{ftools} package) can be used to convolve an input image with a circular Gaussian with a definable standard deviation ($\sigma_{req}$), calculated using Equation \ref{eqn:gaussiananswer}.\\

\begin{equation} \label{eqn:gaussiananswer}
\sigma_{req} = \sqrt{ \sigma_{final}^2 - \sigma_{initial}^2}\\
\end{equation}

As each UKIDSS frame has a different seeing value, it is necessary to break each fits file into its four constituent images. This is not necessary for SDSS images (which are stored in separate files). However, it is necessary to retrieve the SDSS image seeing from the image's tsField file. The SDSS image seeing is stored in the \textit{psf\_width} column of the tsField file. Where an image has a seeing better than our specified value, we use the \texttt{fgauss} utility to convolve our image down to our specified value. Where an image has a seeing worse than our specified value, we copy it without modification using the \texttt{imcopy} utility. Both utilities produce a set of UKIDSS files containing two HDUs: the original instrument header HDU and a single image HDU with seeing greater than or equal to our specified seeing. The output SDSS files contain just a single image HDU. This process takes approximated 2 seconds per frame.

\subsection{Creation of master region images}\label{sec:mosaic}
\begin{figure}
\includegraphics[width=250pt]{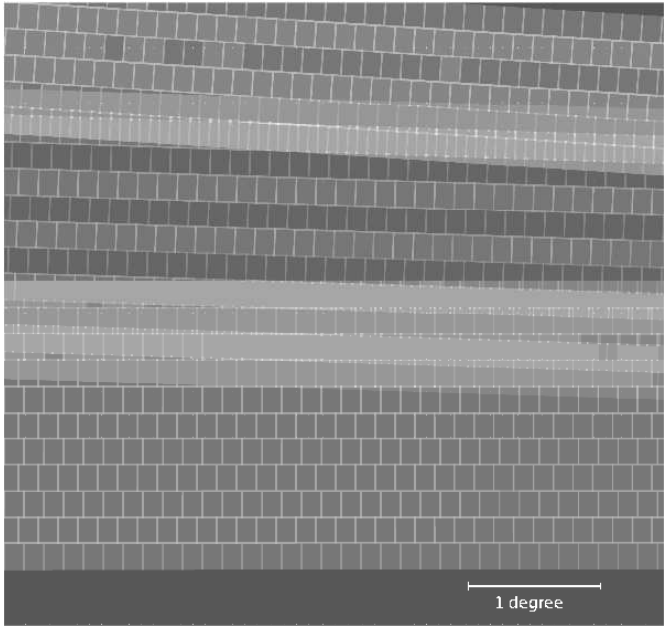}
\caption{The $r$ band weightmap of the 45000x45000 pixel subset region (5x5 deg; defined in Section \ref{sec:sourcecomp}). Joins and overlap between frames are apparent (light grey). The mosaic does not have imaging of the top right corner or the bottom section (dark grey). These areas lie outside the region of interest as the mosaics are slightly larger in Declination than the GAMA regions themselves.}
\label{fig:weightmap}
\end{figure}

\begin{figure}
\includegraphics[width=250pt]{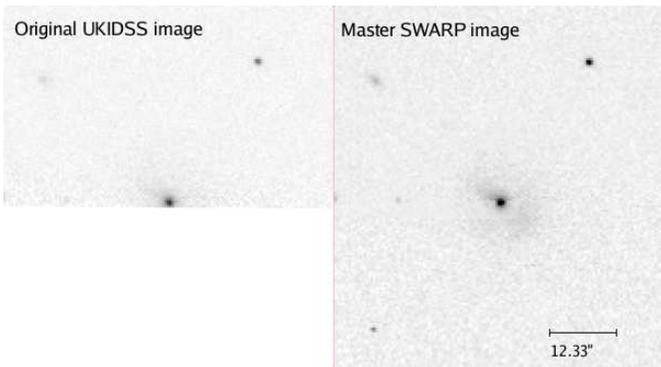}
\caption{A comparison between the original normalised image and the $K$ band mosaic image of a galaxy on the bottom edge of an input UKIDSS frame. The bottom section of the galaxy is not part of this image and it has been stitched together on the mosaic using \texttt{SWARP}.}
\label{fig:joinedgalaxy}
\end{figure}

The \texttt{SWARP} utility is a multi-thread capable co-addition and image-warping tool that is part of the Terapix pipeline \citep{tex:swarp}. We use \texttt{SWARP} to generate complete images of the GAMA regions from the normalised LAS/SDSS fits files. It is vital that the pixel size and area of coverage is the same for each filter, as \texttt{SExtractor}'s dual-image mode requires perfectly matched frames. We define a pixel scale of $0.4 \times 0.4$\,arcsec, and generate $117000 \times 45000$ pixel files centred around 09h00m00.0s, +01d00$'$00.0$''$ (GAMA 9), 12h00m00.0s, +00d00$'$00.0$''$ (GAMA 12), and 14h30m00.0s, +00d00$'$00.0$''$ (GAMA 15). \texttt{SWARP} is set to resample input frames using the default LANCZOS3 algorithm, which the Terapix team found was the optimal option when working with images from the Megacam instrument \citep{tex:swarp}.\\
\texttt{SWARP} produces mosaics that use the TAN WCS projection system. As UKIDSS images are stored in the ZPN projection system, \texttt{SWARP} internally converts the frames to the TAN projection system. There is also an astrometric distortion present in the UKIDSS images that \texttt{SWARP} corrects using the \textit{pv2\_3} and \textit{pv2\_1} fits header parameters\footnote{An analysis of the astrometric distortion can be found in CASU document VDF-TRE-IOA-00009-0002 , currently available from http://www.ast.cam.ac.uk/vdfs/docs/reports/astrom/index.html}.\\ 
\texttt{SWARP} is set to subtract the background from the image, using a background mesh of $256 \times 256$\,pixels ($102 \times 102$\,arcsec) and a back filter size of $3 \times 3$ to calculate the background map. The background calculation follows the same algorithm as \texttt{SExtractor} \citep{tex:sex}. To summarise: it is a bicubic-spline interpolation between the meshes, with a median filter applied to remove bright stars and artifacts.\\ 
Every mosaic contains pixels that are covered by multiple input frames. \texttt{SWARP} is set to use the median pixel value when a number of images overlap. The effects of outlying pixel values, due to cosmic rays, bad pixels or CCD edges, should therefore be reduced. \texttt{SWARP} generates a weight map (Figure \ref{fig:weightmap}) that contains the flux variance in every pixel, calibrated using the background map described above. As the flux variance is affected by overlapping coverage, it is possible to see the survey footprint in the weight map. The weight map can be used within \texttt{SExtractor} to compensate for variations in noise. We do not use it when calculating our photometry for two reasons. Firstly, there is overlap between SDSS fpC frames. This overlap is not from observations, but from the method used to cut the long SDSS stripes into sections. \texttt{SWARP} would not account for this, and the weighting of the overlap regions on the optical mosaics would be calculated incorrectly. Secondly, using the weight maps would alter the effect of mosaic surface brightness limit variations upon our output catalogues. As we intend to model surface brightness effects later, we elect to use an unweighted photometric catalogue.\\ 
A small number of objects will be split between input frames. \texttt{SWARP} can reconstruct them, with only small defects due to CCD edges. One such example is shown in Figure \ref{fig:joinedgalaxy}. We create both seeing-corrected and uncorrected mosaics for each passband and region combination. Each file is 20Gb in size. In total, the mosaics require just over 1 Terabyte of storage space. Each mosaic takes approximately 4 hours to create.\\

\section{Photometry} \label{sec:photomethods}
A major problem with constructing multi-wavelength catalogues is that the definition of what constitutes an object can change across the wavelength range (see Appendix \ref{app:var}, particularly Figure \ref{fig:objvar}). This can be due to internal structure such as dust lanes or star forming regions becoming brighter or fainter in different passbands, causing the extraction software to deblend an object into a number of smaller parts in one filter but not in another. This can lead to large errors in the resulting colours. We cannot be certain that the SDSS object extraction process would produce the same results as the extraction process we use to create our UKIDSS object catalogues. Seeing, deblending and aperture sizes will differ, compromising colours. To create a consistent multi-wavelength sample, the photometry needs to be recalculated consistently across all 9 filters. At the same time we can move from the circular apertures of SDSS and UKIDSS to full elliptical apertures, as well as investigate a variety of photometric methods. To generate our source catalogues we use the \texttt{SExtractor} software \citep{tex:sex}. This is an object extraction utility, and its use is described in subsection \ref{sec:objext}. \\
In this paper we implement four different methods to define our object positions and apertures. We produce three \texttt{SExtractor} catalogues \citep{tex:sex} and one S\'{e}rsic catalogue (based upon \texttt{GALFIT 3}, \citealt{tex:peng}), in addition to the original SDSS dataset. The generation of the three new \texttt{SExtractor} catalogues is detailed in section \ref{sec:objext}. Each of the new \texttt{SExtractor} catalogues contain magnitudes calculated using two different elliptical, extended source apertures: the Kron and the Petrosian magnitude systems. They are briefly described in sections \ref{sec:selfdefap}, \ref{sec:rdefap} and \ref{sec:Kdefap}. We also use a specially designed pipeline (\texttt{SIGMA GAMA}, \citealt{tex:kelvin}, based upon \texttt{GALFIT 3}) to calculate a total magnitude for each galaxy via its best fitting S\'{e}rsic profile. This aperture system, and the process used to generate it, is described in Sections \ref{sec:Sersicap} and \ref{sec:objextSersic}.\\
It is not obvious which photometric method will produce the optimal solution. Whilst the S\'{e}rsic photometric method solves the missing light problem, it requires higher quality data to calculate the set of parameters that best model the galactic light profile. The Kron and Petrosian magnitude systems will work with lower quality data, but may underestimate the flux produced by a galaxy. In this section we describe the photometric systems that we have used. Later, in sections \ref{sec:catprops} and \ref{sec:photomimpact}, we will examine the different results produced by the choice of the photometric system. 

\subsection{Self-defined Kron and Petrosian apertures} \label{sec:selfdefap}
We construct an \textit{independent} catalogue for each filter, containing elliptical Kron and Petrosian apertures. These independent catalogues are then matched across all 9 wavebands using \texttt{STILTS} (see section \ref{sec:catmat} and \citealt{tex:stilts}). The apertures will therefore vary in size, potentially giving inconsistent colours, and as deblending decisions will also change, inconsistent matching between catalogues may occur. However, as the apertures are defined from the image they are used on, there can be no problem with magnitudes being calculated for objects that do not exist, or with missing objects that are not visible in the $r$ band.\\
The self-defined catalogues are generated from the basic mosaics, where no attempt to define a common seeing for the mosaic has been made. This method should generate the optimal list of sources in each band; however, as the precise definition of the source will vary with wavelength, the colours generated using this method will be inaccurate and subject to aperture bias. As the mosaic has variations in seeing, the PSF will also vary across the image.

\subsection{$r$ band-defined Kron and Petrosian apertures} \label{sec:rdefap}
We use \texttt{SExtractor} to define a sample of sources in the $r$ band image, and then use the $r$ band position and aperture information to calculate their luminosity within each filter (using the \texttt{SExtractor} dual image mode). As the aperture definitions do not vary between wavebands this method gives internally consistent colours, and as the list itself does not change source matching between filters is unnecessary. However, where an object has changed in size (see Appendix \ref{app:var}), does not exist (e.g. an artifact in the $r$ band sample) or when the $r$ band aperture definition incorrectly includes multiple objects the output colours may be compromised. Any object that is too faint to be visible within the $r$ band mosaic will also not be detected using this method. However, such objects will be fainter than the GAMA sample's selection criteria, and would not be included within our sample. The $r$ band-defined catalogues are generated from our seeing-degraded mosaics. They provide us with an optically-defined source sample.\\
This method is analogous to the SDSS source catalogues, which define their apertures using the $r$ passband data (unless the object is not detected in $r$, in which case a different filter is chosen). However, the GAMA photometric pipeline has a broader wavelength range as it now includes NIR measurements from the same aperture definition. Furthermore, the SDSS Petrosian magnitudes have not been seeing-standardised. While all data is taken at the same time, the diffraction limit is wavelength dependent and different fractions of light will be missed despite the use of a fixed aperture. SDSS model magnitudes do account for the effects of the PSF.

\subsection{$K$ band-defined Kron and Petrosian apertures} \label{sec:Kdefap}
This method works in the same way as the previous method, but uses the $K$ band image as the detection frame rather than the $r$ band image. We are limited in the total area, as the $K$ band coverage is currently incomplete. However, for samples that require complete colour coverage in all 9 filters, this is not a problem. As with $r$ band-defined catalogues, the $K$ band-defined catalogues are generated from the seeing-corrected mosaics. They provide us with a NIR-defined source sample. The $K$-band defined Kron magnitudes were used in the GAMA input catalogue \citep{tex:gamaic} to calculate the star-galaxy separation $J-K$ colour and the $K$ band target selection.

\subsection{S\'{e}rsic magnitudes} \label{sec:Sersicap}
We use the \texttt{SIGMA} modelling wrapper (see section \ref{sec:objextSersic} and \citealt{tex:kelvin} for more details) which in turn uses the galaxy fitting software \texttt{GALFIT} 3.0 \citep{tex:peng} to fit a single-S\'{e}rsic component to each object independently in 9 filters ($ugrizYJHK$), and recover their S\'{e}rsic magnitudes, indices, effective radii, position angles and ellipticities. Source positions are initially taken from the GAMA input catalogue, as defined in \citet{tex:gamaic}. All S\'{e}rsic magnitudes are self-defined; as each band is modelled independently of the others, the aperture definition will vary and colour may therefore be compromised.\\
Single-S\'{e}rsic fitting is comparable to the SDSS model magnitudes. \texttt{SIGMA} therefore should recover total fluxes for objects that have a S\'{e}rsic index in the range $0.3<n<20$, where model magnitudes force a fit to either an exponential (n=1) or deVaucouleurs (n=4) profile. The systematic magnitude errors that arise when model magnitudes are fit to galaxies that do not follow an exponential or deVaucouleurs profile \citep{tex:graham2001, tex:brown2003} do not occur in \texttt{SIGMA}. The SDSS team developed a composite magnitude system, \textsc{cmodel}, that calculates a magnitude from the combination of the n=1 and n=4 systems, in order to circumvent this issue \citep{tex:sdssdr2}. We compare our S\'{e}rsic magnitudes to their results later. S\'{e}rsic magnitudes do not suffer from the missing-flux issue that affects Kron and Petrosian apertures. Petrosian magnitudes may underestimate a galaxy's luminosity by 0.2\,mag \citep{tex:strauss}, while under certain conditions a Kron aperture may only recover half of a galaxy's total luminosity \citep{tex:andreon}. The S\'{e}rsic catalogues are generated from the seeing-uncorrected mosaics, as the seeing parameters are modelled within \texttt{SIGMA} using the \texttt{PSFEx} software utility (E. Bertin, priv. comm).

\subsection{Object Extraction of Kron and Petrosian apertures} \label{sec:objext}
\begin{figure}
\includegraphics[width=250pt]{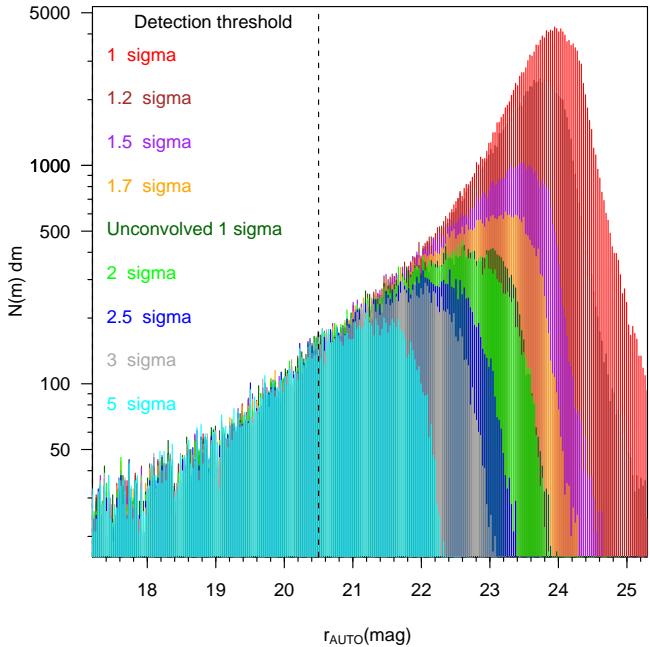}
\caption{The effects of changing the \texttt{SExtractor} \textit{DETECT\_THRESH} parameter on a subset of an $r$ band mosaic. The dotted black line is the deepest $r$ band sample limit of the GAMA survey (for those objects that are $K$ or $z$ selected).}
\label{fig:detectthresh}
\end{figure}

The \texttt{SExtractor} utility \citep{tex:sex} is a program that generates catalogues of source positions and aperture fluxes from an input image. It has the capacity to define the sources and apertures in one frame and calculate the corresponding fluxes in a second frame. This dual image mode is computationally more intensive than the standard \texttt{SExtractor} single image mode (in single image mode, \texttt{SExtractor} can extract a catalogue from a mosaic within a few hours; dual image mode takes a few days per mosaic). Using the $u$, $g$, $r$, $i$, $z$, $Y$, $J$, $H$ and $K$ images created by the \texttt{SWARP} utility, we define our catalogue of sources independently (for the self-defined catalogues), using the $r$ band mosaics (for the $r$ band-defined catalogue) or the $K$ band mosaics (for the $K$ band-defined catalogue) and calculate their flux in all nine bands. The normalisation and \texttt{SWARP} processes removed the image background and standardised the zeropoint; we therefore use a constant \textit{MAG\_ZEROPOINT=30}, and \textit{BACK\_VALUE=0}. \texttt{SExtractor} generates both elliptical Petrosian (2.0 $R_{\rm Petro}$) and Kron-like apertures (2.5 $R_{\rm Kron}$, called \textit{AUTO} magnitudes). SExtractor Petrosian magnitudes are computed using $\frac{1}{\nu_{R{\rm Petro}}}=0.2$, the same parameter as SDSS. As the mosaics have been transformed onto the AB magnitude system, all magnitudes generated by the GAMA photometric pipeline are AB magnitudes.\\
The seeing convolution routine smooths out the background and correlates the read noise of the images (this is apparent in Figure.~\ref{fig:objvar}). As \texttt{SExtractor} detects objects of $> x \sigma$ above the background (where $x$ is a definable parameter, set to $1$ in the default file and for our seeing unconvolved catalogues), this assists the detection process, allowing \texttt{SExtractor} to find objects to a much greater depth, thus increasing the number of sources extracted using the standard setup. However, these new objects are generally much fainter than the photometric limits of the GAMA spectroscopic campaign, many are false detections, and the time required to generate the source catalogues (particularly using \texttt{SExtractor} in dual image mode) is prohibitively large. Using a $10000$x$10000$ pixel subset of the GAMA9 $r$ band mosaic, we have attempted to calculate the \textit{DETECT\_THRESH} parameter that would output a catalogue of approximately the same depth and size as the unconvolved catalogue within our spectroscopic limits (see section \ref{sec:gamasurvey}). The distribution of objects with different \textit{DETECT\_THRESH} sigma parameters, compared to the unconvolved catalogue, is shown in Figure \ref{fig:detectthresh}. We use a \textit{DETECT\_MINAREA} of 9 pixels. As the unconvolved catalogue is slightly deeper than the $2 \sigma$, but not as deep as the $1.7 \sigma$ convolved catalogue, we use a \textit{DETECT\_THRESH} parameter of $1.7 \sigma$ to generate our convolved catalogues. The $1.7 \sigma$ and $1 \sigma$ catalogues have consistent number counts to $r_{\rm{auto}} = 21$\,mag; half a magnitude beyond the $r_{\rm{SDSS}}$ band magnitude limit of the GAMA input catalogue. 

\subsection{Object extraction for S\'{e}rsic magnitudes} \label{sec:objextSersic}
S\'{e}rsic magnitudes are obtained as an output from the galaxy modelling program \texttt{SIGMA} (Structural Investigation of Galaxies via Model Analysis) written in the \textsc{R} programming language \citep{tex:kelvin}.  In brief, \texttt{SIGMA} takes the Right Ascension and Declination of an object that has passed our star-galaxy separation criteria and calculates its pixel position within the appropriate mosaic.  A square region, centred on the object, is cut out from the mosaic containing a minimum of 20 guide stars with which to generate a PSF.  \texttt{SExtractor} then provides a FITS-standard input catalogue to \texttt{PSF Extractor} (E. Bertin, priv. comm.) which generates an empirical PSF for each image.  Ellipticities and position angles are obtained from the \textsc{STSDAS Ellipse} routine within \texttt{IRAF}, and provides an input to \texttt{Galfit}.  The larger cutout is again cut down to a region which contains 90\% of the target object's flux plus a buffer of 10 pixels, and will only deviate from this size if a bright nearby object causes the fitting region to be expanded in order to model any satellites the target may have.\\
\textsc{GALFIT 3} is then used to fit a single-S\'{e}rsic component to each target, and several runs may be attempted if, for example, the previous run crashed, the code reached its maximum number of iterations, the centre has migrated to fit a separate object, the effective radius is too high or low or the S\'{e}rsic index is too large.  \texttt{SIGMA} employs a complex event handler in order to run the code as many times as necessary to fix these problems, however not all problems can be fixed, and so residual quality flags remain to reflect the quality of the final fit. The \texttt{SIGMA} package takes approximately 10 seconds per object. For full details of the \texttt{SIGMA} modelling program, see \citet{tex:kelvin}. 

\subsection{Catalogue matching} \label{sec:catmat}
The definition of the GAMA spectroscopic target selection (herein referred to as the tiling catalogue) is detailed in \citet{tex:gamaic}, and is based on original SDSS DR6 data. We therefore need to relate our revised photometry back to this catalogue in order to connect it to our AAOmega spectra. The tiling catalogue utilises a mask around bright stars that should remove most objects with bad photometry and erroneously bright magnitudes, as well as implementing a revised star-galaxy separation quantified against our spectroscopic results. It has been extensively tested, with sources that are likely to be artifacts, bad deblends or sections of larger galaxies viewed a number of times by different people. By matching our catalogues to the tiling catalogue, we can access the results of this rigorous filtering process, and generate a full, self-consistent set of colours for all of the objects that are within the GAMA sample (and are within regions that have been observed in all nine passbands). As the tiling catalogue is also used when redshift targeting, we will be able to calculate the completeness in all the passbands of the GAMA survey. The GAMA tiling catalogue is a subset of the GAMA master catalogue (herein referred to as the master catalogue). The master catalogue is created using the SDSS DR6 catalogue stored on the CAS\footnote{We use the query SELECT * FROM dr6.PhotoObj as p WHERE ( p.modelmag\_r - p.extinction\_r $<$ 20.5 or p.petromag\_r - p.extinction\_r $<$ 19.8 ) and  ( (p.ra $>$ 129.0 and p.ra $<$ 141.0 and p.dec $>$ -1.0 and p.dec $<$ 3.0) or    (p.ra $>$ 174.0 and p.ra $<$ 186.0 and p.dec $>$ -2.0 and p.dec $<$ 2.0) or    (p.ra $>$ 211.5 and p.ra $<$ 223.5 and p.dec $>$ -2.0 and p.dec $<$ 2.0) ) and  ((p.mode = 1) or (p.mode = 2 and p.ra $<$ 139.939 and p.dec $<$ -0.5  and (p.status \& dbo.fphotostatus('OK\_SCANLINE')) $>$ 0))}. Unlike the master catalogue, the tiling catalogue undertakes star-galaxy separation, and applies surface brightness and magnitude selection. \\
\texttt{STILTS} \citep{tex:stilts} is a catalogue combination tool, with a number of different modes. We use it to join our region catalogues together to create $r$-defined, $K$-defined and self-defined aperture photometry catalogues that cover the entire GAMA area. We also use it to match these catalogues to the GAMA tiling catalogue.
\subsection{Source catalogues} \label{subsec:gencat}
The catalogues that have been generated are listed in Table \ref{tab:cats}. The syntax of the \textit{Key} column is as follows. $X[u]$ means a $u$ band magnitude from an $X$ band-defined aperture, $\lbrace u \rbrace$ means a self-defined $u$ band magnitude and $+$ denotes a \texttt{STILTS} \textit{tskymatch2} $5$ arcsec, unique nearest-object match between two catalogues (see Section \ref{sec:catmat}). Where two datasets are combined together without the $+$ notation (i.e., the final two lines), this denotes a \texttt{STILTS} \textit{tmatch2}, \textit{matcher=``exact``} match using SDSS \textsc{objid} as the primary key. Note that in a set of self-defined samples ($\lbrace ugrizYJHK \rbrace$), each sample must be matched separately (as each contains a different set of sources), and then combined. This is not the case in the aperture defined samples (where each sample contains the same set of sources). Subscripts denote the photometric method used for each catalogue. \\
\begin{table*}
\begin{tabular}{ccc} \hline \hline
Catalogue Name&Key&Abbreviation\\ \hline
$r$-defined catalogue &$r[ugriz]_{SDSS}+r[ugrizYJHK]_{GAMA: Petro, Kron}$&catrdef\\
self-defined catalogue&$r[ugriz]_{SDSS}+\lbrace ugrizYJHK \rbrace _{GAMA: Petro, Kron}$&catsd\\
$K$-defined catalogue&$r[ugriz]_{SDSS}+K[ugrizYJHK]_{GAMA: Petro, Kron}$&catKdef\\
S\'{e}rsic catalogue&$r[ugriz]_{SDSS}\lbrace ugrizYJHK \rbrace_{GAMA: Sersic}$&catsers\\
GAMA master catalogue&$(r[ugriz]_{SDSS}\lbrace r \rbrace_{GAMA: Sersic})+\lbrace ugrizYJHK \rbrace_{GAMA: Petro, Kron}$&catmast\\
\hline
\end{tabular}
\caption{The names of the generated catalogues, the prescription used to create them and their abbreviated filename. The syntax in the Key column is summarised in section \ref{subsec:gencat}.}
\label{tab:cats}
\end{table*}
\section{Testing the GAMA catalogues} \label{sec:sourcecomp}
In order to test the detection and deblending outcomes within the GAMA catalogues, a subsection of $25$\,sq deg has been chosen from near the centre of the GAMA 9 region (the pixels used are 20000--65000 in the x direction of the mosaic, and 0--45000 in the y direction). This region was chosen as it contains some of the issues facing the entire GAMA subset, such as area incompleteness. UKIDSS observations miss a large fraction of the subset area - approximately $3.02$\,sq deg of the region has incomplete NIR coverage. The subset region was also chosen because it partially contains area covered by the Herschel ATLAS science verification region (see \citealt{tex:atlas}). Within this region, we ran \texttt{SExtractor}, and compared our results with the source lists produced by the SDSS and UKIDSS extraction software. Unless otherwise specified, all magnitudes within this section were calculated using $r$-defined apertures.
\subsection{Numerical breakdown}
After generating source catalogues containing self-consistent colours for all objects in the subset region (using the process described in subsections \ref{sec:objext} and \ref{sec:catmat}), we are left with an $r$ band aperture-defined subset region catalogue containing $1810134$ sources and a $K$ band aperture-defined subset region catalogue containing $2298224$ sources (these are hereafter referred to as the $r$ band and $K$ band catalogues). These catalogues contain many sources we are not interested in, such as sources with incomplete colour information, sources that are artifacts within the mosaics (satellite trails, diffraction spikes, etc), sources that are stars and sources that are fainter than our survey limits. \\
The unfiltered $r$ and $K$ band catalogues were matched to the master catalogue with a centroid tolerance of $5$\,arcsec, using the \texttt{STILTS} \textit{tskymatch2} mode (see Section \ref{sec:catmat}). Table \ref{tab:coverage} contains a breakdown of the fraction of matched sources that have credible $X_{AUTO}$ and $X_{PETRO}$ for all nine passbands (sources with incorrect \textit{AUTO} or \textit{PETRO} magnitudes have the value $99$ as a placeholder; we impose a cut at $X=50$ to remove such objects). Generally, the low quality of the $u$ band SDSS images causes problems with calculating extended source magnitudes, and this shows itself in the relatively high fraction of incomplete sources. This problem does not affect the other SDSS filters to anywhere near the same extent. SDSS observations do not cover the complete subset area, but they have nearly complete coverage in both the $r$-defined (which is dependent on SDSS imaging) and $K$-defined (reliant on UKIDSS coverage) catalogues. The UKIDSS observations cover a smaller section of the subset region, with the $Y$ and $J$ observations (taken separately to the $H$ and $K$) covering the least area of sky. This is apparent in the $r$ band catalogue, where at least 16\% of sources lack \textit{PETRO} or \textit{AUTO} magnitudes in one or more passband. By its definition the $K$ band catalogue requires $K$ band observations to be present; as such there is a high level of completeness in the $grizH$ and $K$ passbands. However, the number of matched SDSS sources in the $K$ band catalogue itself is 4.2\% lower than in the $r$ band catalogue.\\

\begin{table}
\begin{center}
\begin{tabular}{cccccc} \hline \hline
Band&\% Cover&Sources (r)&\% (r)&Sources (K)&\% (K)\\ \hline
Total&-&129488& -&123740&-\\
u&100&111403&86&105801&86\\
g&100&129169&100&123317&100\\
r&100&129481&100&123610&100\\
i&100&129358&100&123533&100\\
z&100&128287&99&122479&99\\
Y&88&108167&84&109672&89\\
J&89&109364&84&109816&89\\
H&96&121212&94&121846&98\\
K&94&118224&91&122635&99\\
\hline
\end{tabular}
\end{center}
\caption{Number of sources within the subset region with good \texttt{SExtractor} $X_{Auto}$ and $X_{Petro}$, where $X$ is $ugrizYJHK$, from the $r$ or $K$ band-defined aperture catalogues matched to the GAMA master catalogue. The total number of sources within the GAMA master catalogue for this region of sky is 138233. \% Cover is defined relative to $r$ band cover; where SDSS coverage does not exist there are no GAMA master catalogue sources.}
\label{tab:coverage}
\end{table}

\begin{figure}
\includegraphics[width=220pt]{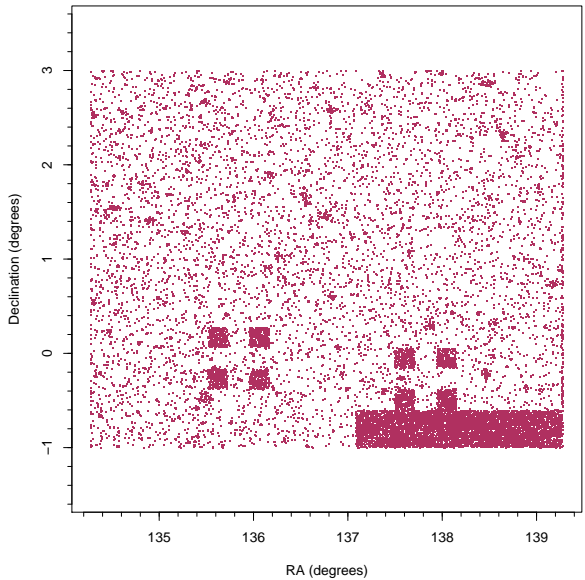}
\caption{The Right Ascension and Declination of SDSS objects that are not in either the $r$ or $K$ band master-cat matched catalogue. The darker areas notes a high density of unmatched objects.}
\label{fig:unmatchedobjid}
\end{figure}

There are $138233$ master catalogue SDSS sources within the subset region. $119330$ SDSS objects have matches (within a $5$\,arcsec tolerance) in both the $r$ band and $K$ band master-cat matched catalogues (this number is found by matching SDSS \textsc{objid} between the catalogues). Those SDSS objects that do not have matches in both master-cat matched catalogues are shown in Figure~\ref{fig:unmatchedobjid}. We detail the reasons for the missing objects in section \ref{sec:notinrK}.

\subsection{SDSS sources missing in the master catalogue- aperture matched catalogues} \label{sec:notinrK}
There are $18903$ SDSS sources that are not found when the master catalogue is matched to either the $r$ or $K$-defined subset region catalogues; 13.7\% of the total number of master catalogue sources within the subset region. Figure~\ref{fig:unmatchedobjid} shows their distribution on the sky. $8745$ sources are not found within the $r$-defined catalogue (6.3\% of the master catalogue sample) and $14493$ are not found within the $K$-defined catalogue (10.5\% of the sample), with $4335$ of the sources unmatched to either the $r$ or the $K$-defined sample (3.1\% of the master catalogue sample). As the SDSS sample is defined by optical data, it is unsurprising that a far larger number of sources are not found within the $K$-defined catalogue. Of the $18903$ unmatched master catalogue sources, only $2367$ have passed star-galaxy separation and are brighter than the GAMA spectroscopic survey magnitude limits ($r<19.8$ or $zK$ selected).\\

Using $r$ band imaging, we have visually inspected all $8745$ SDSS sources where our $r$-defined subset region catalogue cannot find a match within $5$\,arcsec. Table \ref{tab:eyeballing} contains a summary of the reasons we do not find a $r$ band match. Using the \texttt{SExtractor} detection failure rate from the subset region as a guide to the detection failure rate for the entire GAMA region, \texttt{SExtractor} will miss approximately 2.8\% of the objects recovered by SDSS. A second problem was flagged through the inspection process; a further 1.7\% of the master catalogue sources within the subset region were not visible. Either these objects are low surface brightness extended objects, possibly detected in a different band, or the SDSS object extraction algorithm has made a mistake. A further 1.8\% of sources within the GAMA master catalogue will be missed by \texttt{SExtractor} due to differences in deblending decisions (either failing to split two sources or splitting one large object into a number of smaller parts), low SDSS image quality making \texttt{SExtractor} fail to detect any objects, or an artifact in the image being accounted for by \texttt{SExtractor} (such as a saturation spike from a large star being detected as a separate object in SDSS).

\begin{table}
\begin{tabular}{p{3.25cm}p{1.35cm}p{3.4cm}} \hline \hline
Reason for non-detection&Number of objects&\% of GAMA master catalogue subset region sample\\ \hline
Possible deblending mismatch&601&0.4\\
Saturation spike / satellite or asteroid trail&404&0.3\\
\texttt{SExtractor} detection failure&3831&2.8\\
Either a low surface brightness source or no source&2391&1.7\\
Part of a large deblended source&1463&1.1\\
Low image quality making detection difficult&55&0.04\\
\hline
\end{tabular}
\caption{A breakdown of the reasons for faulty detections in the 8745 SDSS objects that are not matched to the $r$ band subset region catalogue. The images of the SDSS objects were generated from the standard $r$ band GAMA mosaics, and all 8745 objects were viewed by one observer (DTH). The criteria selection is as follows. The first category is chosen in those cases where an object has a nearby neighbour or may have been deblended into multiple sources by the SDSS algorithm. The second category is chosen where the position of the object is covered by a spike/trail. The third category is where a source is visible by eye. The fourth category is where a source is not visible above the noise. The fifth category is chosen when a source is obviously part of a larger structure. The sixth category is chosen when the SDSS data is too low quality for visual classification to be undertaken.}
\label{tab:eyeballing}
\end{table}

\subsection{Sources in our $r$ band catalogue that are not in the GAMA master catalogue} \label{sec:rnotsdss}
\begin{figure}
\includegraphics[width=220pt]{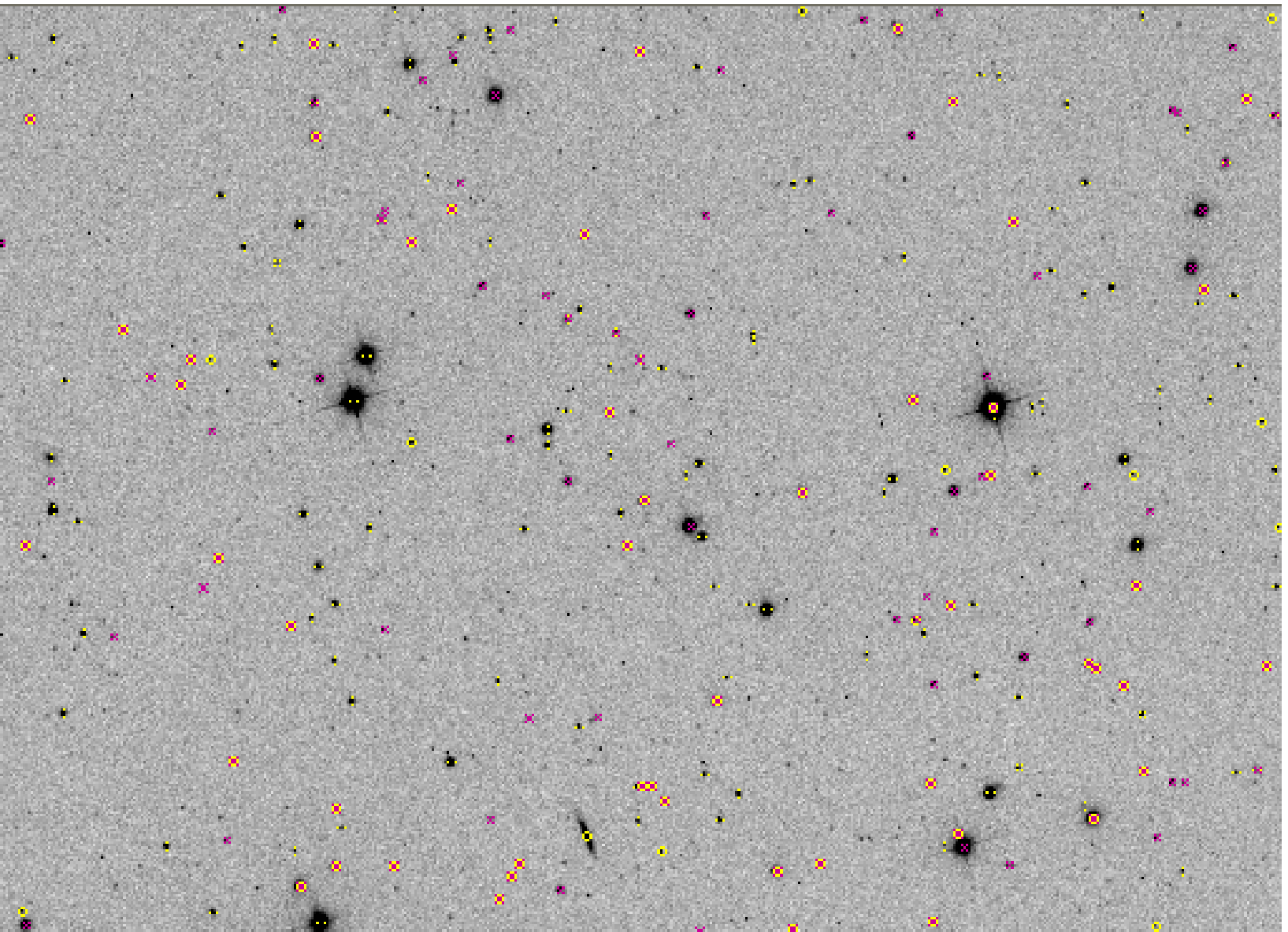}
\caption{A comparison between the objects detected when \texttt{SExtractor} is run over an original SDSS image, and when it is run over the convolved, mosaic imaging. Yellow circles are sources with $r\le20.5$\,mag detected from the GAMA mosaic, red crosses are sources that are detected from the original SDSS data.}
\label{fig:swarpvsorig}
\end{figure}

To be certain that the SDSS extraction software is giving us a complete sample, we check whether our $r$ band subset region catalogue contains sources that should be within the GAMA master catalogue but are not. There are $61351$ sources within the $r$-defined subset region catalogue that have a complete set of credible \textit{AUTO} and \textit{PETRO} magnitudes, and are brighter than the GAMA spectroscopic survey limits. $619$ of these sources do not have SDSS counterparts. We have visually inspected these sources; a breakdown is shown in Table \ref{tab:rnosdsseyeballing}. Similar issues cause missing detections using the SDSS or \texttt{SExtractor} algorithms. However, some of the unseen sources that \texttt{SExtractor} detected may be due to the image convolution process (Section \ref{sec:seeingconv}) gathering up the flux from a region with high background noise and rearranging it so that it overcomes the detection threshold. Figure \ref{fig:swarpvsorig} shows the distribution of $r_{Auto}\le20.5$\,mag sources detected when \texttt{SExtractor} is run upon an original SDSS image file (covering $\sim 0.04$\,deg$^{2}$), and the sources from the same file after it has undergone the image convolution process. 233 sources are found in the original SDSS frame, and 3 additional sources are included within the convolved frame sample. An examination of two sources that are in the convolved frame dataset and not in the original sample shows the effect: these sources have $r_{Auto}$ luminosities of 20.64\,mag and 20.77\,mag pre-convolution, but $r_{Auto}$ luminosities of 20.40\,mag and 20.48\,mag post-convolution.\\
Taking the SDSS non-detection rate within the subset region to be the same as the non-detection rate over the entire GAMA region, we expect that the SDSS algorithm will have failed to detect 0.1\% of sources brighter than the GAMA spectroscopic survey limits; approximately $1000$ sources will not have been included within the master catalogue. 

\begin{table}
\begin{tabular}{cc} \hline \hline
Type of source&Number of objects\\ \hline
Source&171\\
No visible source&274\\
Section of bright star&163\\
Possible deblend mismatch&10\\
Low image quality making detection difficult&1\\
\hline
\end{tabular}
\caption{A breakdown of the 619 $r$-defined subset region catalogue objects brighter than the GAMA sample limits that are not matched to the GAMA master catalogue.The images of the subset region catalogue objects were generated from the standard $r$ band GAMA mosaics, and all 619 objects were viewed by one observer (DTH).}
\label{tab:rnosdsseyeballing}
\end{table}

\subsection{Sources in our $K$ band catalogue that are not in the UKIDSS DR5PLUS database}
We have also tested the UKIDSS DR5 catalogue. We have generated a catalogue from the WSA that selects all UKIDSS objects within the GAMA subset region\footnote{We use the query "SELECT las.ra, las.dec, las.kPetroMag FROM lasSource as las WHERE $las.ra<139.28$ AND $las.ra>134.275$ AND $las.dec>-1$ AND $las.dec<3$"}, and we have matched this catalogue to the $K$ band-defined subset region catalogue. From the $69537$ $K$ band-defined subset region catalogue sources, there are $4548$ sources that have not been matched to an UKIDSS object within a tolerance of $5$\,arcsec. We have visually inspected $K$ band images of those objects that are brighter than the GAMA spectroscopic survey $K$ band limit ($K_{AUTO}\le17.6$\,mag). We find that $29$ of the $117$ unmatched objects are real sources that are missed by the UKIDSS extraction software; a negligible fraction of the entire dataset. A large (but unquantified) fraction of the other $88$ sources are suffering from the convolution flux-redistribution problem discussed in Section \ref{sec:rnotsdss}. The background fluctuations in $K$ band data are greater than in the $r$ band, making this a much greater problem.

\section{Properties of the catalogues} \label{sec:catprops}
\subsection{Constructing a clean sample} \label{sec:photomdist} 
In order to investigate the photometric offsets between different photometric systems, we require a sample of galaxies with a complete set of credible photometry that are unaffected by deblending decisions. This has been created via the following prescription. We match the r-defined aperture catalogue to the GAMA master catalogue with a tolerance of 5\,arcsec. We remove any GAMA objects that have not been matched, or have been matched to multiple objects within that tolerance (when run in \textit{All match} mode \texttt{STILTS} produces a \textit{GroupSize} column, where a \textit{NULL} value signifies no group). We then match to the 9 self-defined object catalogues, in each case removing all unmatched and multiply matched GAMA objects. As our convolution routine will cause problems with those galaxies that contain saturated pixels, we also remove those galaxies that are flagged as saturated by SDSS. This sample is then linked to the S\'{e}rsic pipeline catalogue (using the SDSS \textsc{objid} as the primary key). We remove all those S\'{e}rsic magnitudes where the pipeline has flagged that the model is badly fit or where the photometry has been compromised and match to the $K$ band aperture-defined catalogue, again with unmatched and multiple matched sources removed. This gives us a final population of $18065$ galaxies that have clean $r$-defined, $K$-defined, self-defined and S\'{e}rsic magnitudes, are not saturated and cannot have been mismatched. Having constructed a clean, unambiguous sample of common objects, any photometric offset can only be due to differences between the photometric systems used. As we remove objects that are badly fit by the S\'{e}rsic pipeline, it should be noted that the resulting sample will, by its definition, only contain sources that have a light profile that can be fitted using the S\'{e}rsic function.\\
\subsection{Photometric offset between systems} \label{sec:photomoffset} 
\begin{figure*}
\includegraphics[height=640pt]{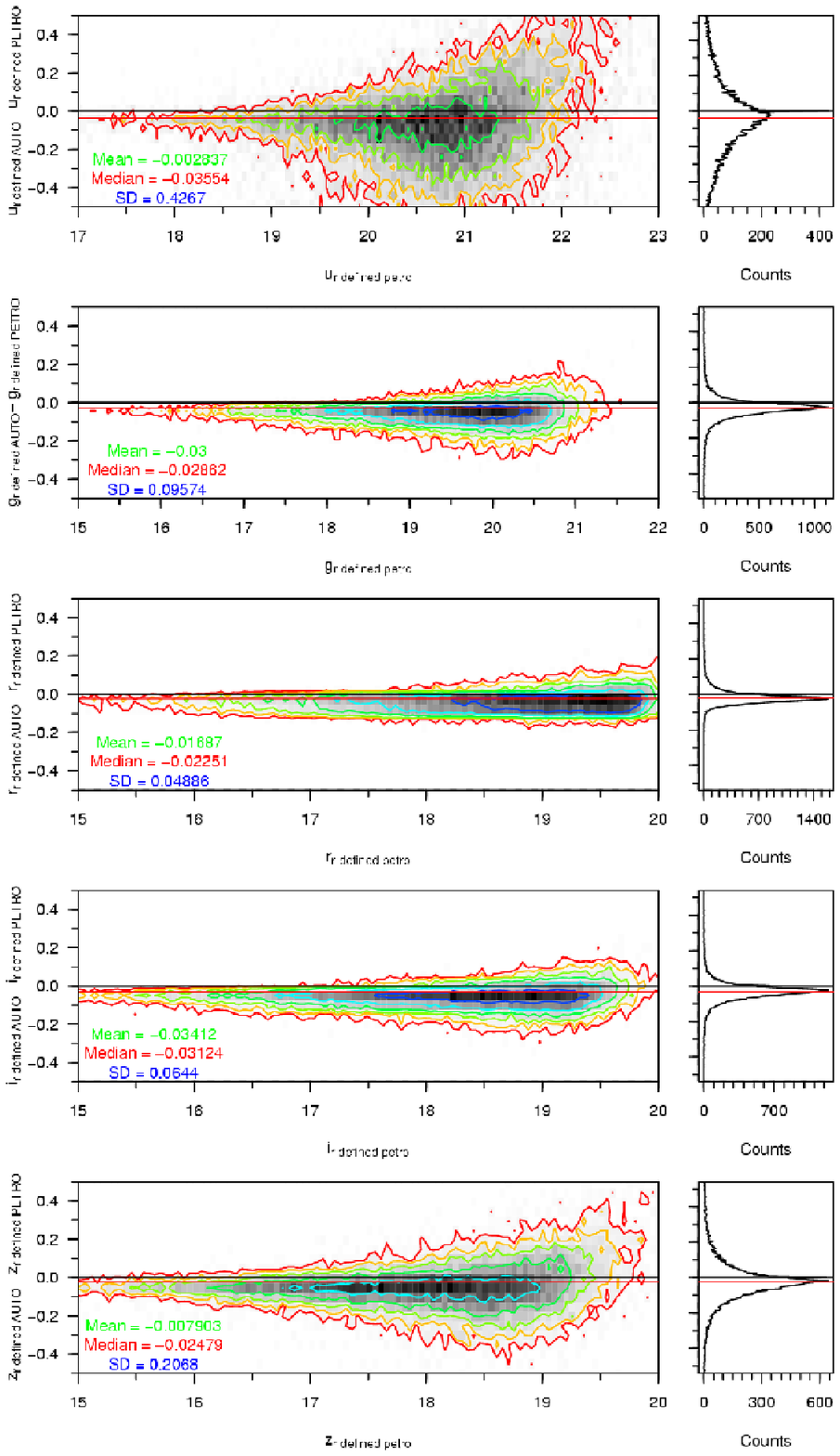}
\caption{GAMA r-defined aperture Petrosian minus Auto magnitudes for a clean sample of galaxies in $ugriz$. Contours are for 4 to 512 galaxies per bin, rising geometrically in powers of 2. Bins are 0.05\,mag$\times$0.05\,mag in size.}
\label{fig:rdpetrdauto}
\end{figure*}

\begin{figure*}
\includegraphics[height=640pt]{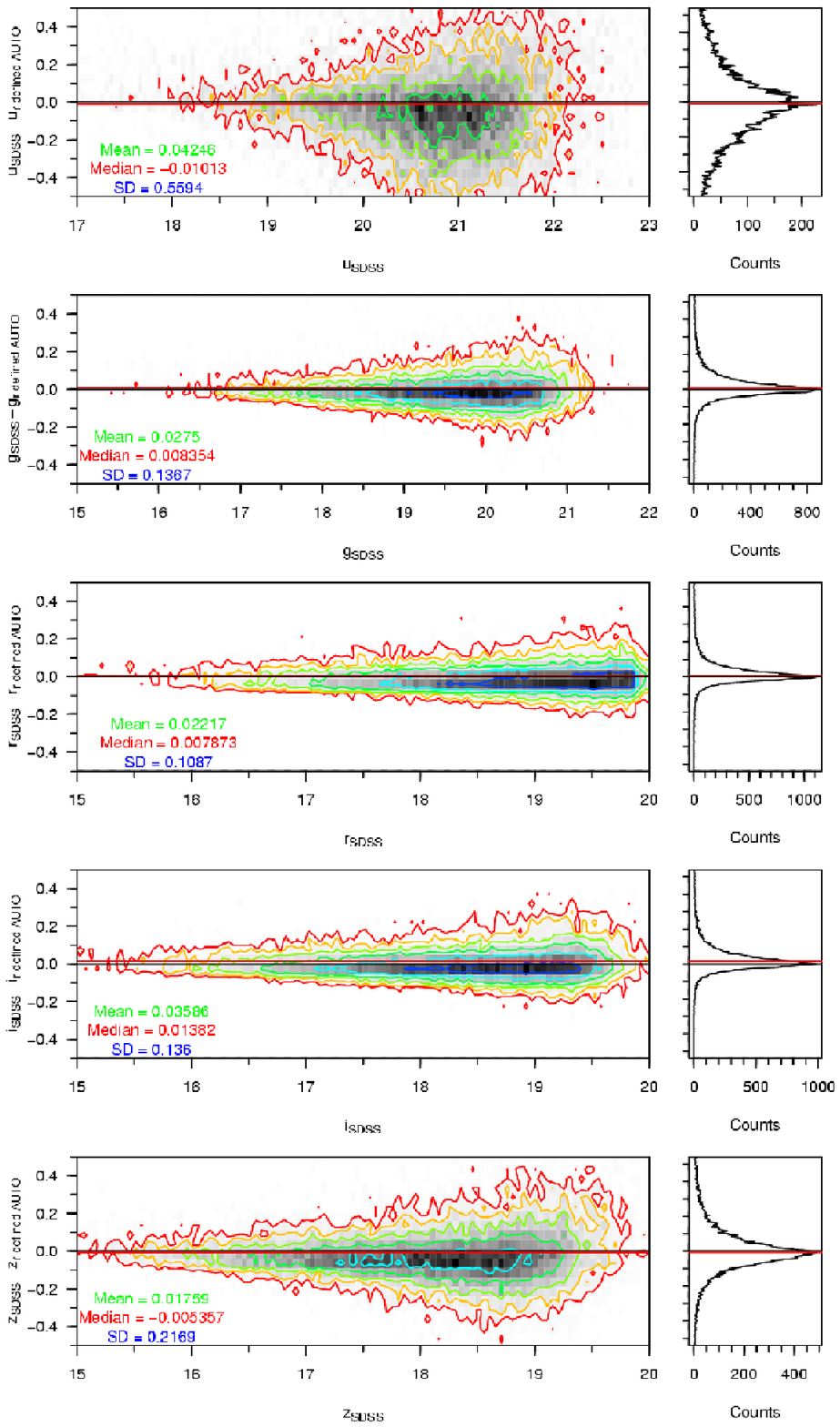}
\caption{SDSS \textsc{petromag} minus GAMA r-defined aperture Auto magnitudes for a clean sample of galaxies in $ugriz$. Contours are for 4 to 512 galaxies per bin, rising geometrically in powers of 2. Bins are 0.05\,mag$\times$0.05\,mag in size.}
\label{fig:sdssauto}
\end{figure*}

\begin{figure*}
\includegraphics[height=640pt]{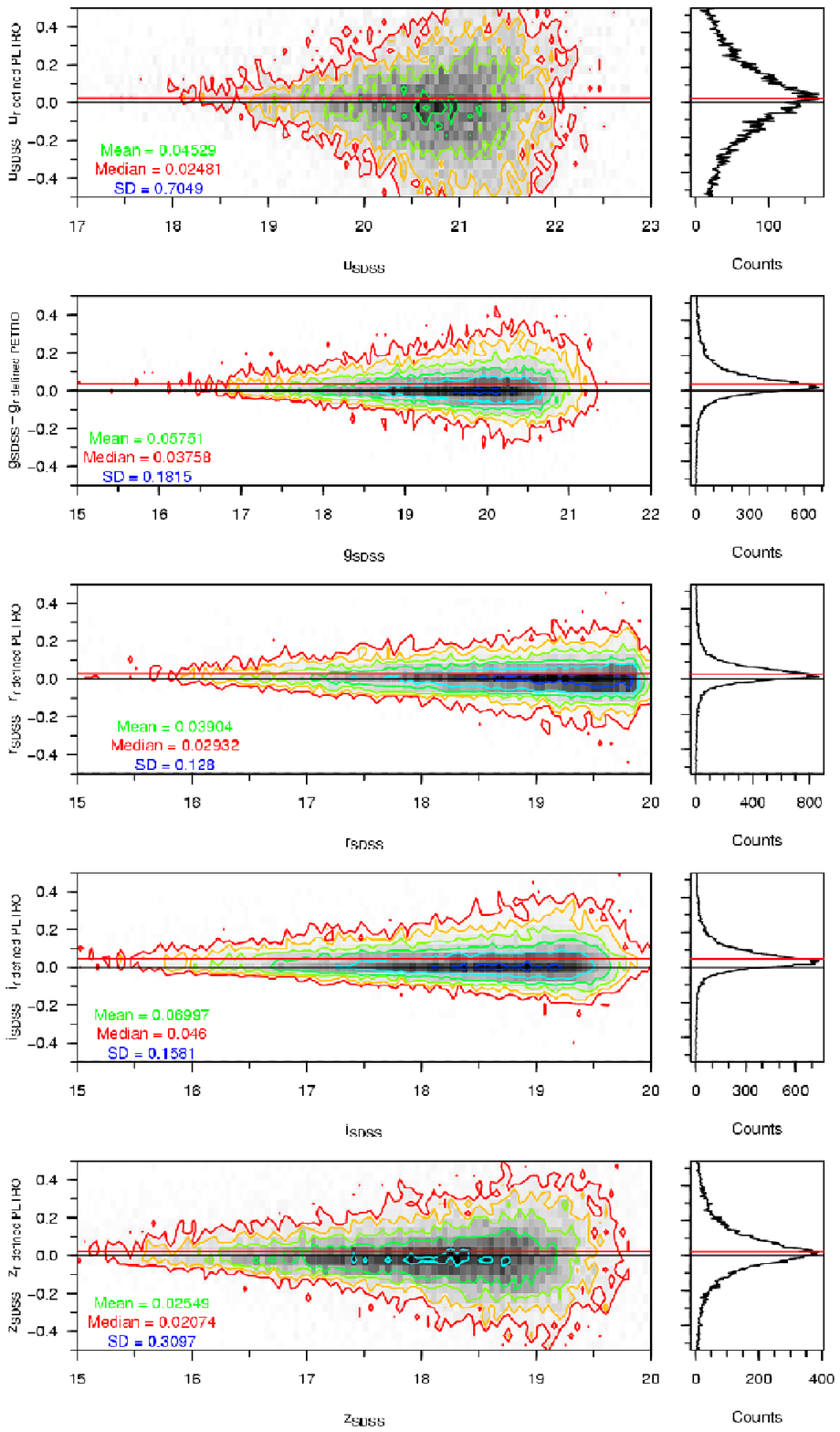}
\caption{SDSS \textsc{petromag} minus GAMA r-defined aperture Petrosian magnitudes for a clean sample of galaxies in $ugriz$. Contours are for 4 to 512 galaxies per bin, rising geometrically in powers of 2. Bins are 0.05\,mag$\times$0.05\,mag in size.}
\label{fig:sdssrpetro}
\end{figure*}

\begin{figure*}
\includegraphics[height=640pt]{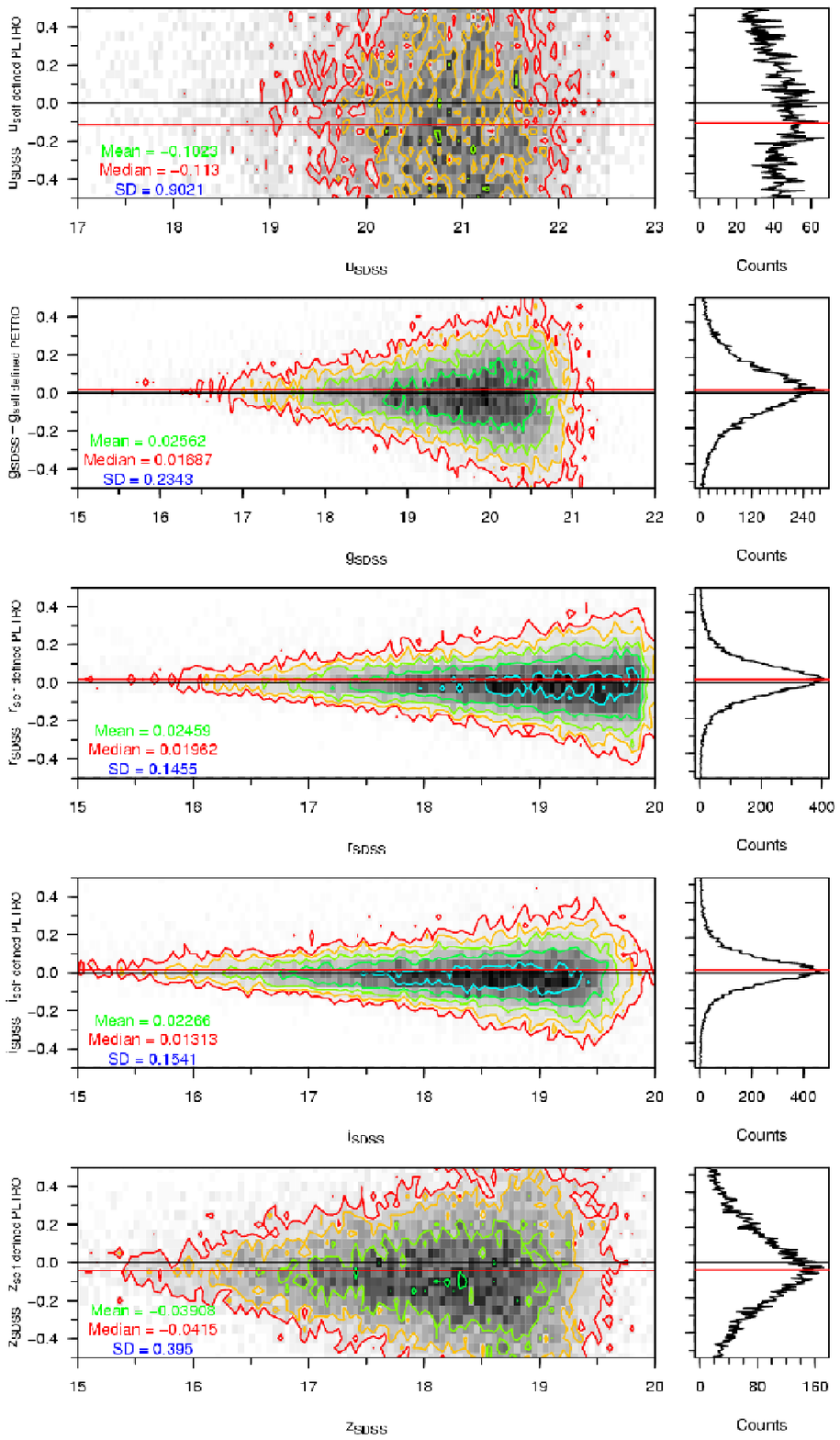}
\caption{SDSS \textsc{petromag} minus GAMA self-defined aperture Petrosian magnitudes for a clean sample of galaxies in $ugriz$. Contours are for 4 to 512 galaxies per bin, rising geometrically in powers of 2. Bins are 0.05\,mag$\times$0.05\,mag in size.}
\label{fig:sdssselfpetro}
\end{figure*}

\begin{figure*}
\includegraphics[height=640pt]{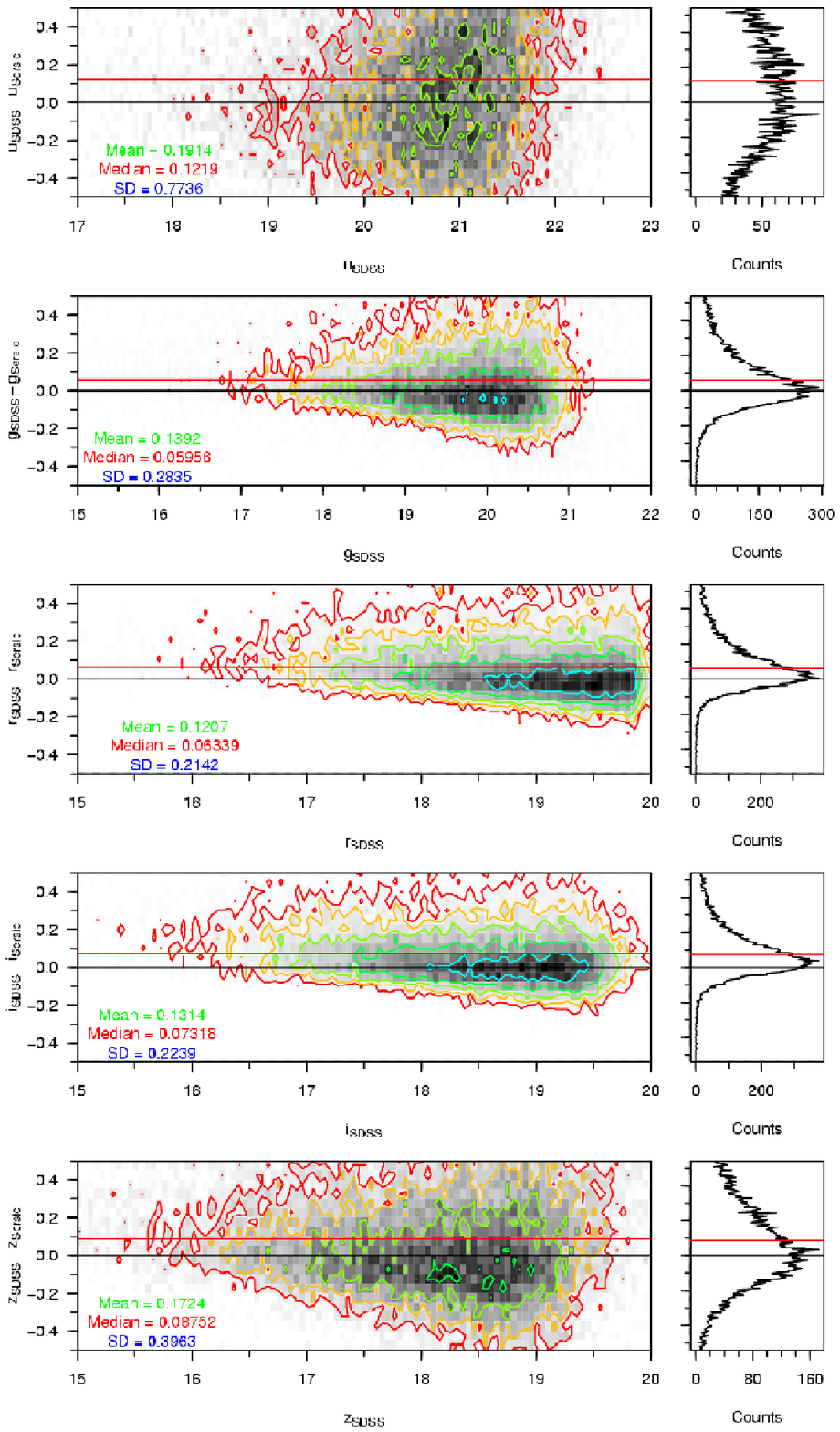}
\caption{SDSS \textsc{petromag} minus GAMA S\'{e}rsic magnitudes for a clean sample of galaxies in $ugriz$. Contours are for 4 to 512 galaxies per bin, rising geometrically in powers of 2. Bins are 0.05\,mag$\times$0.05\,mag in size.}
\label{fig:sdssSersic}
\end{figure*}

\begin{figure*}
\includegraphics[height=640pt]{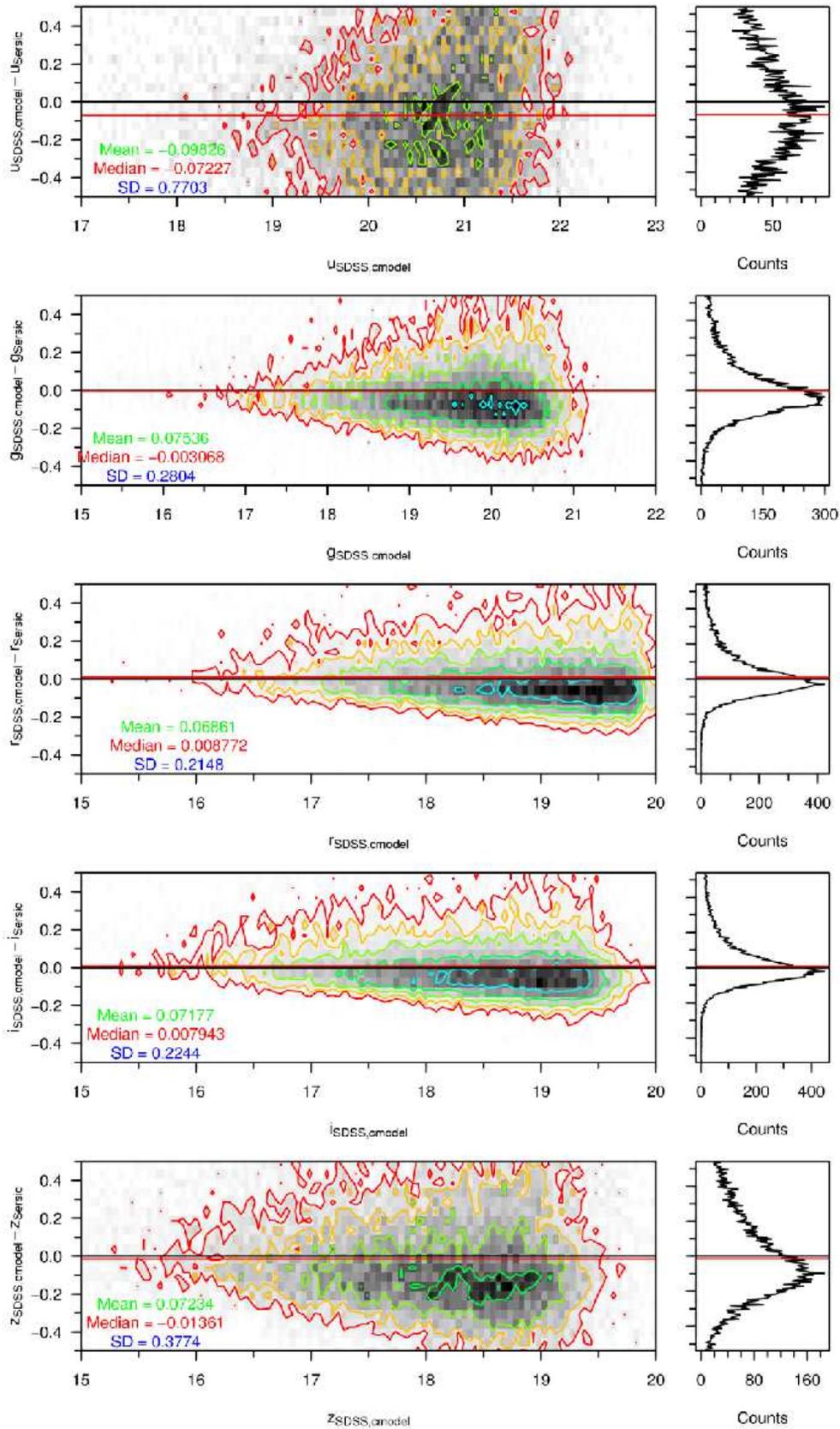}
\caption{SDSS \textsc{cmodel} minus GAMA S\'{e}rsic magnitudes for a clean sample of galaxies in $ugriz$. Contours are for 4 to 512 galaxies per bin, rising geometrically in powers of 2. Bins are 0.05\,mag$\times$0.05\,mag in size.}
\label{fig:cmodSersic}
\end{figure*}

Figures \ref{fig:rdpetrdauto}, \ref{fig:sdssauto}, \ref{fig:sdssrpetro}, \ref{fig:sdssselfpetro} \ref{fig:sdssSersic}, and \ref{fig:cmodSersic} show the dispersion between different photometric systems produced by this sample. In Figure \ref{fig:rdpetrdauto} we compare Kron and Petrosian magnitudes; in all other figures we compare the photometric system to SDSS \textsc{petromag}. In all photometry systems, the $gri$ relationships are tightest, with the $u$ and $z$ relationships subject to a greater scatter, breaking down almost entirely for Figures \ref{fig:sdssselfpetro} and \ref{fig:sdssSersic}. The correlation between the SDSS \textsc{petromag} and the $r$-defined Petrosian magnitude (Figure \ref{fig:sdssrpetro}) looks much tighter than that between the SDSS \textsc{petromag} and the self-defined Petrosian magnitude (Figure \ref{fig:sdssselfpetro}). The standard deviation of the samples are similar, with marginally more scatter in the self-defined sample ($0.129$\,mag against $0.148$\,mag). The median offset between SDSS Petrosian and the $r$-defined Petrosian magnitude, however, is $0.01$\,mag greater.\\
Figure \ref{fig:sdssSersic}, illustrating the relationship between the S\'{e}rsic magnitude and the SDSS \textsc{petromag}, produces median $\Delta m_{SDSS}-m_{Sersic}$ values of $0.12$, $0.06$, $0.06$, $0.07$ and $0.09$\,mag in $ugriz$. These values can be compared to those presented in Figure 13 of \cite{tex:blantonsdsslf} ($-0.14$, $0.00$, $0.06$, $0.09$ and $0.14$\,mag at $z=0.1$, using the $^{0.1}ugriz$ filters), given the variance in the relationship (the standard deviation in our samples are $0.77$, $0.28$, $0.21$, $0.22$ and $0.40$\,mag, respectively). A significant fraction ($\sim28\%$) of the sample has $r_{SDSS}-r_{Sersic}$>0.5\,mag, and therefore lies beyond the boundaries of this image. These offsets are significant, and will be discussed further in Section \ref{sec:gamaphotometry}. We can say that the $r$-defined aperture photometry is the closest match to SDSS \textsc{petromag} photometry. Figure \ref{fig:cmodSersic} shows the relationship between the GAMA S\'{e}rsic magnitude, and the optimal model magnitude provided by SDSS (\textsc{cmodel}). The model magnitudes match closely, with negligible systematic offset between the photometric systems in $gri$.
\subsection{Colour distributions} \label{sec:quality}
\begin{figure*}
\includegraphics[width=420pt]{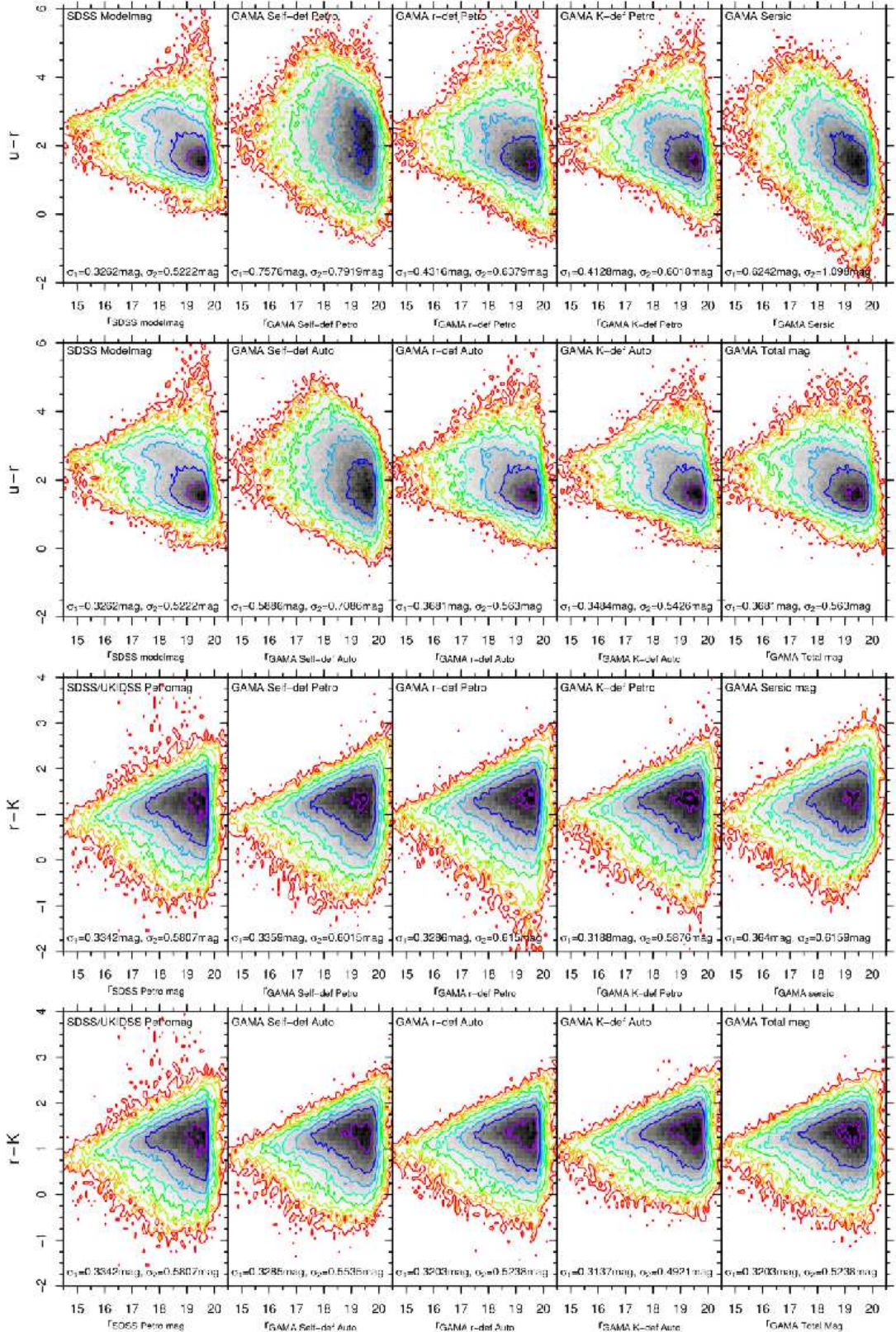}
\caption{A comparison between the $u$ minus $r$ and $r$ minus $K$ colours produced using SDSS \textsc{modelmag}, GAMA self-defined Petrosian magnitudes, GAMA $r$/$K$-defined Petrosian magnitudes, S\'{e}rsic and GAMA Total magnitudes for objects in the subset region. Contours shown are 2 to 512 galaxies per bin, rising geometrically in powers of 2. Bins are 0.1\,mag in width in each axis. The $\sigma$ parameter comes from the best-fitting bivariate-Gaussian distribution, when it is fit to the colour-distribution histogram in each plot.}
\label{fig:rtrumpets}
\end{figure*}

\begin{figure*}
\includegraphics[width=420pt]{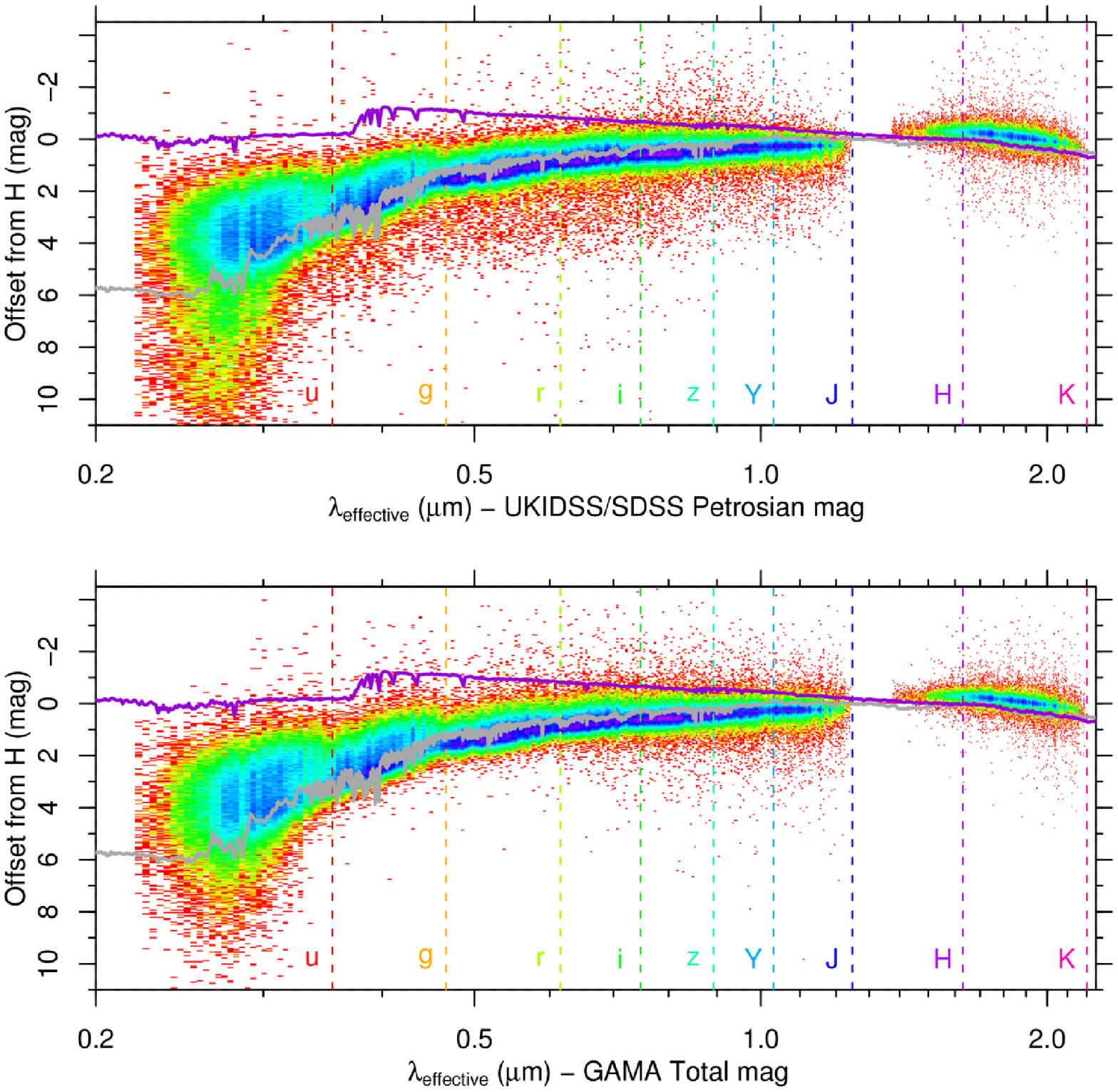}
\caption{A comparison between the $X-H$ colours produced using SDSS magnitudes and GAMA Total magnitudes. Data comes from all GAMA galaxies with good quality redshifts ($0.033<z<0.6$) and complete $ugrizYJHK$ photometry. Effective wavelengths are calculated from the redshift of the galaxy and the filter effective wavelength, and the dataset is binned into a 50$\times$50 bin matrix. Two Bruzual-Charlot 03 SSP instantaneous-burst models are also plotted. Both models use the \citet{tex:chabrier} IMF, with mass cutoffs at 0.1 and 100\,$M_{\astrosun}$. Stellar evolution is undertaken using the Padova 1994 prescription. The dark grey line is a model evolved to 11\,Gyr , with Z=0.05 and Y=0.352. The purple line is a model evolved to 0.25\,Gyr, using Z=0.02 (Z$\astrosun$) and Y=0.28. }
\label{fig:photomsed}
\end{figure*}

In order to identify the optimal photometric system, we assume that intrinsic colour distribution of a population of galaxies can be approximated by a double-Gaussian distribution (the superposition of a pair of Gaussian distributions with different mean and standard deviation parameters). This distribution can model the bimodality of the galaxy population. The presence of noise will broaden the distribution; hence the narrowest colour distribution reveals the optimal photometric system for calculating the colours of galaxies, and therefore deriving accurate SEDs. Figure \ref{fig:rtrumpets} shows the ($u-r$) and ($r-K$) colour distributions for each photometric system, for objects within our subset region. In order to calculate the dispersion in the colour distribution, we generate colour-distribution histograms (with bins of $0.1$\,mag), and find the double-Gaussian distribution parameters that best fit each photometric system. The best-fitting standard deviation parameters for each sample are shown at the bottom of each plot, and are denoted $\sigma_{X,1}$ and $\sigma_{X,2}$ (where $X$ is the photometric system fitted). The sample with the smallest set of $\sigma$ parameters should provide the optimal photometric system.\\ 
The SDSS, GAMA $r$-defined aperture and GAMA $K$-defined distributions (the first, third and fourth diagrams on the top two rows) show a very similar pattern; a tight distribution of objects with a small number of red outliers. As expected, when we use apertures that are defined separately in each filter (the second diagram on the top two rows), the colour distribution of the population is more scattered ($\sigma_{Petro,1} = 0.7576$\,mag, $\sigma_{Petro,2} = 0.7919$\,mag, $\sigma_{Auto,1} = 0.5886$\,mag, $\sigma_{Auto,2} = 0.7086$\,mag) and does not show the bimodality visible in the matched aperture photometry (at the bright end of the distribution there are two distinct sub-populations; one sub-population above $u-r=2$\,mag, the other below). For the same reason, and probably because of the low quality of the observations, the ($u-r$) plot using the S\'{e}rsic magnitudes (the final diagram on the top row) has the broadest colour distribution ($\sigma_{Sersic,1} = 0.6242$\,mag, $\sigma_{Sersic,2}=1.098$\,mag), although it is well behaved in ($r$-$K$).\\
To generate a series of ($r-K$) colours using the UKIDSS survey (leftmost plot on the bottom two rows), we have taken all galaxies within the UKIDSS catalogue\footnote{We run a query at the WSA on \textsc{UKIDSSDR5PLUS} looking for all objects within our subset region with $lasSource.pGalaxy>0.9 \text{ \& } lasSource.kPetroMag<20$ - equivalent to $K_{AB}<21.9$\,mag} and match them (with a maximum tolerance of 5\,arcsec) to a copy of the tiling catalogue that had previously been matched with the $K$ band aperture-defined catalogue. The distribution of ($r-K$) colours taken from the SDSS and UKIDSS survey catalogues is the first diagram on the bottom two rows of the image. As the apertures used to define the UKIDSS and SDSS sources are not consistent, we find that the tightest ($r-K$) distribution comes from the GAMA $K$-defined aperture sample (fourth from the left on the bottom row, with $\sigma_{Auto,1} = 0.3137$\,mag, $\sigma_{Auto,2} =0.4921$\,mag). The GAMA sample that relies on matching objects between self-defined object catalogues (the second diagram on the bottom two rows) has the broadest distribution ($\sigma_{Petro,1} = 0.3359$\,mag, $\sigma_{Petro,2} = 0.6015$\,mag). The distribution of sources in the S\'{e}rsic ($r-K$) colour plot is much tighter than in the ($u-r$), though still not as tight as the distribution in the fixed aperture photometric systems ($\sigma_{Sersic,1} = 0.364$\,mag, $\sigma_{Sersic,2} = 0.6159$\,mag). Figure \ref{fig:rtrumpets} confirms the utility of the GAMA method: by redoing the object extraction ourselves, we have generated self-consistent colour distributions based on data taken by multiple instruments that has a far smaller scatter than a match between the survey source catalogues ($\sigma_{SDSS+UKIDSS,1} = 0.3342$\,mag, $\sigma_{SDSS+UKIDSS,2} = 0.5807$\,mag).\\
We provide one more comparison between our colour distribution and that provided by SDSS and UKIDSS survey data. Figure \ref{fig:photomsed} displays the $X-H$ distribution produced by the GAMA galaxies with complete $ugrizYJHK$ photometry and good quality redshifts within $0.033<z<0.6$. The effective wavelengths of the filter set for each galaxy are shifted using the redshift of the galaxy. The colour distribution provided by the GAMA photometry produces fewer outliers than the SDSS/UKIDSS survey data sample, and is well constrained by the \citet{tex:bc03} models.

\section{Final GAMA photometry} \label{sec:gamaphotometry}
Sections \ref{sec:sourcecomp} and \ref{sec:catprops} show that the optimal deblending outcome is produced by the original SDSS data, but the best colours come from our $r$-defined aperture photometry (Section \ref{sec:quality}). We see that our $r$-defined aperture photometry agrees with the SDSS \textsc{petromag} photometry. However, we have also demonstrated that SDSS \textsc{petromag} misses flux when compared to our S\'{e}rsic total magnitude. Here we combine these datasets to arrive at our final photometry. We combine the SDSS deblending outcome with our $r$-defined aperture colours and the S\'{e}rsic total-magnitude to produce our best photometric solution.

\subsection{Sersic magnitudes} \label{sec:sersiccheck}
\begin{figure*}
\includegraphics[width=440pt]{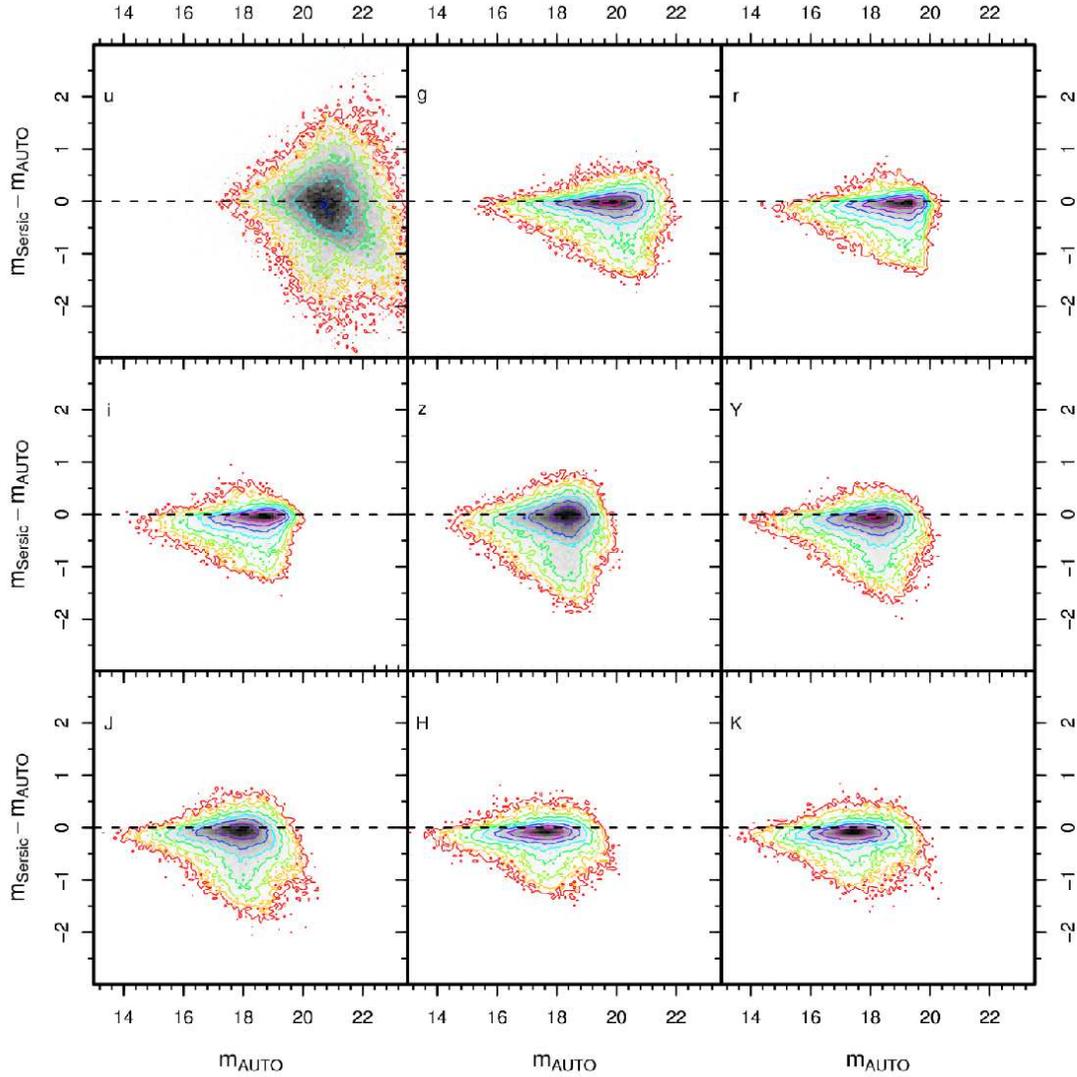}
\caption{S\'{e}rsic minus GAMA $r$-defined Auto magnitude against $r$-defined Auto magnitude, in all nine bands, for all objects in the GAMA sample that pass our star-galaxy separation criteria, and have credible $ugrizYJHK$ $r$-defined Auto magnitudes. Contours increase geometrically in powers of 2, from 4 to 512. Bins are 0.1\,mag (x axis) $\times$ 0.05\,mag (y axis) in size.}
\label{fig:sersallobjsallbands}
\end{figure*}

\begin{figure*}
\includegraphics[width=440pt]{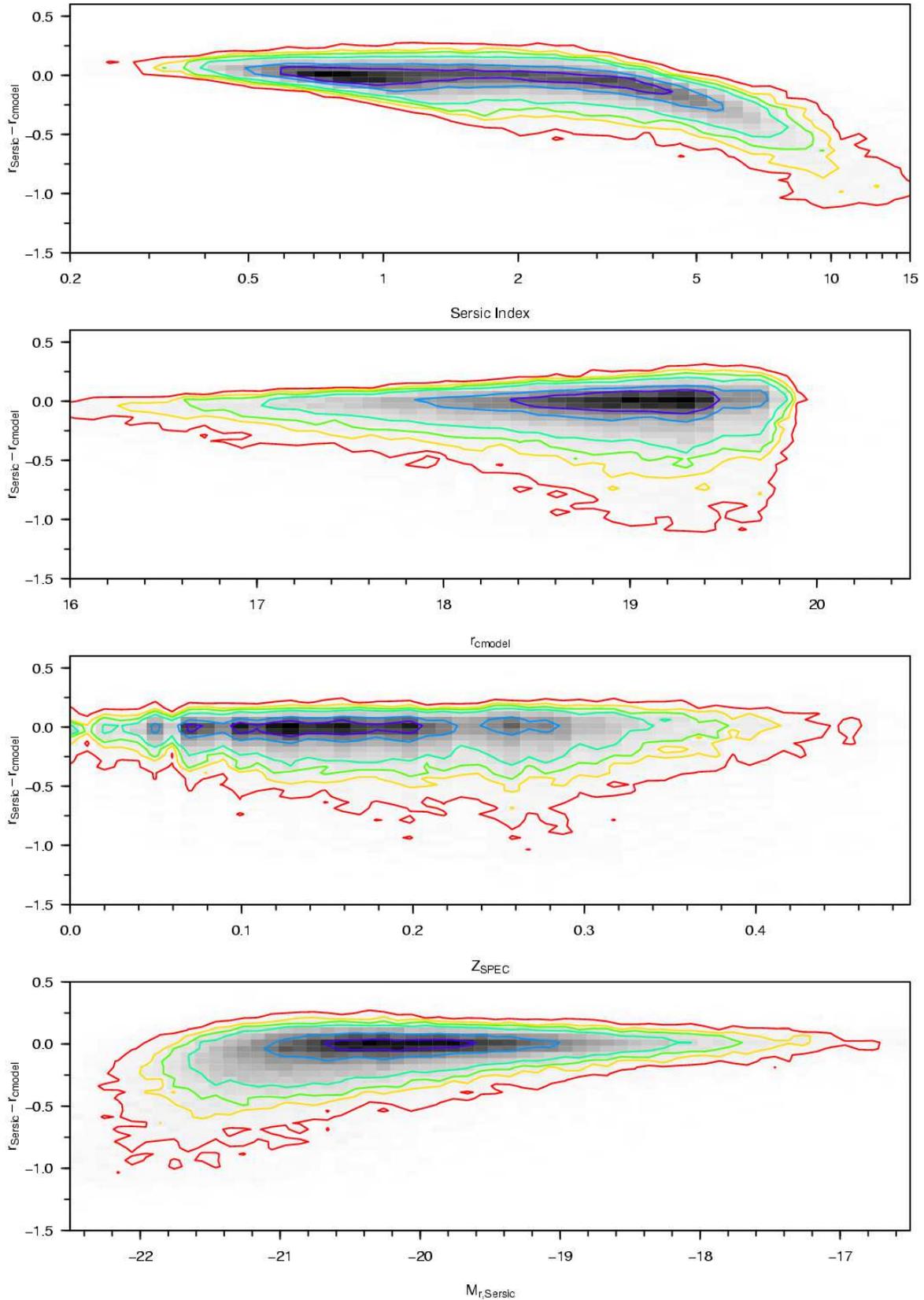}
\caption{S\'{e}rsic $r$ - SDSS \textsc{cmodel} $r$ magnitude against S\'{e}rsic index, SDSS $r$ \textsc{cmodel} magnitude, z, $M_{r, Sersic}$ for all objects in the GAMA sample that have passed our star-galaxy separation criteria, and have credible $urK$ $r$-defined Petrosian magnitudes. Contours increase geometrically in powers of 2, from 4 to 512.}
\label{fig:cmodelallobjsr}
\end{figure*}

\begin{figure*}
\includegraphics[width=440pt]{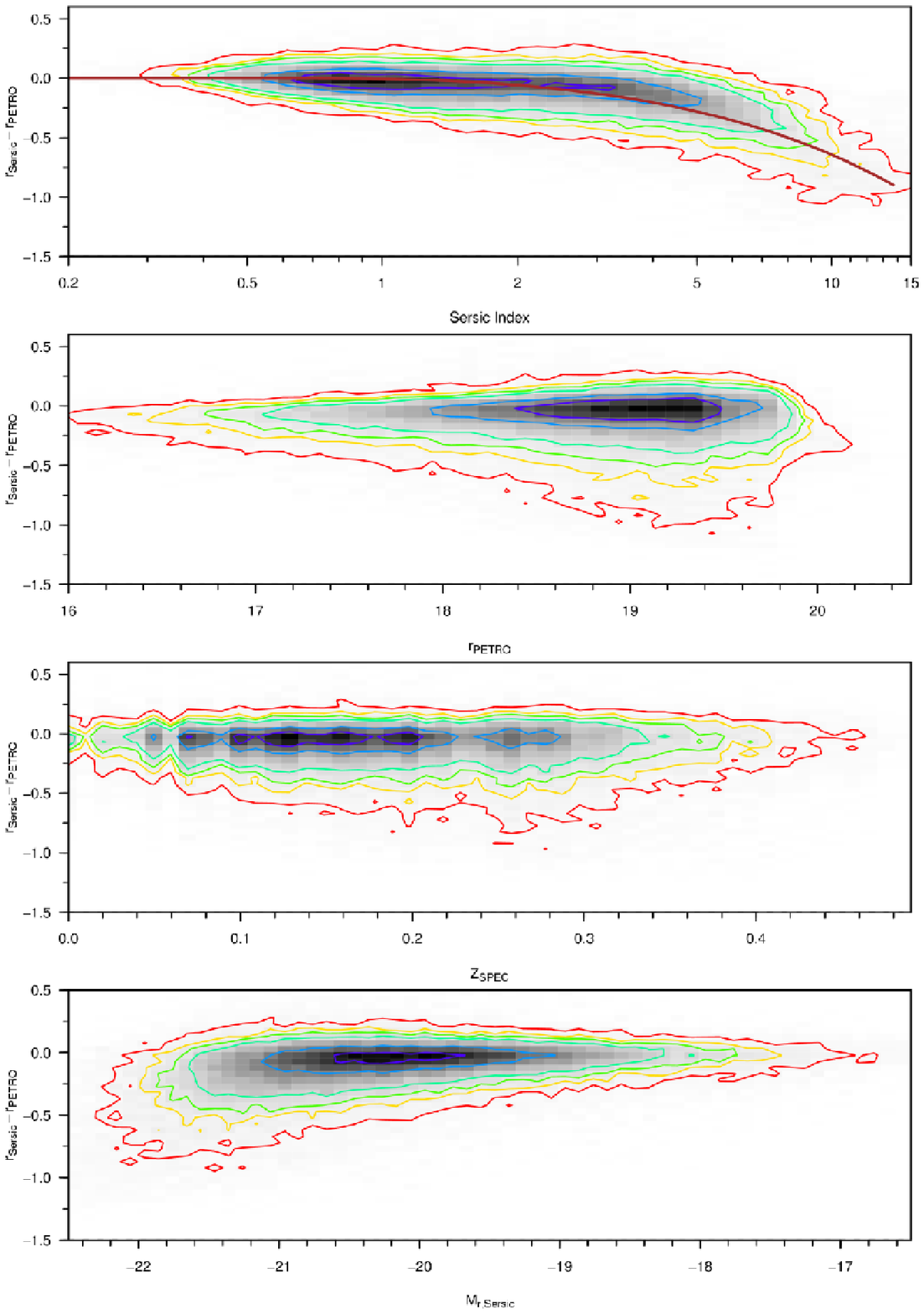}
\caption{S\'{e}rsic $r$ - GAMA $r$-defined Petrosian $r$ against S\'{e}rsic index, $r$-defined $r$ band Petrosian magnitude, z, $M_{r, Sersic}$ for all objects in the GAMA sample that have passed our star-galaxy separation criteria, and have credible $urK$ $r$-defined Petrosian magnitudes. Contours increase geometrically in powers of 2, from 4 to 512. The brown function plotted in the S\'{e}rsic $r$ - GAMA $r$-defined Petrosian $r$ against S\'{e}rsic index plot is taken from Figure.~2 (upper panel) of \citet{tex:grahampetrosian}.}
\label{fig:sersallobjsr}
\end{figure*}

To check the reliability of the S\'{e}rsic photometry pipeline, we must examine its distribution against a photometric system we consider reliable. We examine the distribution of the S\'{e}rsic photometry against our $r$-defined AUTO photometry. Figure \ref{fig:sersallobjsallbands} shows the distribution of S\'{e}rsic - GAMA $r$-defined AUTO magnitude against $r$-defined AUTO magnitude for all objects in the GAMA sample that have passed our star-galaxy separation criteria and have credible AUTO magnitudes. Whilst there is generally a tight distribution, the scatter in the $u$ band, in particular, is a cause for concern.\\ 
\citet{tex:grahampetrosian} analytically calculate how the ratio of S\'{e}rsic flux to Petrosian flux changes with the S\'{e}rsic index of the object. The fraction of light missed by a Petrosian aperture is dependent upon the light profile of source. Figure \ref{fig:sersallobjsr} shows the distribution of S\'{e}rsic - GAMA $r$-defined Petrosian magnitude against S\'{e}rsic index, redshift, absolute and apparent magnitude for all $r$-band objects in the GAMA sample that have passed our star-galaxy separation criteria, and have credible $r$, $u$ and $K$ $r$-defined PETRO magnitudes. \citeauthor{tex:grahampetrosian} report a $0.20$\,mag offset for an $n=4$ profile, and a $0.50$\,mag offset for an $n=8$ profile. The median $r_{Sersic}-r_{Petrosian}$ offset for objects with $3.9<n<4.1$ in this sample is $-0.115$\,mag, with rms scatter of $0.212$\,mag, and $-0.408$\,mag, for objects with $7.9<n<8.1$, with rms scatter of $0.292$\,mag. Both results agree with the reported values. We have plotted the magnitude offset with S\'{e}rsic index function from Figure.~2 (their Panel a) of \citet{tex:grahampetrosian} in the uppermost plot of Figure \ref{fig:sersallobjsr}. The function is an extremely good match to our photometry. Figure \ref{fig:cmodelallobjsr} shows the distribution of S\'{e}rsic - SDSS \textsc{cmodel} magnitude against S\'{e}rsic index, redshift, absolute and apparent magnitude for all $r$-band objects in the GAMA sample that have passed our star-galaxy separation criteria, and have credible $r$, $u$ and $K$ $r$-defined PETRO magnitudes. The distributions are very similar to those produced by the S\'{e}rsic-Petrosian colours in Figure \ref{fig:sersallobjsr}. An exception is the distribution with S\'{e}rsic index, where the S\'{e}rsic - \textsc{cmodel} offset is distributed closer to 0\,mag, until n=4, at which point the S\'{e}rsic magnitude detects more flux. As the \textsc{cmodel} magnitude is defined as a combination of n=1 and n=4 profiles, it is unsurprising that it cannot model high n profile sources as well as the GAMA S\'{e}rsic magnitude, which allows the n parameter greater freedom.\\ 
The $r$ band S\'{e}rsic magnitude shows no anomalous behaviour. S\'{e}rsic profiling is reliable when undertaken using the higher quality SDSS imaging (particularly $gri$), but not when using the noisier $u$ band data. It is clear that the $u$ band S\'{e}rsic magnitude is not robust enough to support detailed scientific investigations. In order to access a S\'{e}rsic-style total magnitude in the $u$ band, we are therefore forced to create one from existing, reliable data. We devise such an approach in Section \ref{sec:fixedapers}.

\subsection{`Fixed aperture' S\'{e}rsic magnitudes} \label{sec:fixedapers}
\begin{figure*}
\begin{center}
\includegraphics[height=640pt]{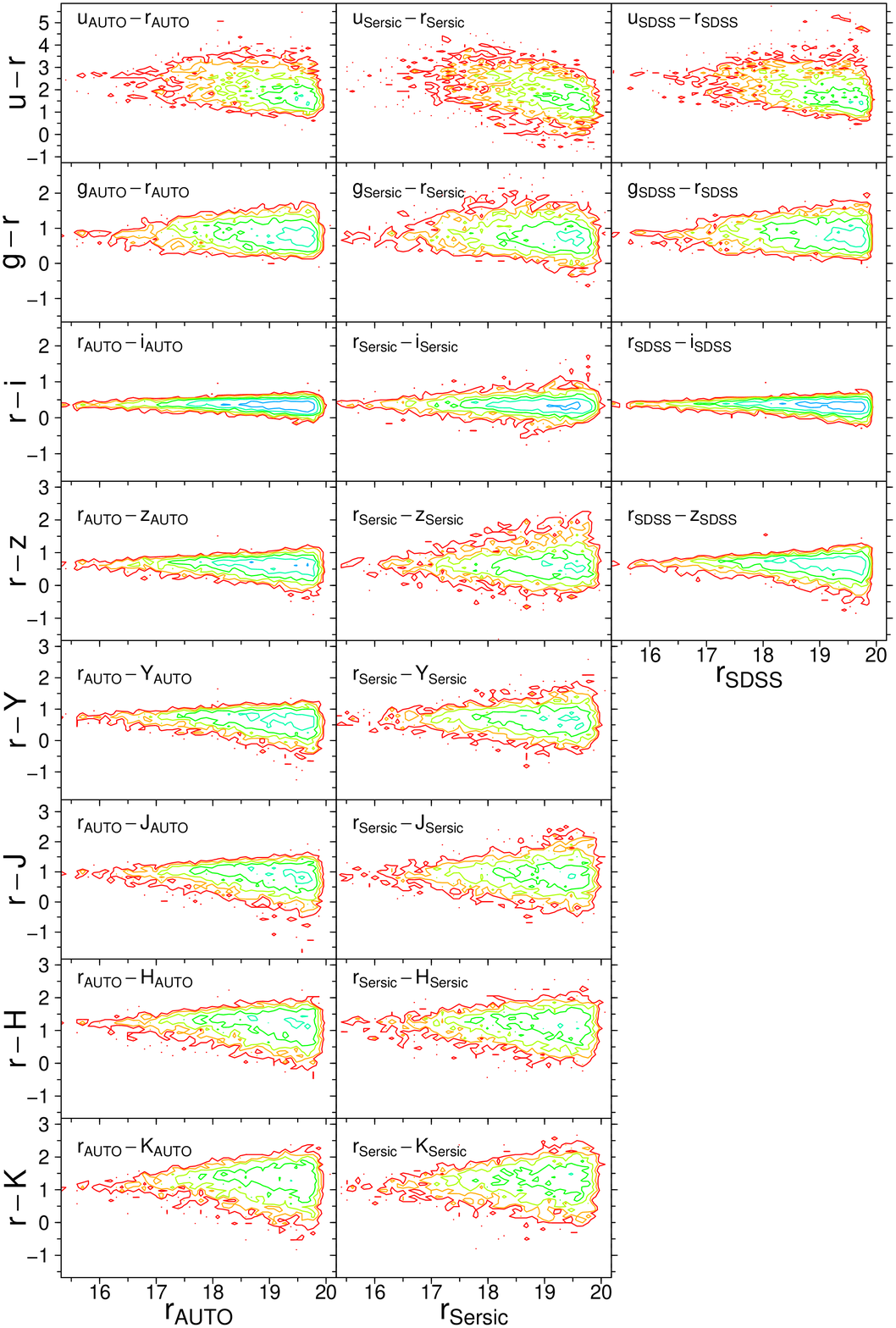}
\caption{$X$-$r$ distributions from GAMA $r$-defined Auto magnitudes, GAMA S\'{e}rsic magnitudes and SDSS \textsc{petromag}s (left to right), against $r$, for all $X=u$,$g$,$i$,$z$,$Y$,$J$,$H$ and $K$. Sources used are those within the subset region with credible Auto and S\'{e}rsic magnitudes, and without the SDSS saturated object bit set. Contours increase geometrically in powers of 2, from 2 to 512 galaxies bin$^{-1}$. Bins are 0.1\,mag $\times$ 0.1\,mag in size}
\label{fig:rcolours}
\end{center}
\end{figure*}

As mentioned in Section \ref{sec:Sersicap}, the S\'{e}rsic magnitude is taken from a different aperture in each band. We therefore cannot use S\'{e}rsic magnitudes to generate accurate colours (compare the scatter in the S\'{e}rsic colours and the AUTO colours in Figure \ref{fig:rcolours}). We also do not consider the $u$ band S\'{e}rsic magnitudes to be credible (see Section \ref{sec:sersiccheck}). However, we also believe that the $r$ band S\'{e}rsic luminosity function may be more desirable than the light-distribution defined aperture $r$ band luminosity functions. The calculation of the total luminosity density using a non-S\'{e}rsic aperture system may underestimate the parameter. We require a system that accounts for the additional light found by the S\'{e}rsic magnitude, but also provides a credible set of colours.\\ 
We derive a further magnitude $X_{total}$, using the equation $X_{total}=(X_{auto}-r_{auto}) + r_{Sersic}$, where $auto$ is the $r$-defined \textit{AUTO} magnitude. In effect, this creates a measure that combines the total $r$ band flux with optimal colours, using SDSS deblending to give us the most accurate catalogue of sources (by matching to the GAMA master catalogue); the best of all possibilities. We accept that this assumes that the colour from the $r$-defined AUTO aperture would be the same as the colour from a $r$-defined S\'{e}rsic aperture, however, this is the closest estimation to a fixed S\'{e}rsic aperture we can make at this time.

\subsection{Uncertainties within the photometry} \label{sec:magerrors}
\begin{figure}
\includegraphics[width=220pt]{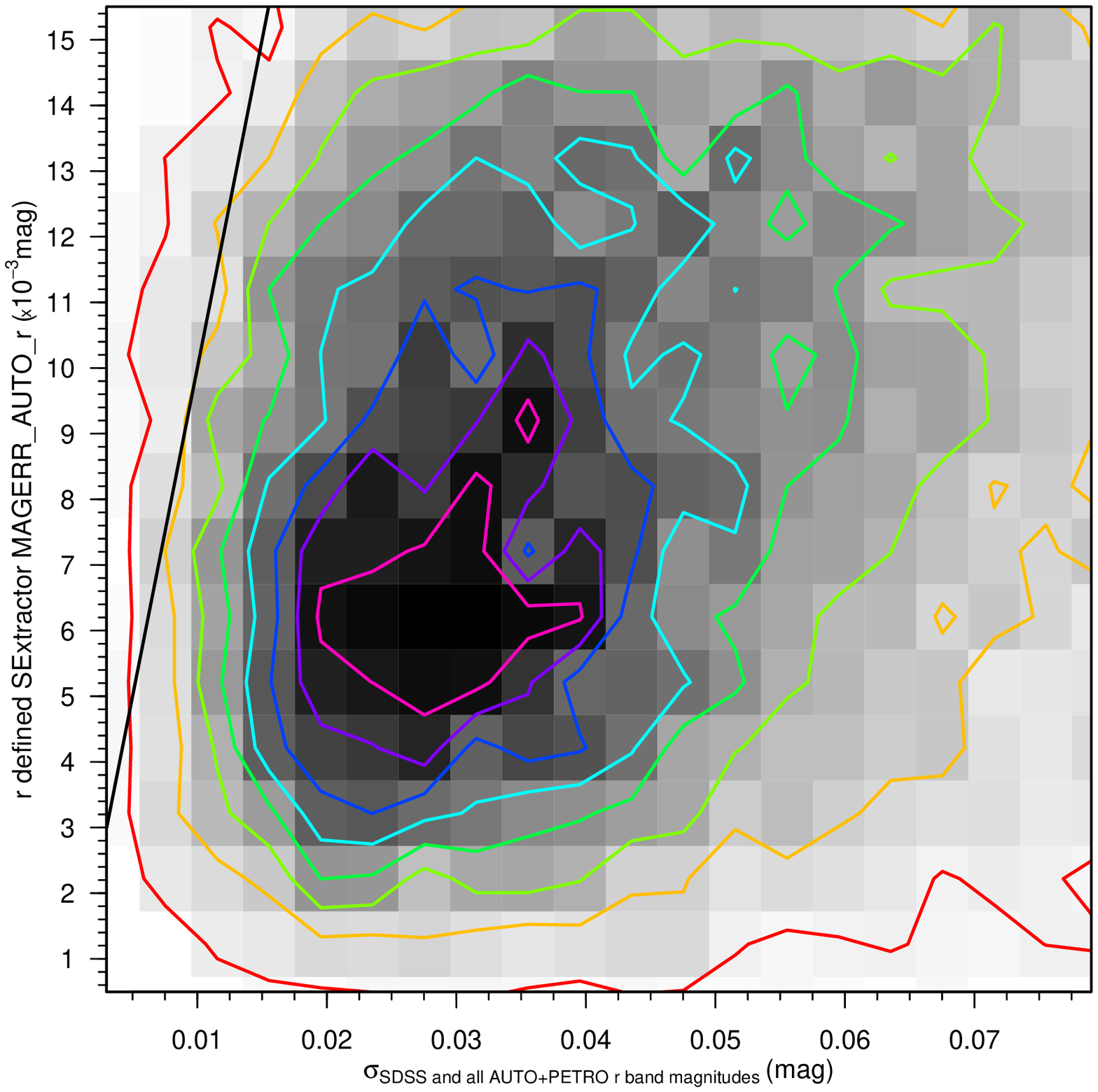}
\caption{The distribution of standard deviation in $r$ band apparent magnitude against \texttt{SExtractor}'s calculated magnitude error (using the first quartile gain from the gain distribution of the mosaic's input images) for our clean sample of galaxies, using SDSS, $r$-defined, $K$-defined and self-defined AUTO and PETRO magnitudes to calculate the standard deviation. Contours rise linearly by 16 galaxies bin$^{-1}$, ranging from 8 to 120 galaxies bin$^{-1}$. Bins are 0.004\,mag (x axis) $\times$ 0.001\,mag (y axis) in size.}
\label{fig:magerrsd}
\end{figure}

\begin{figure}
\includegraphics[width=220pt]{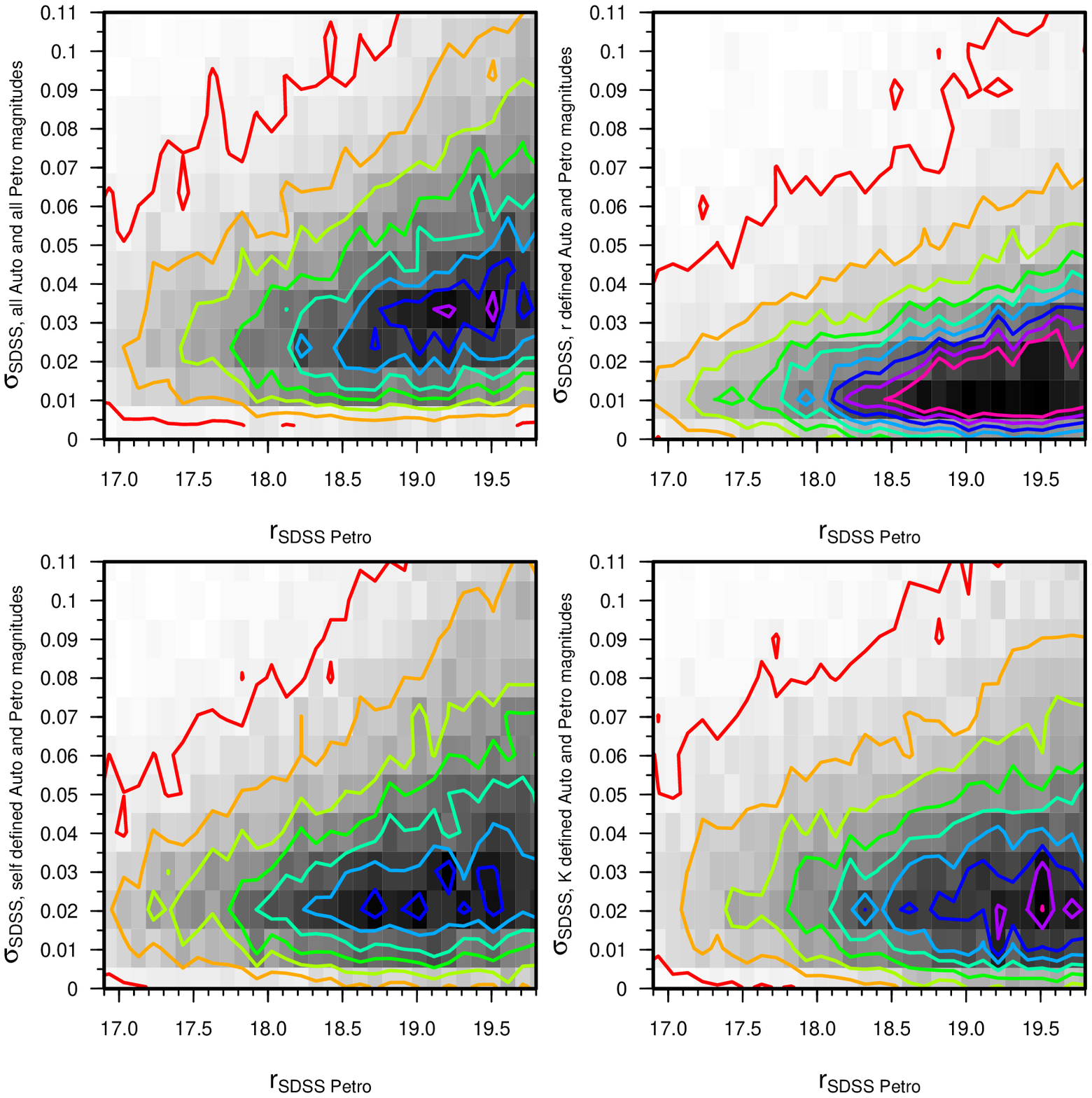}
\caption{The distribution of standard deviation in $r$ band apparent magnitude against apparent magnitude for our clean sample of galaxies, using four different sets of magnitudes to calculate the standard deviation in each case. Contours rise linearly by 20 galaxies bin$^{-1}$, ranging from 10 to 170 galaxies bin$^{-1}$. Bins are 0.1\,mag (x axis) $\times$ 0.01\,mag (y axis) in size.}
\label{fig:magstandev}
\end{figure}

The gain value in SDSS data is constant within each stripe but varies between stripes. The SDSS mosaic creation process that is detailed in sections \ref{sec:sdren} and \ref{sec:mosaic} combines images from a number of different stripes to generate the master mosaic. As the mosaics are transformed from different zeropoints, the relationship between electrons and pixel counts will be different for each image. This mosaic must suffer from variations in gain. The \texttt{SExtractor} utility can be set up to deal with this anomaly, by using the weightmaps generated by \texttt{SWARP}. However, this may introduce a level of surface brightness bias into the resulting catalogue that would be difficult to quantify. We calculate the \texttt{SExtractor} magnitude error via the first quartile value, taken from the distribution of gain parameters used to create the mosaic. The Gain used in the SDSS calculation is the average for the strip. The \texttt{SExtractor} error is calculated using Equation \ref{eqn:sexerror}, where $A$ as is the area of the aperture, $\sigma$ is the standard deviation in noise and $F$ is the total flux within the aperture. By using the first quartile gain value, we may be slightly overestimating the $\frac{F}{\rm gain}$ component of the magnitude uncertainty calculation. However, given the amount of background noise in the mosaic, this component will constitute only a small fraction towards the error in the fainter galaxies, and in the brighter galaxies the uncertainty in magnitude due to the aperture definition will be much greater than the \texttt{SExtractor} magnitude error itself. The \texttt{SExtractor} magnitude error is calculated separately for each aperture type, and is available within the GAMA photometric catalogues.\\
\begin{equation} \label{eqn:sexerror}
\Delta m= \frac{1.0857 \sqrt{A \sigma^{2} + \frac{F}{\rm gain}} }{F}\\
\end{equation} 
We have attempted to quantify the uncertainty due to the aperture definition, in order to calculate its extent relative to the \texttt{SExtractor} magnitude error. We use the cleaned sample defined in section \ref{sec:photomdist}. The dispersion in calculated magnitude between our different photometric methods for this sample are shown in Figures \ref{fig:rdpetrdauto}, \ref{fig:sdssauto}, \ref{fig:sdssrpetro}, \ref{fig:sdssselfpetro} and \ref{fig:sdssSersic}. Figure \ref{fig:magerrsd} shows the relative scales of the uncertainty due to a galaxy's aperture definition (calculated from the standard deviation in AUTO/PETRO luminosities from the SDSS survey and our $r$/$K$/self-defined catalogues) and the error generated by \texttt{SExtractor} in the $r$ band. The aperture definition uncertainty is generally much greater than that due to background variation and $\frac{S}{N}$ that \texttt{SExtractor} derives. Figure \ref{fig:magstandev} shows how this standard deviation in a galaxy's $r$ band magnitude changes with apparent magnitude. Whilst this uncertainty is larger than the \texttt{SExtractor} magnitude error, it is fundamentally a more consistent judgement of the uncertainty in a given galaxy's brightness as it does not assume that any particular extended-source aperture definition is correct. Whilst the dispersion of the relationship increases with apparent magnitude (along with the number of galaxies), the modal standard deviation is approximately constant. Taking this to be a good estimate of the average uncertainty in the apparent magnitude of a galaxy within our sample, we have confidence in our published apparent magnitudes to within $\pm 0.03$\,mag in $gri$, $\pm 0.06$\,mag in $z$, and $\pm 0.20$\,mag in $u$. We calculate the same statistics in the NIR passbands (though without SDSS \textsc{Petromag}). We have confidence in our published apparent magnitudes within $\pm 0.05$\,mag in $YJHK$; approximately two and a half times the size of the photometric rms error UKIDSS was designed to have ($\pm0.02$\,mag, \citealt{tex:ukirt}).

\subsection{Number counts} \label{sec:mlim}
In order to construct a unbiased dataset, it is necessary for us to calculate the apparent magnitude where the GAMA sample ceases to be complete. 
\subsubsection{Definition of GAMA galaxy sample used in this section}
The GAMA sample used in this section is defined as those SDSS objects that are within the area that has complete $ugrizYJHK$ colour coverage, and have  passed the star-galaxy separation criteria. Of the $908022$ objects in the GAMA master catalogue, only $124622$ fulfil this criteria. The area of sky that has complete GAMA $ugrizYJHK$ coverage is $129.1232 \pm 0.0008$\,sq deg; 89.7\% of the entire GAMA region. All magnitudes in this section are $r$-defined \textit{AUTO} magnitudes, unless otherwise defined.
\subsubsection{Determination of apparent magnitude limits}
\begin{figure*}
\includegraphics[width=440pt]{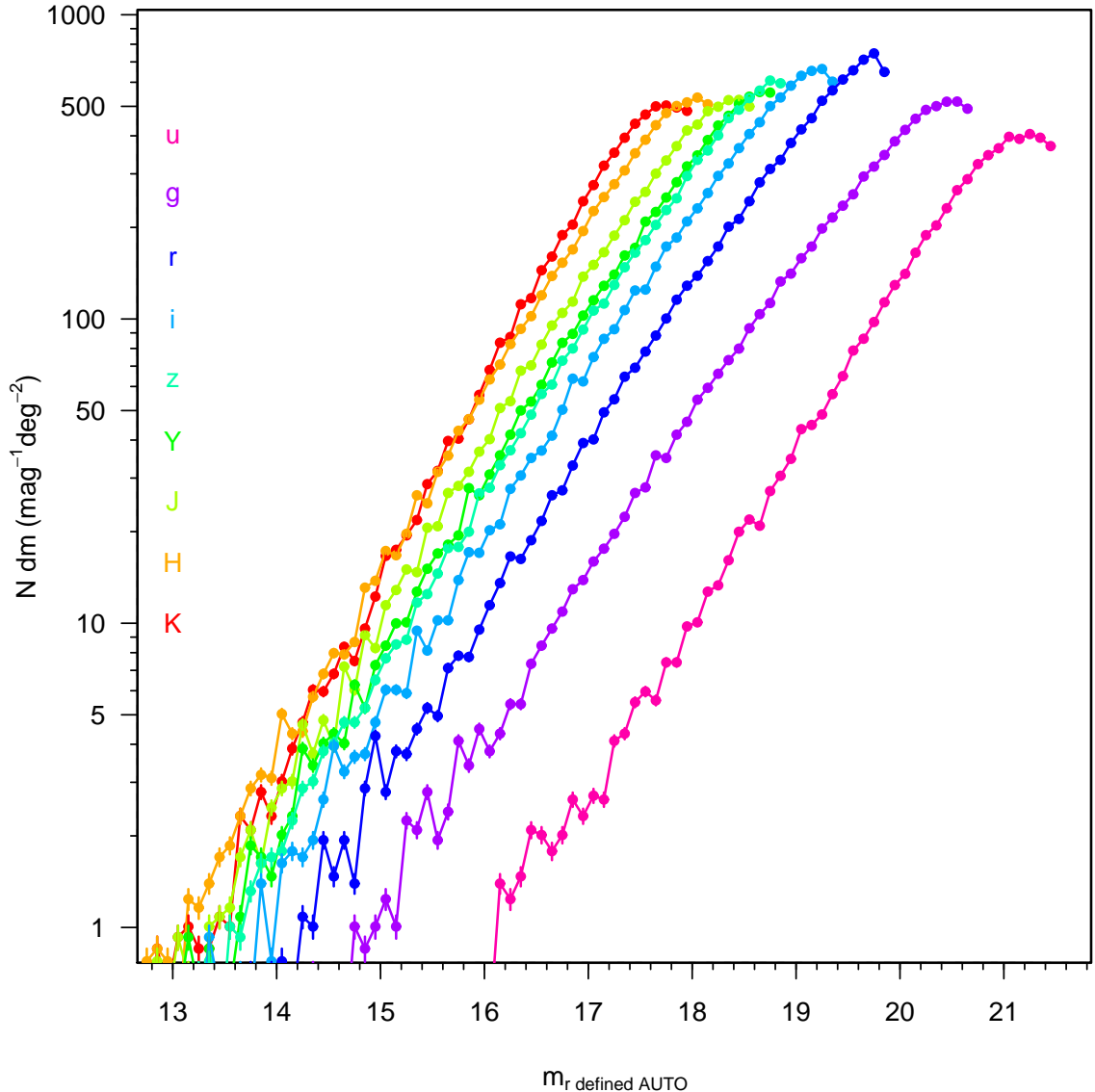}
\caption{Number counts of GAMA galaxies (sources that have passed our star-galaxy separation criteria) with good $ugrizYJHK$ colours, split into 0.1 magnitude bins and divided by the total area they cover. Error bars shown are for Poissonian number counts.}
\label{fig:numcounts}
\end{figure*}

Figure \ref{fig:numcounts} shows how the sky density of GAMA galaxies in the nine passbands varies with apparent magnitude. The distributions peak in the 0.1 magnitude bins centred at $u=21.25$, $g=20.55$, $r=19.75$, $i=19.25$, $z=18.75$, $Y=18.65$, $J=18.45$, $H=18.05$ and $K=17.75$\,mag. Tables \ref{tab:ugrcounts}, \ref{tab:izYcounts} and \ref{tab:JHKcounts} contain the number counts of GAMA galaxies in $ugrizYJHK$, this time using 0.25 magnitude bins. Poissonian uncertainties are also included. Both sets of data have been converted to deg$^{-2}$ mag$^{-1}$ units.\\
The $r$ band number count drop off, despite hitting the \textsc{petromag\_r}$=19.8$\,mag GAMA main sample magnitude limit, is not absolute because the SDSS limit was extended to \textsc{petromag\_r}$=20.5$\,mag in the GAMA 12 region so that useful filler objects could be selected, and because radio/$K$/$z$ band selected objects in G9 and G15 will also be included within the catalogue. Objects that are fainter than $r_{model}=20.5$\,mag (722 sources; 0.5\% of the sample) will be due to differences in object extraction between SDSS and \texttt{SExtractor}, as mentioned in previous sections.\\
The turnovers in Figure \ref{fig:numcounts} will occur where the $r=19.8$\,mag limit is reached for galaxies with the median $passband-r$ colour.  We are within the domain where the number of galaxies within a magnitude bin increases linearly with increasing apparent magnitude, but a deviation from this relationship is visible in the figure approximately 3 magnitude bins before the turnover occurs in all bands except $r$. This effect is due to colour incompleteness becoming a factor. Unfortunately, despite our radio/$K$/$z$ selection, there will be a population of objects that are bright in other passbands, but too faint in $r$ to be included within our sample. Assuming the $passband - r$ colour distribution is approximately Gaussian, this population will feature predominantly in the apparent magnitude bins near the turnover, causing the characteristic flattening we see. Accounting for this effect, we define the apparent magnitude sample limits of our sample to be a few bins brighter than this turnover, where the linear relationship still holds. Our apparent magnitude limits are set to $u=21.0$, $g=20.3$, $r=19.8$, $i=19.0$, $z=18.5$, $Y=18.4$, $J=18.2$, $H=17.8$ and $K=17.6$\,mag.

\begin{table}
\begin{tabular}{p{2.9cm}ccc} \hline \hline
Band&Sources&Redshifts&\% Redshifts\\ \hline
Star-galaxy separation criteria only&124622& 82926&66.5\\
u$\le$21.0&46006&39767&86.4\\
g$\le$20.3&67913&58956&86.8\\
r$\le$19.8&106032&79672&75.1\\
i$\le$19.0&74885&66981&89.4\\
z$\le$18.5&59470&55202&92.8\\
Y$\le$18.4&57739&53339&92.4\\
J$\le$18.2&60213&54264&90.1\\
H$\le$17.8&55734&51033&91.6\\
K$\le$17.5&46424&43252&93.2\\
\hline
\end{tabular}
\caption{The number of sources within the star-galaxy separation and apparent magnitude limited GAMA samples that have a complete set of good $ugrizYJHK$ r-defined magnitudes, the number of those sources that have redshifts from first and second year data and the percentage redshift completeness. Apparent magnitudes are r-defined magnitudes, using the AB magnitude system.}
\label{tab:appmaglimsamps}
\end{table}

\begin{table*}
 \begin{tabular}{p{0.9cm}p{2.8cm}p{0.9cm}p{2.8cm}p{0.9cm}p{2.8cm}}\hline\hline
u (mag)&$N_{u}(m)$$\pm$$\sigma_{N_{u}(m)}$ ($deg^{-2}$ $(mag)^{-1}$)&g (mag)&$N_{g}(m)$$\pm$$\sigma_{N_{g}(m)}$ ($deg^{-2}$ ($mag)^{-1}$)&r (mag)&$N_{r}(m)$$\pm$$\sigma_{N_{r}(m)}$ $(deg^{-2}$ $(mag)^{-1}$)\\
\hline
12.125&0.031 $\pm$ 0.015&10.125&0 $\pm$ 0&10.125&0.031 $\pm$ 0.015\\
12.375&0 $\pm$ 0&10.375&0 $\pm$ 0&10.375&0 $\pm$ 0\\
12.625&0 $\pm$ 0&10.625&0 $\pm$ 0&10.625&0 $\pm$ 0\\
12.875&0 $\pm$ 0&10.875&0.031 $\pm$ 0.015&10.875&0.031 $\pm$ 0.015\\
13.125&0 $\pm$ 0&11.125&0 $\pm$ 0&11.125&0 $\pm$ 0\\
13.375&0.031 $\pm$ 0.015&11.375&0 $\pm$ 0&11.375&0.124 $\pm$ 0.031\\
13.625&0.093 $\pm$ 0.027&11.625&0.031 $\pm$ 0.015&11.625&0.062 $\pm$ 0.022\\
13.875&0.062 $\pm$ 0.022&11.875&0 $\pm$ 0&11.875&0 $\pm$ 0\\
14.125&0.093 $\pm$ 0.027&12.125&0.062 $\pm$ 0.022&12.125&0.031 $\pm$ 0.015\\
14.375&0.062 $\pm$ 0.022&12.375&0.124 $\pm$ 0.031&12.375&0.031 $\pm$ 0.015\\
14.625&0.062 $\pm$ 0.022&12.625&0.031 $\pm$ 0.015&12.625&0.062 $\pm$ 0.022\\
14.875&0.062 $\pm$ 0.022&12.875&0.031 $\pm$ 0.015&12.875&0.155 $\pm$ 0.035\\
15.125&0.124 $\pm$ 0.031&13.125&0.031 $\pm$ 0.015&13.125&0.217 $\pm$ 0.041\\
15.375&0.248 $\pm$ 0.044&13.375&0.062 $\pm$ 0.022&13.375&0.186 $\pm$ 0.038\\
15.625&0.279 $\pm$ 0.046&13.625&0.062 $\pm$ 0.022&13.625&0.558 $\pm$ 0.066\\
15.875&0.527 $\pm$ 0.064&13.875&0.248 $\pm$ 0.044&13.875&0.589 $\pm$ 0.068\\
16.125&0.929 $\pm$ 0.085&14.125&0.248 $\pm$ 0.044&14.125&0.712 $\pm$ 0.074\\
16.375&1.735 $\pm$ 0.116&14.375&0.712 $\pm$ 0.074&14.375&1.425 $\pm$ 0.105\\
16.625&1.828 $\pm$ 0.119&14.625&0.62 $\pm$ 0.069&14.625&1.611 $\pm$ 0.112\\
16.875&2.478 $\pm$ 0.139&14.875&0.836 $\pm$ 0.08&14.875&3.16 $\pm$ 0.156\\
17.125&3.098 $\pm$ 0.155&15.125&1.27 $\pm$ 0.099&15.125&3.253 $\pm$ 0.159\\
17.375&4.616 $\pm$ 0.189&15.375&2.478 $\pm$ 0.139&15.375&4.771 $\pm$ 0.192\\
17.625&6.01 $\pm$ 0.216&15.625&2.447 $\pm$ 0.138&15.625&6.536 $\pm$ 0.225\\
17.875&8.457 $\pm$ 0.256&15.875&4.089 $\pm$ 0.178&15.875&8.333 $\pm$ 0.254\\
18.125&11.772 $\pm$ 0.302&16.125&4.213 $\pm$ 0.181&16.125&13.352 $\pm$ 0.322\\
18.375&17.1 $\pm$ 0.364&16.375&6.32 $\pm$ 0.221&16.375&17.286 $\pm$ 0.366\\
18.625&22.273 $\pm$ 0.415&16.625&9.139 $\pm$ 0.266&16.625&24.783 $\pm$ 0.438\\
18.875&31.815 $\pm$ 0.496&16.875&13.166 $\pm$ 0.319&16.875&34.2 $\pm$ 0.515\\
19.125&44.763 $\pm$ 0.589&17.125&16.728 $\pm$ 0.36&17.125&46.808 $\pm$ 0.602\\
19.375&58.58 $\pm$ 0.674&17.375&24.225 $\pm$ 0.433&17.375&64.28 $\pm$ 0.706\\
19.625&84.973 $\pm$ 0.811&17.625&31.722 $\pm$ 0.496&17.625&85.933 $\pm$ 0.816\\
19.875&117.159 $\pm$ 0.953&17.875&42.719 $\pm$ 0.575&17.875&118.491 $\pm$ 0.958\\
20.125&159.754 $\pm$ 1.112&18.125&58.673 $\pm$ 0.674&18.125&151.452 $\pm$ 1.083\\
20.375&211.736 $\pm$ 1.281&18.375&74.502 $\pm$ 0.76&18.375&200.955 $\pm$ 1.248\\
20.625&282.831 $\pm$ 1.48&18.625&101.299 $\pm$ 0.886&18.625&271.524 $\pm$ 1.45\\
20.875&351.602 $\pm$ 1.65&18.875&132.122 $\pm$ 1.012&18.875&347.885 $\pm$ 1.641\\
21.125&396.304 $\pm$ 1.752&19.125&170.256 $\pm$ 1.148&19.125&454.357 $\pm$ 1.876\\
21.375&386.731 $\pm$ 1.731&19.375&222.175 $\pm$ 1.312&19.375&575.729 $\pm$ 2.112\\
 &  &19.625&282.428 $\pm$ 1.479&19.625&695.832 $\pm$ 2.321\\
 &  &19.875&356.559 $\pm$ 1.662&19.875&540.445 $\pm$ 2.046\\
 &  &20.125&446.148 $\pm$ 1.859& &   \\
 &  &20.375&505.471 $\pm$ 1.979& &  \\
 &  &20.625&492.894 $\pm$ 1.954& &  \\
\hline
\end{tabular}
\caption{Number counts for the $ugr$ filters, using r-defined AUTO photometry, with Poissonian uncertainties.}
\label{tab:ugrcounts}
\end{table*}

\begin{table*}
 \begin{tabular}{p{0.9cm}p{2.8cm}p{0.9cm}p{2.8cm}p{0.9cm}p{2.8cm}}\hline\hline
i (mag)&$N_{i}(m)$$\pm$$\sigma_{N_{i}(m)}$ ($deg^{-2}$ $(mag)^{-1}$)&z (mag)&$N_{z}(m)$$\pm$$\sigma_{N_{z}(m)}$ ($deg^{-2}$ $(mag)^{-1}$)&Y (mag)&$N_{Y}(m)$ $\pm$$\sigma_{N_{Y}(m)}$ ($deg^{-2}$ $(mag)^{-1}$)\\
\hline
9.375&0 $\pm$ 0&9.375&0 $\pm$ 0&9.375&0.031 $\pm$ 0.015\\
9.625&0 $\pm$ 0&9.625&0.031 $\pm$ 0.015&9.625&0 $\pm$ 0\\
9.875&0.031 $\pm$ 0.015&9.875&0 $\pm$ 0&9.875&0.031 $\pm$ 0.015\\
10.125&0 $\pm$ 0&10.125&0.031 $\pm$ 0.015&10.125&0 $\pm$ 0\\
10.375&0.031 $\pm$ 0.015&10.375&0 $\pm$ 0&10.375&0 $\pm$ 0\\
10.625&0 $\pm$ 0&10.625&0.124 $\pm$ 0.031&10.625&0.124 $\pm$ 0.031\\
10.875&0.124 $\pm$ 0.031&10.875&0.031 $\pm$ 0.015&10.875&0.031 $\pm$ 0.015\\
11.125&0.031 $\pm$ 0.015&11.125&0.031 $\pm$ 0.015&11.125&0.031 $\pm$ 0.015\\
11.375&0.031 $\pm$ 0.015&11.375&0 $\pm$ 0&11.375&0 $\pm$ 0\\
11.625&0 $\pm$ 0&11.625&0.031 $\pm$ 0.015&11.625&0.062 $\pm$ 0.022\\
11.875&0.031 $\pm$ 0.015&11.875&0.031 $\pm$ 0.015&11.875&0.062 $\pm$ 0.022\\
12.125&0.031 $\pm$ 0.015&12.125&0.093 $\pm$ 0.027&12.125&0.093 $\pm$ 0.027\\
12.375&0.186 $\pm$ 0.038&12.375&0.217 $\pm$ 0.041&12.375&0.186 $\pm$ 0.038\\
12.625&0.186 $\pm$ 0.038&12.625&0.248 $\pm$ 0.044&12.625&0.217 $\pm$ 0.041\\
12.875&0.124 $\pm$ 0.031&12.875&0.31 $\pm$ 0.049&12.875&0.403 $\pm$ 0.056\\
13.125&0.372 $\pm$ 0.054&13.125&0.558 $\pm$ 0.066&13.125&0.651 $\pm$ 0.071\\
13.375&0.682 $\pm$ 0.073&13.375&0.62 $\pm$ 0.069&13.375&0.743 $\pm$ 0.076\\
13.625&0.62 $\pm$ 0.069&13.625&1.022 $\pm$ 0.089&13.625&1.022 $\pm$ 0.089\\
13.875&0.96 $\pm$ 0.086&13.875&1.611 $\pm$ 0.112&13.875&1.673 $\pm$ 0.114\\
14.125&1.704 $\pm$ 0.115&14.125&1.983 $\pm$ 0.124&14.125&2.478 $\pm$ 0.139\\
14.375&2.168 $\pm$ 0.13&14.375&3.501 $\pm$ 0.165&14.375&3.779 $\pm$ 0.171\\
14.625&3.748 $\pm$ 0.17&14.625&4.554 $\pm$ 0.188&14.625&4.585 $\pm$ 0.188\\
14.875&3.965 $\pm$ 0.175&14.875&5.545 $\pm$ 0.207&14.875&6.289 $\pm$ 0.221\\
15.125&5.917 $\pm$ 0.214&15.125&8.116 $\pm$ 0.251&15.125&9.015 $\pm$ 0.264\\
15.375&8.302 $\pm$ 0.254&15.375&11.555 $\pm$ 0.299&15.375&13.506 $\pm$ 0.323\\
15.625&10.749 $\pm$ 0.289&15.625&16.697 $\pm$ 0.36&15.625&17.689 $\pm$ 0.37\\
15.875&16.635 $\pm$ 0.359&15.875&21.994 $\pm$ 0.413&15.875&25.774 $\pm$ 0.447\\
16.125&22.211 $\pm$ 0.415&16.125&32.031 $\pm$ 0.498&16.125&34.727 $\pm$ 0.519\\
16.375&31.598 $\pm$ 0.495&16.375&43.4 $\pm$ 0.58&16.375&49.937 $\pm$ 0.622\\
16.625&40.984 $\pm$ 0.563&16.625&61.43 $\pm$ 0.69&16.625&69.484 $\pm$ 0.734\\
16.875&60.872 $\pm$ 0.687&16.875&83.703 $\pm$ 0.805&16.875&93.802 $\pm$ 0.852\\
17.125&82.464 $\pm$ 0.799&17.125&113.318 $\pm$ 0.937&17.125&124.873 $\pm$ 0.983\\
17.375&111.367 $\pm$ 0.929&17.375&151.266 $\pm$ 1.082&17.375&161.83 $\pm$ 1.12\\
17.625&142.995 $\pm$ 1.052&17.625&198.849 $\pm$ 1.241&17.625&221.277 $\pm$ 1.309\\
17.875&193.397 $\pm$ 1.224&17.875&264.213 $\pm$ 1.43&17.875&292.124 $\pm$ 1.504\\
18.125&253.432 $\pm$ 1.401&18.125&356.156 $\pm$ 1.661&18.125&375.424 $\pm$ 1.705\\
18.375&336.423 $\pm$ 1.614&18.375&458.973 $\pm$ 1.885&18.375&480.285 $\pm$ 1.929\\
18.625&438.093 $\pm$ 1.842&18.625&561.2 $\pm$ 2.085&18.625&549.831 $\pm$ 2.064\\
18.875&549.336 $\pm$ 2.063&18.875&582.39 $\pm$ 2.124& & \\
19.125&649.457 $\pm$ 2.243& &  & &    \\
19.375&564.205 $\pm$ 2.09& &  & &  \\
\hline
\end{tabular}
\caption{Number counts for the $izY$ filters, using r-defined AUTO photometry, with Poissonian uncertainties.}
\label{tab:izYcounts}
\end{table*}

\begin{table*}
 \begin{tabular}{p{0.9cm}p{2.9cm}p{0.9cm}p{2.9cm}p{0.9cm}p{2.9cm}}\hline\hline
J (mag)&$N_{J}(m)$$\pm$$\sigma_{N_{J}(m)}$ ($deg^{-2}$ $(mag)^{-1}$)&H (mag)&$N_{H}(m)$$\pm$$\sigma_{N_{H}(m)}$ $(deg^{-2}$ $(mag)^{-1})$&K (mag)&$N_{K}(m)$$\pm$$\sigma_{N_{K}(m)}$ $(deg^{-2}$ $(mag)^{-1})$\\
\hline
9.125&0 $\pm$ 0&9.125&0.031 $\pm$ 0.015&9.125&0 $\pm$ 0\\
9.375&0 $\pm$ 0&9.375&0 $\pm$ 0&9.375&0.031 $\pm$ 0.015\\
9.625&0.031 $\pm$ 0.015&9.625&0.031 $\pm$ 0.015&9.625&0 $\pm$ 0\\
9.875&0.031 $\pm$ 0.015&9.875&0 $\pm$ 0&9.875&0.031 $\pm$ 0.015\\
10.125&0 $\pm$ 0&10.125&0.093 $\pm$ 0.027&10.125&0.031 $\pm$ 0.015\\
10.375&0.062 $\pm$ 0.022&10.375&0.062 $\pm$ 0.022&10.375&0.062 $\pm$ 0.022\\
10.625&0.062 $\pm$ 0.022&10.625&0.031 $\pm$ 0.015&10.625&0.031 $\pm$ 0.015\\
10.875&0.062 $\pm$ 0.022&10.875&0 $\pm$ 0&10.875&0.062 $\pm$ 0.022\\
11.125&0 $\pm$ 0&11.125&0 $\pm$ 0&11.125&0 $\pm$ 0\\
11.375&0 $\pm$ 0&11.375&0.031 $\pm$ 0.015&11.375&0 $\pm$ 0\\
11.625&0.031 $\pm$ 0.015&11.625&0.124 $\pm$ 0.031&11.625&0.062 $\pm$ 0.022\\
11.875&0.124 $\pm$ 0.031&11.875&0.155 $\pm$ 0.035&11.875&0.062 $\pm$ 0.022\\
12.125&0.155 $\pm$ 0.035&12.125&0.248 $\pm$ 0.044&12.125&0.186 $\pm$ 0.038\\
12.375&0.217 $\pm$ 0.041&12.375&0.372 $\pm$ 0.054&12.375&0.217 $\pm$ 0.041\\
12.625&0.31 $\pm$ 0.049&12.625&0.527 $\pm$ 0.064&12.625&0.31 $\pm$ 0.049\\
12.875&0.527 $\pm$ 0.064&12.875&0.774 $\pm$ 0.077&12.875&0.682 $\pm$ 0.073\\
13.125&0.712 $\pm$ 0.074&13.125&0.867 $\pm$ 0.082&13.125&0.867 $\pm$ 0.082\\
13.375&0.898 $\pm$ 0.083&13.375&1.518 $\pm$ 0.108&13.375&1.022 $\pm$ 0.089\\
13.625&1.549 $\pm$ 0.11&13.625&2.23 $\pm$ 0.131&13.625&1.735 $\pm$ 0.116\\
13.875&2.076 $\pm$ 0.127&13.875&3.098 $\pm$ 0.155&13.875&2.478 $\pm$ 0.139\\
14.125&3.191 $\pm$ 0.157&14.125&4.492 $\pm$ 0.187&14.125&3.903 $\pm$ 0.174\\
14.375&4.43 $\pm$ 0.185&14.375&6.041 $\pm$ 0.216&14.375&5.545 $\pm$ 0.207\\
14.625&5.7 $\pm$ 0.21&14.625&7.806 $\pm$ 0.246&14.625&7.621 $\pm$ 0.243\\
14.875&8.147 $\pm$ 0.251&14.875&12.763 $\pm$ 0.314&14.875&10.192 $\pm$ 0.281\\
15.125&12.639 $\pm$ 0.313&15.125&17.72 $\pm$ 0.37&15.125&17.689 $\pm$ 0.37\\
15.375&17.224 $\pm$ 0.365&15.375&24.194 $\pm$ 0.433&15.375&23.915 $\pm$ 0.43\\
15.625&24.163 $\pm$ 0.433&15.625&35.036 $\pm$ 0.521&15.625&35.966 $\pm$ 0.528\\
15.875&33.425 $\pm$ 0.509&15.875&49.379 $\pm$ 0.618&15.875&49.999 $\pm$ 0.622\\
16.125&46.715 $\pm$ 0.601&16.125&70.413 $\pm$ 0.738&16.125&77.879 $\pm$ 0.777\\
16.375&66.448 $\pm$ 0.717&16.375&94.298 $\pm$ 0.855&16.375&109.198 $\pm$ 0.92\\
16.625&92.222 $\pm$ 0.845&16.625&133.237 $\pm$ 1.016&16.625&158.422 $\pm$ 1.108\\
16.875&121.434 $\pm$ 0.97&16.875&176.792 $\pm$ 1.17&16.875&218.272 $\pm$ 1.3\\
17.125&162.976 $\pm$ 1.123&17.125&245.781 $\pm$ 1.38&17.125&305.631 $\pm$ 1.538\\
17.375&220.348 $\pm$ 1.306&17.375&319.602 $\pm$ 1.573&17.375&406.031 $\pm$ 1.773\\
17.625&288.159 $\pm$ 1.494&17.625&420.931 $\pm$ 1.806&17.625&488.99 $\pm$ 1.946\\
17.875&384.191 $\pm$ 1.725&17.875&504.448 $\pm$ 1.977&17.875&491.716 $\pm$ 1.951\\
18.125&464.796 $\pm$ 1.897&18.125&515.229 $\pm$ 1.998& &  \\
18.375&521.394 $\pm$ 2.009& &  & &   \\
\hline
\end{tabular}
\caption{Number counts for the $JHK$ filters, using r-defined AUTO photometry, with Poissonian uncertainties.}
\label{tab:JHKcounts}
\end{table*}

\subsubsection{GAMA apparent-magnitude limited catalogues} \label{sec:gamcat}
Table \ref{tab:appmaglimsamps} contains the sizes of the apparent magnitude limited samples, and their current redshift completeness. Our magnitude limited optical samples contain approximately forty thousand less galaxies than the equivalent samples in \citet{tex:blantonsdsslf}, which covers the SDSS DR2 region, but extend two magnitudes deeper. When we compare our number counts to 6dFGS \citep{tex:hj}; our samples are also smaller in area coverage, but similar in size and much deeper in magnitude completeness. \citet{tex:hill}, our previous attempt at defining a sample across the optical and NIR (combining MGC, SDSS and UKIDSS data to form a $B$ band selected $ugrizYJHK$ catalogue), was just one tenth of the size and was $0.2-1.8$\,mag shallower. Currently, a large fraction of our samples have not been spectroscopically sampled. After the completion of the 2008-2009  allocations of AAOmega spectroscopy, our apparent magnitude samples have $\ge75$\% completeness. It is anticipated that this statistic will have raised to $\ge95$\% after the completion of the 2010 allocation.

\subsection{Incorporating GALEX data}
We have also combined the GAMA sample with UV data. The GAMA master catalogue has been matched to GALEX photometry \citep{tex:galex}. As GALEX observations are low resolution (typical imaging FWHM $\sim 10$\,arcsec), the matching is complex compared to the simple UKIDSS/SDSS matching described within this paper as a number of separate SDSS objects may be matched to one larger GALEX object.  The precise method of generating the GALEX matches is described in Robotham et al (in prep). In summary, all SDSS objects within the 90\% Petrosian radius of a GALEX source are considered to be contributing flux to that source. The flux of the GALEX object is then apportioned to the SDSS objects, with the alloted fraction calculated via the distance between the SDSS and GALEX object. If no other nearby source is within $2.5$\,mag (in $g$) of the closest match, all flux is assigned to the closest match. GALEX has two distinct filters $NUV$ and $FUV$. The generated magnitudes are stored within columns labelled \textit{MAG\_AUTO\_FUV} and \textit{MAG\_AUTO\_NUV}.

\subsection{SED fits using GAMA data} \label{sec:sedcomp}
\begin{figure*}
\includegraphics[width=430pt]{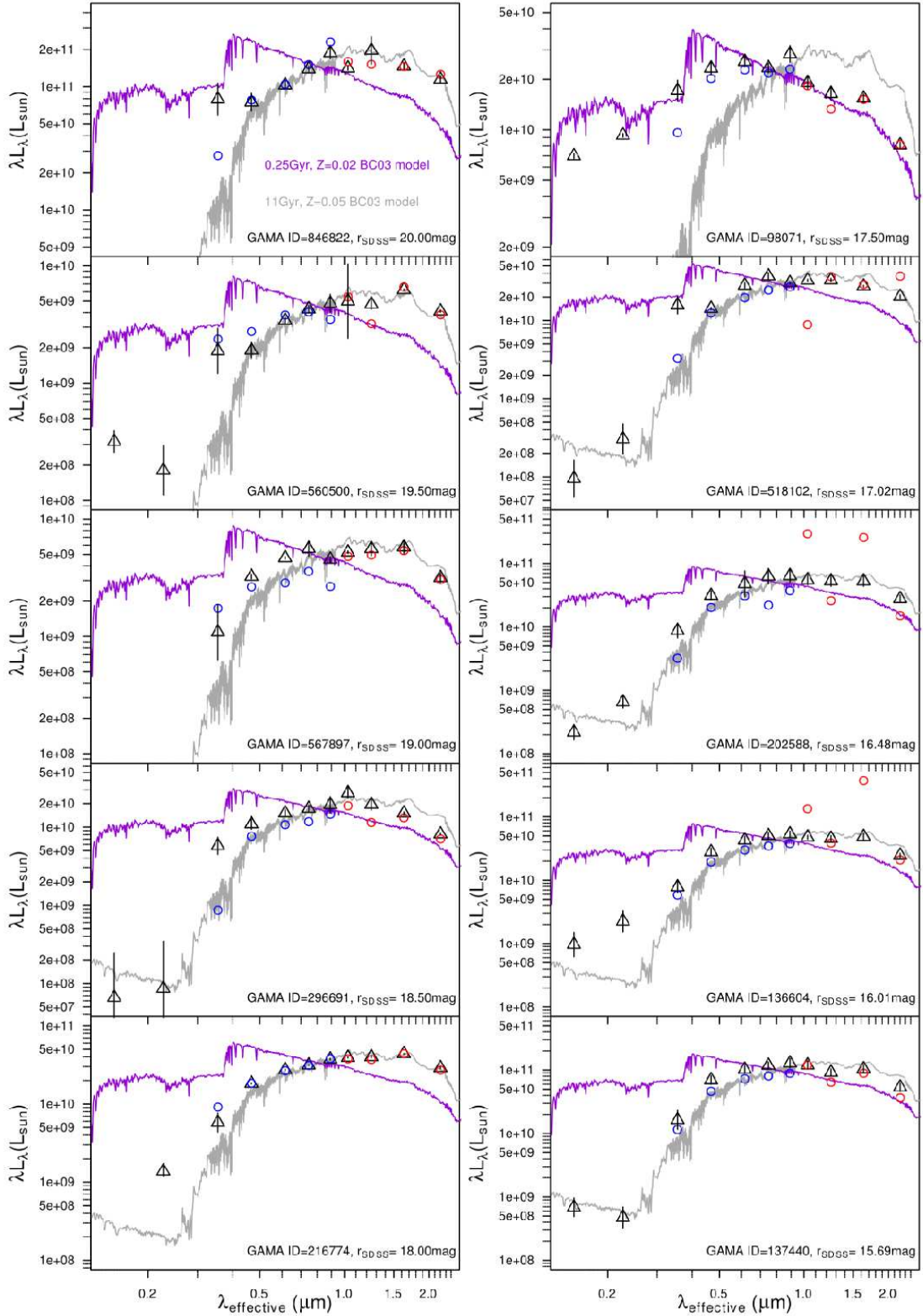}
\caption{SEDs of 10 GAMA galaxies using GAMA matched GALEX photometry and $r$ defined AUTO $ugrizYJHK$ photometry (black triangles), and the comparable SDSS (blue circles) and UKIDSS (red circles) \textsc{PETROMAG} photometry. Uncertainties shown for GAMA $ugrizYJHK$ points are calculated from the standard deviation in the photometry (as in Section \ref{sec:magerrors}). GAMA-GALEX uncertainties are \texttt{SExtractor} errors from the GALEX pipeline catalogues. Two Bruzual-Charlot 03 models are also plotted: the grey line is a 11\,Gyr model using Z=0.05 and the purple line is a 0.25Gyr model using Z=0.02 (Z$\astrosun$). The models shown are the same as those in Figure \ref{fig:photomsed}.}
\label{fig:seds}
\end{figure*}
The SEDs of 10 galaxies selected at random from the GAMA sample are shown in Figure \ref{fig:seds}. We show the GAMA-Galex UV luminosities, GAMA Total luminosities, Petrosian luminosities taken from the UKIDSS and SDSS surveys, and 2 Bruzual-Charlot \citep{tex:bc03} galaxy models with different ages and metallicities. The models are normalised via least squares best-fitting to the 9 GAMA datapoints. For image clarity, we do not show the uncertainties on SDSS and UKIDSS datapoints. GAMA UV uncertainties are taken from \texttt{SExtractor} magnitude errors. GAMA optical and NIR uncertainties are calculated using the standard deviation in the luminosity when different photometric methods are used (following the method described in Section \ref{sec:magerrors}). In some cases, the photometry provided by Survey-catalogues and the GAMA photometry are near identical, and match the galaxy models well (see $216774$ and $137440$ on Figure \ref{fig:seds}). In other cases, where there is a discrepancy between our derived luminosities and the Survey-catalogue parameters, the GAMA photometry is a better fit to the models (e.g., $202588$, $518102$ on Figure \ref{fig:seds}). We therefore judge our $r$-defined AUTO colours to be a significant improvement.
\subsection{Released GAMA photometry}
The GAMA photometry described in this paper provides the \textit{GamaPhotom} catalogue. This catalogue is filtered and combined with the other GAMA catalogues to produce the first GAMA data release, defined in Driver et al, 2010.

\section{The impact of the photometric method on the observed luminosity distribution} \label{sec:photomimpact}
\subsection{Comparison between $r$ band luminosity functions} \label{sec:rlf}
In order to illustrate the effect that photometric methods have on statistical measurements of the galaxy population, we derive the $r$ band luminosity function using 9 different photometric methods. We use all our photometric systems and the original SDSS photometry, and derive luminosity functions from the same population of galaxies. This should provide a consistent analysis for each method, removing all systematic effects except for that produced by the photometric method.

\subsection{Luminosity distribution and function measurement}
A number of techniques exist for measuring the galaxy luminosity distribution (see \citealt{tex:willmer}), and functions to parameterise it. We follow the methodology described in \citet{tex:hill}. This utilises the stepwise maximum-likelihood method (SWML), originally described in detail in \citet{tex:eep}, and the standard functional form, the Schechter luminosity function \citep{tex:schechter}. In \citet{tex:hill}, we developed a unique flux limit for each object based upon the spectroscopic limit and a colour limit. As we are working only in the $r$ band, the colour limit is now used to calculate the change in the magnitude limit between the studied photometric method and SDSS \textsc{petromag}. We now set the apparent magnitude threshold for each object using $r_{aperture, \rm{limit}}=19.4$\,mag.  An unfortunate side effect of the SWML method is that it requires normalisation to calculate the luminosity density. We use the method of luminosity density scaling described in sections 3.1.2 and 3.1.4 of \citet{tex:hill}. This involves calculating the number density of galaxies within a 1 magnitude range containing the $M^{*}$mag galaxies, and using this to work out the required scaling multiplier. We also account for the cosmic variance within the GAMA regions. We calculate the source density of galaxies within a 5150 sq deg section of the SDSS survey (large enough for cosmic variance to be negligible) with dereddened $-21.09<M_r - 5 log_{10} h<-20.09$ (i.e. $M^{*} - 5 log_{10} h \pm 0.5$\,mag, taking $M^* - 5 log_{10} h$ from the $r$-defined $r_{AUTO}$ photometry) and $0.023<z<0.1$, and compare this with the source density calculated (using the same catalogue) from the GAMA regions of sky. We find that the GAMA regions are $95.2$\% as dense as the SDSS superpopulation\footnote{In essence, the GAMA survey is a post-stratified sampling of the SDSS, with the GAMA regions a stratum of the entire SDSS area. The SDSS source density is a universal parameter of our superpopulation, and can be used to improve the accuracy of the total luminosity density estimation we make from the GAMA dataset.}, and therefore scale our $\phi^*$ parameters upwards by a factor of $\frac{1}{0.952}$. The area incompleteness of the $K$ band-defined sample is accounted for by calculating the normalisation volume with Area=$133.5$\,sq deg (the total coverage of the GAMA regions by $K$ band UKIDSS data), rather than the Area=$143.9$\,sq deg used for the other samples.

\subsection{Sample selection}
We limit our sample using our star-galaxy separation criteria and an apparent magnitude limit of $r \le 19.4$\,mag (imposed on the dereddened magnitude system used to calculate the luminosity function). We use a brighter apparent magnitude cut than that defined in Section \ref{sec:mlim} because $19.4$\,mag is the GAMA sample's target completeness limit over all three regions. Brighter than this limit our samples are 91.3\% spectroscopically complete (using $r_{AUTO}$). Our samples suffer greatly from spectroscopic incompleteness fainter than this magnitude limit. We impose a limit based on the spectroscopic limit and a colour limit (i.e. $19.4-(r_{SDSS}-r_{aperture})$). To remove the necessity of modelling the K or E corrections for each galaxy, we also impose a redshift limit of $0.0033 < z \le 0.1$. We adopt an evolution $\beta=0$ (where $E(z)=2.5\beta\log_{10}(1+z)$, setting $\beta=0$ denotes no evolution in this redshift range), and $K(z)=0.95z$ (following the $r$ band in \citealt{tex:hill}). We use the SDSS \textsc{EXTINCTION\_R} parameter to deredden all our photometric methods. We combine the data from the three GAMA regions, and treat them as one sample. Column 2 of Table \ref{tab:lfmagsamp} contains our sample sizes.

\begin{table*}
\begin{tabular}{cccccc} \hline \hline
Magnitude system&Sources&$M^{*} - 5 log_{10} h$&$\alpha$&$\phi^{*}$ ($\rm h^3 \text{} Mpc^{-3}$)&j ($\times$ $10^{8}$ $h$ $L_{ \astrosun }$ $Mpc^{-3}$)\\ \hline
\citet{tex:monterodorta}&-&-20.71&-1.26&0.0093&1.78\\
\citet{tex:blantonsdsslf}&-&-20.44&-1.05&0.0149&1.85\\
\citet{tex:hill}&-&-20.81&-1.18&0.0124&2.29\\
\hline
SDSS \textsc{petromag}&12599&-20.612$^{+0.031} _{-0.021}$&-1.076$^{+0.013} _{-0.010}$&0.0130$^{+0.0005} _{-0.0003}$&1.84$^{+0.12} _{-0.11}$\\
SDSS \textsc{modelmag}&12740&-20.812$^{+0.029} _{-0.023}$&-1.146$^{+0.011} _{-0.009}$&0.0111$^{+0.0004} _{-0.0003}$&1.99$^{+0.13} _{-0.12}$\\
r-defined AUTO&12292&-20.789$^{+0.035} _{-0.024}$&-1.111$^{+0.015} _{-0.009}$&0.0114$^{+0.0005} _{-0.0003}$&1.95$^{+0.15} _{-0.13}$\\
r-defined PETRO&12268&-20.818$^{+0.026} _{-0.034}$&-1.112$^{+0.010} _{-0.012}$&0.0113$^{+0.0003} _{-0.0004}$&1.98$^{+0.14} _{-0.13}$\\
K-defined AUTO&10855&-20.596$^{+0.029} _{-0.031}$&-1.063$^{+0.012} _{-0.013}$&0.0126$^{+0.0004} _{-0.0004}$&1.74$^{+0.12} _{-0.11}$\\
K-defined PETRO&11265&-20.699$^{+0.034} _{-0.029}$&-1.087$^{+0.013} _{-0.011}$&0.0123$^{+0.0005} _{-0.0004}$&1.90$^{+0.15} _{-0.14}$\\
self-defined AUTO&12284&-20.734$^{+0.033} _{-0.028}$&-1.097$^{+0.013} _{-0.011}$&0.0119$^{+0.0005} _{-0.0004}$&1.91$^{+0.15} _{-0.14}$\\
self-defined PETRO&12247&-20.781$^{+0.031} _{-0.028}$&-1.100$^{+0.012} _{-0.011}$&0.0117$^{+0.0004} _{-0.0004}$&1.97$^{+0.14} _{-0.14}$\\
S\'{e}rsic (TOTAL)&12711&-21.142$^{+0.038} _{-0.030}$&-1.203$^{+0.011} _{-0.009}$&0.0090$^{+0.0004} _{-0.0003}$&2.30$^{+0.19} _{-0.18}$\\
\hline
\end{tabular}
\caption{The number of sources that pass our star-galaxy separation criteria, redshift limit and $r \le 19.4$\,mag limit, depending on which magnitude system is used to define the $r$ band magnitude, with comparison luminosity function parameters from SDSS (\citealt{tex:monterodorta}, \citealt{tex:blantonsdsslf}) and SDSS+MGC \citep{tex:hill} defined samples. All magnitudes use the AB magnitude system, and have been dereddened using the \textsc{EXTINCTION\_R} SDSS parameter. $j$ statistics are calculated using $M_{\astrosun,r}$=4.71 from Table 1 of \citet{tex:hill}. Note that the comparison study samples have much brighter magnitude limits; 17.77\,mag in \citeauthor{tex:monterodorta}, 17.79\,mag in \citeauthor{tex:blantonsdsslf} and 18.76\,mag in \citeauthor{tex:hill}}
\label{tab:lfmagsamp}
\end{table*}

\subsection{The effects of surface brightness bias on the presented luminosity distributions}
Aperture selection can systematically bias the calculation of the luminosity distribution, particularly where a sample has a high surface brightness constraint (see \citealt{tex:nickbbd}, particularly their Figure 5, and \citealt{tex:cameronkron}). The SDSS photometric pipeline unfortunately is incomplete for $\mu_{r,50}>23$\,mag\,arcsec$^{-2}$ (see Section 3.4 of \citealt{tex:gamaic} and references therein). It follows that any spectroscopic survey that bases itself upon SDSS photometry, such as GAMA, will suffer from the same flaw. \citet{tex:nickbbd} have quantified the surface brightness dependency that the luminosity distribution inherently suffers from, and advise that a bivariate brightness distribution (BBD) is the best way to quantify, and remove, SB bias. That is beyond the scope of this paper, but this shall be explored in subsequent work. \citet{tex:nickbbd} point out that a sample that is complete to $\mu_{lim} \ge24$\,mag\,arcsec$^{-2}$ has very little uncertainty in its Schechter parameters due to SB selection effects, as the $L^{*}$ population that define the fitting are fully covered (see also section 4.1.2 of \citealt{tex:mgcbbd}). VST KIDS should provide such a catalogue. For now, however, we accept that the \textit{SDSS} input catalogue will not contain all faint, low surface brightness galaxies. The luminosity functions we present in this section are for samples that are surface brightness complete to $\mu_{r,50}<23$\,mag\,arcsec$^{-2}$, and suffer from varying levels of completeness between $23<\mu_{r,50}<26$\,mag\,arcsec$^{-2}$. As these luminosity functions are for a specifically low redshift sample, however, the effects of the surface brightness selection bias should be minimised.

\subsection{The effects of the aperture definition system on output Schechter parameters} \label{sec:lf}
\begin{figure*}
\includegraphics[width=440pt]{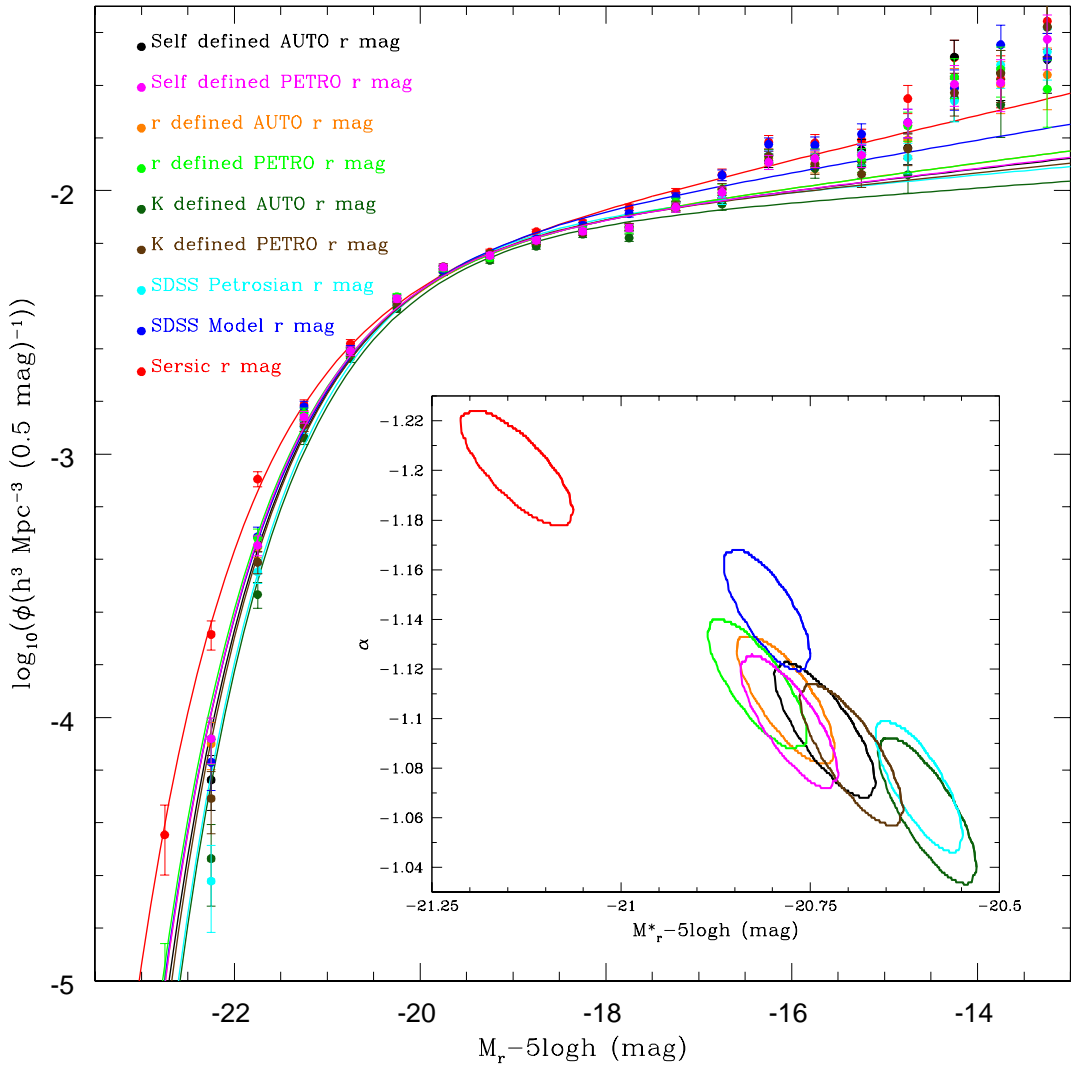}
\caption{Luminosity distributions, and the best fitting Schechter functions, calculated using different aperture definitions. Inset: 1 sigma chi-squared best-fit contours in the $M^{*}$-$\alpha$ plane. Errors on luminosity distribution points are Poissonian errors.}
\label{fig:lfs}
\end{figure*}

Figure \ref{fig:lfs} shows the luminosity distributions generated from different aperture systems, and illustrates how dependent the best-fitting luminosity function parameters are on the choice of aperture definition. The best fitting Schechter function parameters (calculated via $\chi^2$ minimisation) are shown in Table \ref{tab:lfmagsamp}.\\
The proximity of the $r$ and self-defined mag luminosity distributions signify that changing the \texttt{SExtractor} detection threshold (these catalogues utilise a detection threshold of 1.7 $\sigma$ and 1 $\sigma$ respectively) has a limited effect on the properties of a large sample ($M^{*} - 5 log_{10} h \pm 0.055$\,mag, $\alpha \pm 0.014$, $\phi^{*} \pm 0.0005$\,$\rm h^3 \text{} Mpc^{-3}$). There is an offset between the $K$ and $r$-defined best fitting luminosity functions. This is not caused by the cosmic variance in the missing area of the $K$ band sample; the best fitting Schechter function parameters vary only slightly when this is accounted for. By using the \textit{COVER\_BITWISE} flag, we can define a population of galaxies that are covered by $K$ band imaging. The best fitting Schechter parameters for an $r_{Auto}$ sample within area covered by $K$ band imaging (and normalised to the smaller volume) are $M^{*} - 5 log_{10} h$=-20.791\,mag, $\alpha$=-1.115, and $\phi^{*}$=0.0114 $\rm h^3 \text{} Mpc^{-3}$ - consistent with the area-complete LF within the uncertainty. It may be caused by a systematic alteration in the definition of the apertures used to calculate the flux of the galaxy population. \\
The best-fitting elliptical Kron and Petrosian aperture luminosity functions are similarly distributed in the $r$, $K$ and $self$-defined samples, indicating that the choice of light-distribution defined aperture does produce an offset that can be quantified. Whether the aperture is circular or elliptical is important. The SDSS \textsc{petromag} luminosity distribution should be similar to the $r$ and self-defined elliptical \textit{PETRO} distributions, but there is a noticeable $M^{*} - 5 log_{10} h$ offset ($0.20$\,mag; inset of Figure \ref{fig:lfs}). There is also a marked discrepancy between luminosity distributions calculated using total magnitude apertures (S\'{e}rsic and SDSS \textsc{modelmag}), and light-distribution defined apertures. The luminosity distributions of the former are overdense for faint galaxies ($M_{r} - 5 log_{10} h \ge -16$), and their best fitting power-law slopes are thus flatter.\\
The S\'{e}rsic luminosity distribution also measures higher densities of the brightest galaxies. This result is real, we have visually inspected the 139 galaxies that are distributed in the $-22 \le M_{r}- 5 log_{10} h \le -21.5$ magnitude bin and only 2 have suffered catastrophic failures\footnote{These profiles are viewable at http://star-www.st-and.ac.uk/$\sim$dth4/139eye/}. The remainder are generally well fit, though prominent spiral features do pose difficulties for the fitting algorithm. Of the 527 galaxies with $M_{r,Sersic} - 5 log_{10} h<-21$ mag within our redshift limited, apparent magnitude cut sample, only 8 have $m_{r,SDSS}-m_{r,Sersic}>0$. The marked discrepancy between the S\'{e}rsic $M^* - 5 log_{10} h$ parameter and that generated with the other samples ($0.33$\,mag brighter) is indicative of a scenario where galaxies are moved out of the magnitude bins near $M^* - 5 log_{10} h$, and into the brighter bins. $M^{*}$ is not an independent parameter, it is correlated with the other Schechter parameters, and accordingly the $\phi^{*}$ parameter has declined. The total luminosity density ($j$ in Table \ref{tab:lfmagsamp}), whilst $15\%$ higher, is consistent with that generated by the SDSS model magnitude within uncertainties. As our S\'{e}rsic magnitudes are not truncated, and the SDSS model magnitudes are truncated at $7 R_{e}$ for a de Vaucolouers profile / $3 R_{e}$ for an elliptical profile, the $M^{*}$ offset between these photometric systems is expected. We note that the residuals generated by this S\'{e}rsic-fitting process expose the requirement of multi-component galaxy decomposition; many galaxies have some structure within their central bulges that the S\'{e}rsic pipeline cannot model. A multi-component extension to this pipeline is detailed in \citet{tex:kelvin}, and that paper also examines any discrepancies between the S\'{e}rsic and multiple-component fitting results. No matter which aperture system we use, the luminosity distribution is overdense in the $M_{r}- 5 log_{10} h>-16$ magnitude bins when compared to the best fitting luminosity function. This indicates an upturn in the space density of galaxies at the dwarf-giant boundary, and the limitations of the single Schechter function fit.\\
As noted in the introduction, the Schechter parameters generated are for a sample that will suffer surface brightness incompleteness fainter than $\mu=23$\,mag\,arcsec$^{-2}$. In a future paper, we intend to account for this effect by undertaking a complete bivariate brightness analysis of the sample. The total luminosity densities we show here may therefore be systematically underdense due to surface brightness limitations (c.f. \citealt{tex:nickbbd}). It is also apparent that the simple Schechter function parameterisation is no longer a good fit for the luminosity distribution of galaxies at fainter magnitudes; there is an obvious upturn in each sample that is not being modelled. As the S\'{e}rsic photometric system is the only system that accounts for missing light, it is the most effective way of calculating the luminosity distribution.

\section{Summary}
The GAMA photometric pipeline has been designed to combine photometric data from a number of sources in a scientific and consistent manner. We have generated a series of large mosaics from imaging data taken from the SDSS and UKIDSS instruments using the \texttt{SWARP} utility, and undertaken optical and NIR photometry on an $r$ and $K$ band-defined sample of sources using \texttt{SExtractor}. We have also used a \texttt{Galfit} based pipeline (\texttt{SIGMA}) to generate S\'{e}rsic $ugrizYJHK$ magnitudes for all sources within the GAMA sample that pass our star-galaxy separation criteria.\\
We have created a set of r-defined source catalogues in the $ugrizYJHK$ passbands, calculated the apparent magnitude limits at which these samples are complete, and estimated their current redshift completeness. Whilst these samples do not contain as many sources as those used to calculate the SDSS luminosity functions (\citealt{tex:blantonsdsslf}, \citealt{tex:monterodorta}), they are complete to a greater depth. Our NIR apparent magnitude limited catalogues are of comparable size to those produced by the 6dFGS (\citealt{tex:hj}), with the decrease in coverage area matched by increasing depth. Our source catalogues have the advantage that they can be used to accurately compare the optical and NIR luminosity of any source within our sample. As the aperture definition is constant we can calculate exact colours. We have attempted to quantify the level of uncertainty in source flux due to the definition of the aperture. We find that the uncertainty is generally $\sim400$\% of the \texttt{SExtractor} output uncertainty (produced by background noise variations and $\frac{S}{N}$).\\
We have attempted to quantify the percentage of mismatches between our catalogues and the SDSS catalogues. We have taken a subset region of our images and tested their results against the SDSS detection software. We have detailed reasons for mismatches between these catalogues. They are generally an issue of different deblending decisions, artifacts, background noise or proximity to saturated stars. We find that there do seem to be some low surface brightness, compact objects that are missed by the SDSS detection software that are found by \texttt{SExtractor}, and some false SDSS detections. These faulty sources would be removed from the spectroscopic observing list by the extensive visual classification process undertaken by the GAMA team.\\
As our S\'{e}rsic magnitudes are generated from the images that the profile is calculated on, the colours produced from the S\'{e}rsic magnitudes are liable to be inaccurate and suffer from bias due to the aperture definition. We have defined a set of 'Fixed aperture' S\'{e}rsic magnitudes using the S\'{e}rsic $r$ band magnitude and colours from the $r$-defined \textit{AUTO} catalogue. We recommend the use of these magnitudes, called $X_{total}$. These magnitudes will provide both an accurate estimate of the luminosity of each galaxy, and optimal colours.\\
The GAMA photometric pipeline catalogue does not have the breadth of focus that the existing products from the SDSS and UKIDSS surveys have. As both the instrument gain (in SDSS only) and seeing change from frame to frame, our mosaics have variations within them that would cause errors in the calculation of stellar photometry. We have not attempted to deal with saturated objects, which causes uncertainty at the bright end of our catalogue, or masking of artifacts. These problems primarily affect stars, or are removed following the extensive visual classification program we have undertaken. For the task our catalogues were developed for, the accurate calculation of extended source apertures across instruments for a specific sample of extended sources, they are more adept than the existing tools. \\
Finally, we have generated $r$ band luminosity distributions and best-fitting luminosity functions from our population of galaxies using 9 different aperture definitions: SDSS circular Petrosian and Model magnitudes, $r$-defined elliptical Petrosian and Auto magnitudes, $K$-defined elliptical Petrosian and Auto magnitudes, $self$-defined elliptical Petrosian and Auto magnitudes, and an elliptical S\'{e}rsic total magnitude. We find that the $r$ and $self$-defined elliptical, light-distribution defined apertures produce similar results, indicating that the choice of detection threshold is unimportant. We find that there is a similar AUTO-PETRO offset in the $r$, $self$ and $K$-defined samples. We find that the use of circular apertures does have an effect on the best-fitting Schechter fit, with the SDSS \textsc{petromag} having a fainter $M^*$ parameter than the \texttt{SExtractor} samples. We also find that the use of total magnitude systems affects the slope of the luminosity function, with both the S\'{e}rsic and SDSS \textsc{modelmag} luminosity functions having a steeper $\alpha$ parameter. When we calculate the total luminosity density for each sample we find that using the S\'{e}rsic magnitude system gives us a higher value, approximately 15\% higher than samples that use other aperture definition systems. Following visual classification of a subsection of our sample, it is clear that this is not due to errors within our S\'{e}rsic magnitude calculation. We also note that the Schechter luminosity function does not provide a good fit at the faint end of the luminosity distribution, and a clear upturn at the dwarf-giant boundary is seen.

\section{Acknowledgements} 
DTH and LSK are funded by STFC studentships. ASGR is funded by a STFC grant. AMH acknowledges support provided by the Australian Research Council
through a QEII Fellowship (DP0557850).\\

Funding for the SDSS and SDSS-II has been provided by the Alfred P. Sloan Foundation, the Participating Institutions, the National Science Foundation, the U.S. Department of Energy, the National Aeronautics and Space Administration, the Japanese Monbukagakusho, the Max Planck Society, and the Higher Education Funding Council for England. The SDSS Web Site is http://www.sdss.org/. The SDSS is managed by the Astrophysical Research Consortium for the Participating Institutions. The Participating Institutions are the American Museum of Natural History, Astrophysical Institute Potsdam, University of Basel, University of Cambridge, Case Western Reserve University, University of Chicago, Drexel University, Fermilab, the Institute for Advanced Study, the Japan Participation Group, Johns Hopkins University, the Joint Institute for Nuclear Astrophysics, the Kavli Institute for Particle Astrophysics and Cosmology, the Korean Scientist Group, the Chinese Academy of Sciences (LAMOST), Los Alamos National Laboratory, the Max-Planck-Institute for Astronomy (MPIA), the Max-Planck-Institute for Astrophysics (MPA), New Mexico State University, Ohio State University, University of Pittsburgh, University of Portsmouth, Princeton University, the United States Naval Observatory, and the University of Washington.\\

The UKIDSS project is defined in \citealt{tex:ukirt}. UKIDSS uses the UKIRT Wide Field Camera (WFCAM; \citealt{tex:casali}) and a photometric system described in \citealt{tex:hewettpassband}. The pipeline processing and science archive are described in Irwin et al (2010, in preparation) and \citealt{tex:wfcam}.\\

\texttt{STILTS} is a set of command-line based catalogue matching tools, based upon the STIL library. It is currently being supported by its author (Mark Taylor), and can be downloaded from http://www.star.bris.ac.uk/$\sim$mbt/stilts/.\\

We thank Robert Lupton for his useful comments and suggestions when refereeing this paper.

\bibliography{gamav2}

\appendix 
\section{Variation between passbands} \label{app:var} 
\begin{figure*}
\includegraphics[width=440pt]{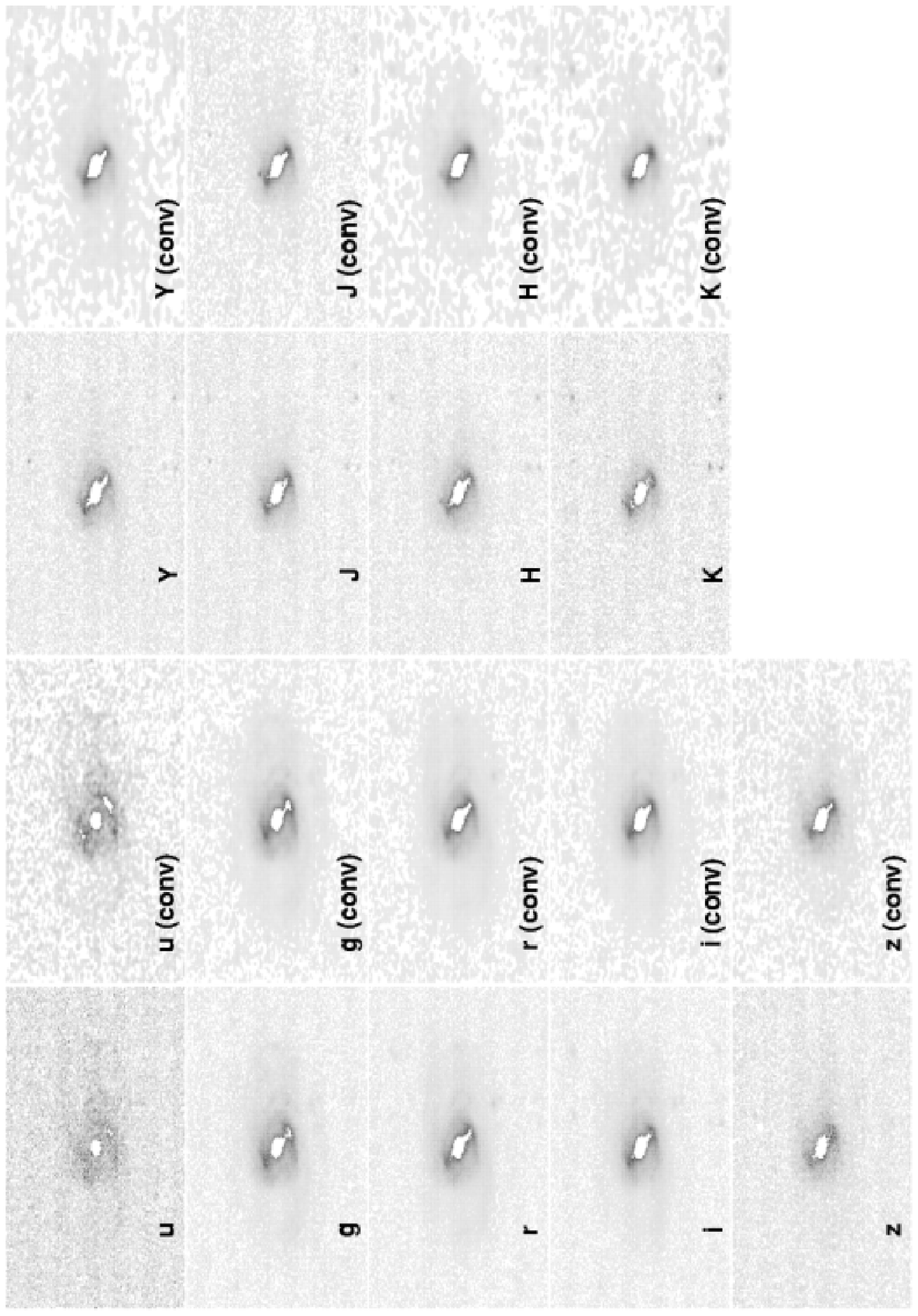}
\caption{The effects of convolution and the change in passband of observations of SDSS object 588848900968480848.}
\label{fig:objvar}
\end{figure*}

Figure \ref{fig:objvar} shows the 18 200x200 pixel images of the piece of sky containing SDSS object 588848900968480848; 9 cutouts from the standard image mosaics, and 9 from the convolved image mosaics. What is easily noticeable is that not only does the ability to see features of the object change dramatically between the $u$ (top left) and $K$ (bottom right) wavebands (spiral arms are visible in the optical, but in the $K$ band there only seems to be a bar and a bulge component), but that objects around it appear and disappear (a small blip to the SE in the $r$ band that may or may not be part of the object itself, at least 5 faint objects in the E of the frame in the NIR). The size of the object seems to halve from the $g$ band to the $J$ band, though this may be an effect of the image quality (the SDSS $g$ band should have a much smoother background than the UKIDSS $J$). The apparent magnitude of the object itself changes by 2.8 magnitudes from u to its peak in H (\texttt{SExtractor} calculates AB magnitudes using an $r$ band-defined AUTO aperture of $16.67$, $15.44$, $14.85$, $14.49$, $14.32$, $14.23$, $14.04$, $13.83$, $14.11$\,mag in $ugrizYJHK$). This is probably due to the decrease in dust opacity from the UV to the NIR.\\
The convolved images also show greater variation between the object and the background (these images all use a linear scale between the 99.5\% quantile pixel and 0, the background). For instance, the extended spiral arm to the left of the bulge in the $u$ band becomes slightly more apparent in the convolved $u$ band image. The size of the object in the convolved images generally looks larger than the standard images, though this again is probably due to the smoothing of the background making flux overdensities more apparent in the convolved images.

\label{lastpage}
\end{document}